\documentclass{article}
\usepackage{arxiv}
\usepackage{graphicx}
\usepackage{mathptmx}
\usepackage[subrefformat=parens,labelformat=parens]{subcaption}
\usepackage{authblk}
\usepackage{comment}
\usepackage[titletoc]{appendix}

\def\E{{\mathbb E}}

\usepackage{amsfonts}
\usepackage{algorithm}
\usepackage{amsbsy}
\usepackage{amsthm}
\usepackage{amsmath}
\usepackage{amssymb}

\def\E{{\mathbb E}}
\usepackage{algpseudocode}
\usepackage{bbm}
\usepackage{wrapfig}
\usepackage{booktabs}
\usepackage{multirow}
\usepackage{makecell}
\usepackage{wrapfig,lipsum,booktabs}
\usepackage{hyperref}
\allowdisplaybreaks
\usepackage{colortbl}
\usepackage{seqsplit}
\usepackage{xcolor}
\definecolor{KWgreen}{RGB}{112,173,71} 
\definecolor{KWblue}{RGB}{0,112,192} 
\definecolor{KWred}{RGB}{192,0,0} 
\definecolor{KWpurple}{RGB}{112,48,160} 
\usepackage{float}
\usepackage{siunitx}

\def\argmin{\mathop{\rm arg\,min}}%

\title{A Modular Mechanistic 
\textit{In Silico} Model for \textit{In Vitro} Transcription Process Yield and Product Quality Prediction
}

\begin{document}

\author[1]{Keqi Wang}
\author[1]{Keilung Choy}
\author[2]{Eli Reiser}
\author[1]{Jinxiang Pei}
\author[1]{Hua Zheng}
\author[3]{Aparajita Dasgupta}
\author[1]{Fuqiang Cheng}
\author[ ~,2]{Guogang Dong \thanks{Guogang.Dong@pfizer.com}}
\author[ ~,2]{Bhanu Chandra Mulukutla \thanks{BhanuChandra.Mulukutla@pfizer.com}}
\author[2]{Joshua Mannheimer}
\author[2]{Carolyn Huang}
\author[2]{Hooman Farsani}
\author[ ~,1]{Wei Xie \thanks{w.xie@northeastern.edu}}

\affil[1]{Department of Mechanical and Industrial Engineering, Northeastern University, 
360 Huntington Ave, Boston, MA 02115, USA}
\affil[2]{Bioprocess Research and Development, Pfizer Research and Development, 
1 Burtt Road, Andover, MA 01810, USA}
\affil[3]{Manufacturing Sciences and Technology, Pfizer Global Supply,
1 Burtt Road, Andover, MA 01810, USA}

\maketitle

\begin{abstract}
\textit{In vitro} transcription (IVT) plays a critical role in the manufacture of mRNA vaccines and therapeutics. Optimizing mRNA yield and ensuring product quality, such as capping efficiency and integrity, are essential but mechanistically complex. This study presents a modular mechanistic model of the IVT process to advance scientific understanding and improve predictive capability. The IVT reaction network is decomposed into interconnected modules describing (1) initiation and capping, (2) elongation and truncation, (3) termination and read-through, (4) mRNA degradation, (5) magnesium pyrophosphate precipitation, and (6) enzymatic degradation of pyrophosphate. 
Guided by biochemical principles and experimental data, kinetic models were developed for each module, accounting for mass balances, molecular complexation, and enzyme activity, and were subsequently assembled to capture coupled IVT dynamics.
Multivariate residual analysis and Shapley value-based sensitivity analysis—guided by domain knowledge—were applied to iteratively improve model fidelity. These machine learning–driven analytics enabled identification of key mechanisms, supported \textit{in silico} experimentation, and facilitated root-cause analysis. 
Combined with Gaussian-process-based batch Bayesian optimization for efficient parameter estimation, this framework establishes a scalable hybrid (mechanistic $+$ machine-learning) modeling platform that integrates heterogeneous data, accelerates model calibration, and supports rational design and optimization of mRNA manufacturing processes.

\end{abstract}

\keywords{
in vitro transcription (IVT), mRNA manufacturing, mechanistic modeling, modular kinetic model, Bayesian optimization, mRNA product quality attributes
}


\section{Introduction}
\label{sec:introduction}





The COVID-19 pandemic highlighted the transformative potential of messenger ribonucleic acid (mRNA)-based vaccines, which were the first to receive authorization in both Europe and the United States for combating SARS-CoV-2 \cite{barbier2022clinical}. Comirnaty and Spikevax, the two leading mRNA vaccines, represent a landmark achievement in biomedical science, playing a central role in mitigating the COVID-19 pandemic and reshaping the global response to infectious disease \cite{walsh2022biopharmaceutical}.
Compared to traditional vaccine production methods, which can be slow to respond to emerging viral outbreaks, the mRNA synthesis process via \textit{in vitro} transcription (IVT) is relatively simple and allows for flexible, rapid production of large quantities of mRNAs \cite{kis2020rapid,rosa2021mrna,gebre2021novel,boman2024quality,schmidt2022process}. 

The large-scale manufacturing of mRNA vaccines typically begins with plasmid generation, where the DNA template is generated via microbial fermentation, 
purified, and linearized. This is followed by the upstream process, in which mRNA is synthesized through the IVT reaction. 
The manufacturing process concludes with a downstream purification platform comprising of multiple steps, which may include DNase digestion, precipitation, chromatography, or tangential flow filtration \cite{kis2020rapid,rosa2021mrna}.
The IVT process, along with the quality of the DNA template, is pivotal in ensuring mRNA product integrity, stability, and the efficiency of antigen protein translation. While downstream purification steps effectively remove impurities and improve product quality, 
excessive impurities or degraded RNA species can complicate purification. 
This underscores the importance of improving the IVT process that minimizes impurity burdens and alleviates challenges to downstream purification \cite{dousis2023engineered}. 

This study focuses on developing a mechanistic model with a modular design 
to improve the characterization of the IVT processes for different mRNAs and enhance the prediction of yield and product quality attributes (PQAs). 
The mRNA synthesis is based on a DNA template, typically a linearized plasmid, containing a promoter region that is specifically recognized by an RNA polymerase (RNAP). RNAP synthesizes mRNA by incorporating nucleotides (NTPs) complementary to the coding strand of the DNA template. The most widely used RNAP for IVT is T7 RNAP, a magnesium-dependent enzyme \cite{martin1987kinetic,tabor1985bacteriophage,lodish2008molecular}.  
The incorporation of chemically modified bases, such as pseudouridine ($\Psi$), N1-methyl-pseudouridine (m$^1\Psi$), and 2-thiouridine (s$^2$U), into synthetic mRNAs has been shown to suppress the formation of double-stranded RNA (dsRNA), mitigate immune responses, and enhance the translational efficiency of the resulting mRNA \cite{mu2018origin,kariko2008incorporation,andries2015n1,chen2022n,baiersdorfer2019facile}.
IVT reaction mixtures often include additional components such as magnesium salts, typically in the form of magnesium chloride or magnesium acetate, spermidine, and dithiothreitol (DTT). These components stabilize the enzyme, maintain its activity, and enhance synthesis efficiency \cite{arnold2001kinetic,young1997modeling,boman2024quality,rosa2022maximizing}. Buffers like Tris (tris(hydroxymethyl)aminomethane) or HEPES (4-(2-hydroxyethyl)-1-piperazineethanesulfonic acid), titrated to the desired pH, are commonly used to neutralize hydrogen ions produced during the IVT reaction \cite{kern1997application}. 
To enhance stability and translatability and reduce immunogenicity of mRNA product, a 5$^\prime$ capping reagent, such as CleanCap (TriLink Biotechnologies, San Diego, CA, USA), may be added to create a cap structure at the 5$^\prime$ terminus \cite{vlatkovic2022ribozyme,ramanathan2016mrna}.

Creating a flexible modeling platform that can leverage prior knowledge, faithfully represent industrial-scale IVT production processes, and provide reliable predictions, has been an active area of development in recent years \cite{dousis2023engineered}. 
A deeper understanding of the underlying mechanisms 
on how key process parameters, such as reaction mixture composition, temperature, pH, and processing time, affect yield and mRNA product quality attributes, including capping efficiency and mRNA integrity, is essential to optimize the IVT processes. These efforts are crucial for enhancing flexibility, scalability, and reliability of mRNA manufacturing.


Numerous modeling approaches 
have been developed in literature focusing on different
steps of the IVT reaction, providing valuable insight into underlying mechanisms. These include initiation and abortive cycling, where RNAP repeatedly synthesizes and releases short RNA fragments without successfully transitioning to productive elongation  \cite{mcclure1980rate,xue2008kinetic,friedman2012mechanism}, as well as elongation \cite{bai2004sequence,douglas2020bayesian,tadigotla2006thermodynamic}, and termination \cite{von1991transcript,gusarov1999mechanism}. The existing studies on whole IVT process modeling can be broadly categorized into two primary streams: mechanistic models \cite{young1997modeling,arnold2001kinetic,akama2012multiphysics,van2021quality,wang2023stochastic,stover2024mechanistic,hengelbrock2024digital}, and purely data-driven approaches, such as Gaussian process (GP) surrogate models \cite{rosa2022maximizing} and multiple linear regression models \cite{boman2024quality}.

Young et al. (1997) \cite{young1997modeling} developed a mechanistic model for the IVT process that accounted for aborts—RNA molecules shorter than the desired product. The model effectively predicted the dynamic trajectories of critical states, including pH, NTP concentrations, and pyrophosphate-magnesium complexes.
Arnold et al. (2001) \cite{arnold2001kinetic} created a mechanistic model for the IVT process, deriving the transcript synthesis rate as a function of: (a) essential reactant concentrations, including T7 RNAP, its promoter, substrate nucleotides, and inhibitory byproducts such as inorganic pyrophosphate; (b) DNA plasmid characteristics, such as recognition sequences governing transcription initiation and termination; and (c) nucleotide sequence properties, including transcript length and composition. 
Building on these earlier studies, Akama et al. (2012) \cite{akama2012multiphysics} and Stover et al. (2024) \cite{stover2024mechanistic} refined the mechanistic model by further investigating magnesium pyrophosphate precipitation, enzymatic pyrophosphate degradation, and their impact on the IVT process.
Additionally, Van de Berg et al. (2021) \cite{van2021quality} and Hengelbrock et al. (2024) \cite{hengelbrock2024digital} proposed bioprocess mechanistic models to quantify and map the effects of process parameters on product quality attributes and effective product yield. This work further optimized the production process by analyzing the impact of reactant concentrations and temperature. 

Likewise, Rosa et al. (2022) \cite{rosa2022maximizing} and Boman et al. (2024) \cite{boman2024quality} utilized purely data-driven approaches to assess how process parameters—such as the concentrations of NTPs, magnesium, DNA templates, and RNAP—influence yield and dsRNA content, providing valuable insights into the underlying mechanisms of the IVT process.


However, compared to overall production levels, product quality attributes—such as mRNA integrity and 5$^\prime$ capping efficiency, which significantly influence the efficacy, stability, translational capacity, and immunogenicity of the mRNA product—have been relatively underexplored in terms of analysis, modeling, and prediction in existing literature.
Therefore, this study presents an IVT process mechanistic model  with a modular design, where each module characterizes key steps of IVT process, including (1) initiation, (2) elongation, and (3) termination, as well as parallel reactions such as (4) mRNA degradation, (5) magnesium pyrophosphate (Mg$_2$PPi) precipitation, and (6) enzymatic pyrophosphate (PPi) degradation. These modules are systematically interconnected to comprehensively represent the critical pathways leading to the synthesis of truncated, full-length, and extended transcripts, each with either capped or uncapped structures.
The developed model can advance the scientific 
understanding of the critical factors that determine the yield, integrity, and capping efficiency of the IVT product. 

The modular design 
makes it inherently straightforward to extend and incorporate additional process steps, parameters and quality attributes, enabling flexible assembly of IVT process models for manufacturing different mRNA vaccines and therapeutics. In addition, the developed mechanistic learning framework with a modular design enables the integration of different heterogeneous data, e.g., discrete and time-course data from batch and fed-batch experiments, as well as specific experimental data on each step studying molecule-to-molecule interactions and certain regulatory mechanisms. Each module can be independently refined and calibrated based on new insights or data from specifically designed experiments. For instance, the module characterizing enzymatic degradation of pyrophosphate can be updated using targeted specific experimental data.

Investigation of model prediction error patterns of the multivariate IVT process provides a valuable learning framework to iterative model improvements. An initial process model was constructed based on existing domain knowledge and subsequently refined through machine learning (ML)–driven data analytics, guided by biological understanding.
To improve the model’s representation of underlying mechanisms, systematic multivariate residual analysis and Shapley value sensitivity analysis \cite{shapley1953value} were employed.
Key dynamic features of the IVT process were identified by integrating domain expertise with data-driven modeling and Shapley value-based sensitivity analysis.
The model is then iteratively refined through training on heterogeneous datasets, including both discrete and time-course batch process data collected under varying conditions.
Additionally, systematic residual analysis is performed to ensure robust predictive performance while maintaining interpretability. Residual analysis involves examining the differences between observed and predicted values to identify systematic biases and, more importantly, to detect missing key features and mechanisms that significantly impact system dynamic prediction performance, guiding further model refinement. 

The developed IVT process mechanistic model is presented in Section~\ref{sec:IVTmodeling}. 
To address the challenges of parameter optimization—namely the high computational cost of IVT simulations, the high dimensionality of the parameter space, and the non‑convexity of the objective function when incorporating equilibrium assumptions for fast reactions—we implement a Gaussian process (GP)-based batch Bayesian optimization (BO) approach \cite{hunt2020batch}.
This strategy enables efficient exploration of the modular model design, leverages parallel computing, and accelerates the search for optimal parameters. Details on model fitting are provided in Section~\ref{sec:inference}.


The model prediction performance was assessed in Section~\ref{sec:results}. The mRNA yield predictions achieved a mean absolute error (MAE) of 0.95 g/L and a Spearman’s rank correlation coefficient of 0.94. Predictions for integrity and capping efficiency yielded MAE values of 4.01\% and 7.06\%, with corresponding correlation coefficients of 0.86 and 0.84, respectively.
Afterwards, comprehensive \textit{in silico} experiments were conducted to enhance the understanding of underlying mechanisms and assess the effects of process parameters on IVT production yield and quality attributes, the results of which are presented in Section~\ref{sec:sensivitiy}. 

In summary, the developed mechanistic model with a modular design was validated using well-designed laboratory experiments. Experimental descriptions are provided in Section~\ref{sec:datadescription}. This modeling strategy advances scientific understanding and demonstrates reliable and robust prediction performance. This modeling and analytics framework can guide experiments and data-driven learning focusing on the bottleneck, i.e., the regulatory mechanisms of the IVT reaction network and critical pathways with less knowledge. 

\section{Data Description and Analysis}
\label{sec:datadescription}

\subsection{Materials}
\label{subsec:materials}
Template DNA was provided internally, with DNA concentration confirmed by UV–Vis spectroscopy and DNA linearity percentages determined by agarose gel electrophoresis (AGE). T7 RNAP, ATP, CTP, GTP, UTP, p-UTP (pseudouridine-5$^\prime$-triphosphate), pyrophosphatase, RNase-free water, DNase I, Proteinase K, DMSO, and RNase inhibitor were purchased from Thermo Fisher. CleanCap AG was obtained from Trilink. 1 M Tris buffer (various pH values) and lithium chloride precipitation solution (pH 7.5) were obtained from Invitrogen. 1 M magnesium acetate, spermidine, phosphonoacetic acid TraceCert, DTT, RNase-free EDTA, and 3-(trimethylsilyl)propionic-2,2,3,3-d$_4$ acid sodium salt (TSP-d$_4$) were purchased from Sigma.

\subsection{IVT Methods}
\label{subsec:IVTmethods}
Batch \textit{in vitro} transcription (IVT) of RNA was performed using T7 RNAP to catalyze run-off reactions with a linear DNA template. Co-transcriptional capping was carried out using the CleanCap AG cap analog (Trilink) and sequence-modified templates \cite{henderson2021cap}. Vendors report activity as units per volume of enzyme solution. IVT reactions were initiated at defined starting activities (U/mL), obtained by diluting the vendor-supplied enzyme preparations to the desired concentration in the reaction mixture.

To support model development, data was collated and generated from a variety of experiments with ranges of reaction parameters, described in Table~\ref{tab:RangePara} in Appendix~\ref{appendix:RangePara} with the full list of IVT input parameters and the corresponding value ranges. 
The IVTs were performed across a range of scales and reaction systems, including (1) Tube reactions in Eppendorf tubes agitated and temperature controlled on a ThermoMixer; (2) 96 Well Plate reactions on the Tecan Fluent Liquid Handler with agitation and temperature controlled by the Bioshake unit; (3) Sartorius AMBR\textsuperscript{\textregistered} 15 and AMBR\textsuperscript{\textregistered} 250 single-use bioreactor systems which control agitation and temperature; and (4) Mettler-Toledo EasyMax Synthesis Workstation 100 mL reactors, with system pH and temperature monitoring and control. In some experiments, flow-NMR monitoring \cite{sarkar2024flow} was incorporated.

Manual pipetting of IVT reaction components, buffer, and reaction cofactors is required prior to reaction start in the AMBR\textsuperscript{\textregistered} 250 and EasyMax systems and is commonly completed for the AMBR\textsuperscript{\textregistered} 15. Automated preparation by the liquid handler is possible in the AMBR\textsuperscript{\textregistered} 15 if desired and is always completed in the Tecan Fluent method. The addition of T7 RNAP  and pyrophosphatase can be done manually or by the liquid handler; After addition to each reaction, the IVT progresses for the full incubation time. The reactions may then be processed through DNase I and ProK digestion steps before being harvested from the system. 

Some IVT conditions had time-course sampling in which reaction volume was removed and quenched with the EDTA chelating agent which would sequester the co-factor magnesium thereby halting transcription. RNA was precipitated by lithium chloride/ethanol precipitation, the mixture is centrifuged at 4$^\circ$C, the pellet then resuspended in nuclease free water.  

\subsection{Analytical Methods}
\label{subsec:analyticalmethod}

RNA concentration measured using either the ThermoFisher NanoDrop UV–Vis spectrophotometer or the Repligen CTech SoloVPE  variable pathlength UV–Vis spectrophotometer. RNA has a characteristic absorbance maximum at 260 nm, and according to Beer’s Law, the absorbance of an RNA solution is directly correlated with its concentration.

Separation of RNA species and subsequent measurement of mRNA integrity were performed using the Agilent Fragment Analyzer, an array-based capillary gel electrophoresis (CGE) system, using Agilent’s High Sensitivity RNA Kit \cite{patel2023characterization}. The smear analysis function in the Agilent ProSize software was used to integrate electropherograms quantitating \% area measurements, including (1) RNA integrity, defined as the \% area corresponding to full-length RNA transcripts, and (2) Late-migrating species (LMS), defined as the \% area of all species migrating after the full-length transcript.

Quantitation of 5$^\prime$-capping efficiency was performed following established methods \cite{patel2023characterization,beverly2016label}. The produced mRNA was annealed to a biotinylated probe complementary to the 5$^\prime$ end. Then treating the material with RNase H, the resulting duplex was separated from the larger RNA remnant using biotin-streptavidin affinity purification. The 5$^\prime$ end related species were chromatographically separated by reversed-phase high-performance liquid chromatography (RP-HPLC) using an Agilent 1260 system with quaternary pumps and DAD detector - while monitoring UV absorbance at 260 nm. An ACQUITY UPLC Oligonucleotide BEH C18 Column (130\AA, 1.7 $\mu$m, 2.1 mm $\times$ 150 mm, 1K - 30K) (Waters) held at 75 \r{C} was used. Relative quantification of the 5$^\prime$ end cap level (5$^\prime$-Cap) was determined from relative peak areas. The percentage of species that are capped was determined by dividing the percent area of the capped species by the total area of all 5$^\prime$ end related species. 

The concentration of residual NTPs and/or CleanCap AG were measured using anion exchange chromatography on an HPLC-UV system and the ProPac\textsuperscript{TM} SAX–10 HPLC Column. Component concentrations were quantitated by correlating relative area of each component with a standard curve. In some cases, reaction component concentrations were measured by flow-NMR quantitation \cite{sarkar2024flow}.  
Phosphate concentrations were measured using the Cedex Bio HT Analyzer from Roche and the Phosphate Bio HT kit. 



\subsection{IVT Process Performance Indicators (Output)}
\label{subsec:outputsMeasure}

In this study, five performance metrics are evaluated to assess IVT process, as summarized in Table~\ref{tab:IVT_inputs_outputs}. These are mRNA yield (g/L) and four PQAs: capping efficiency (\%), integrity (\%), LMS proportion (\%), and truncation proportion (\%).

\textbf{(1) Yield} represents the total concentration of mRNA transcripts, expressed in g/L, including truncated, full-length, and late-migrating species, regardless of their capping status. Since abortive transcripts are undetectable by the mRNA quantification method in this study, they are excluded from the yield measurement.

\textbf{(2) Capping Efficiency (CE)} measures the percentage of mRNA molecules that was capped, i.e., 
\begin{equation*}
\mathrm{CE}~(\%) =
\frac{
\#\text{ of capped (full-length, truncated, LMS)}
}{
\#\text{ of (full-length, truncated, LMS)}
}
\times 100.
\end{equation*}

\textbf{(3) Integrity} represents the percentage of mRNA that consists of the desired full-length species, regardless of whether it is capped or uncapped, as measured using CGE. It can be expressed as:
\begin{equation*}
   \text{mRNA Integrity (\%)} = \frac{\text{Mass of full-length mRNA}}{\text{Mass of total mRNA}} \times 100, 
\end{equation*}
where $\text{Mass of total mRNA} 
= \text{(Mass of full-length mRNA)} 
+ \text{(Mass of truncated transcripts)} 
+ \text{(Mass of LMS)}.$

\textbf{(4) Late Migrating Species Proportion} represents the percentage of mRNA that consists of larger mass species than the target mRNA, whether capped or uncapped, as measured using CGE. It can be expressed as:
\begin{equation*}
\text{LMS Proportion (\%)} = \frac{\text{Mass of LMS}}{\text{Mass of total mRNA}} \times 100.
\end{equation*}

\textbf{(5) Truncation Proportion} represents the percentage of mRNA that consists of truncated transcripts, whether capped or uncapped. This is calculated based on its relationship with Integrity and LMS Proportion as:
$\text{Truncation Proportion} (\%) = 100- \text{Integrity} (\%) - \text{LMS Proportion} (\%).$

\section{Modular Mechanistic Model Development for IVT Process}
\label{sec:IVTmodeling}

Beginning with an introduction to the IVT process reaction network in Section~\ref{subsec:IVT} and leveraging the inherent advantages of a modular design, the IVT process is decomposed into several interdependent steps, referred to as modules. 
Through the developed hybrid (mechanistic + machine learning) framework presented in Section~\ref{subsec:HybridFramework}, the mechanisms underlying each module are thoroughly explored and iteratively improved. The resulting kinetic rate models for each module are presented in detail in Section~\ref{subsec:DynamicModel}. To accurately capture and predict the dynamics of the IVT process, the key molecular species, including both free and complexed components, are identified. Their concentration relationships are constrained through mass balance and equilibrium equations, as detailed in Section~\ref{subsec:MassBalance}.

\subsection{
IVT Process Reaction Network}
\label{subsec:IVT}
\begin{figure*}[h]
    \centering
    \includegraphics[width=0.9\linewidth]{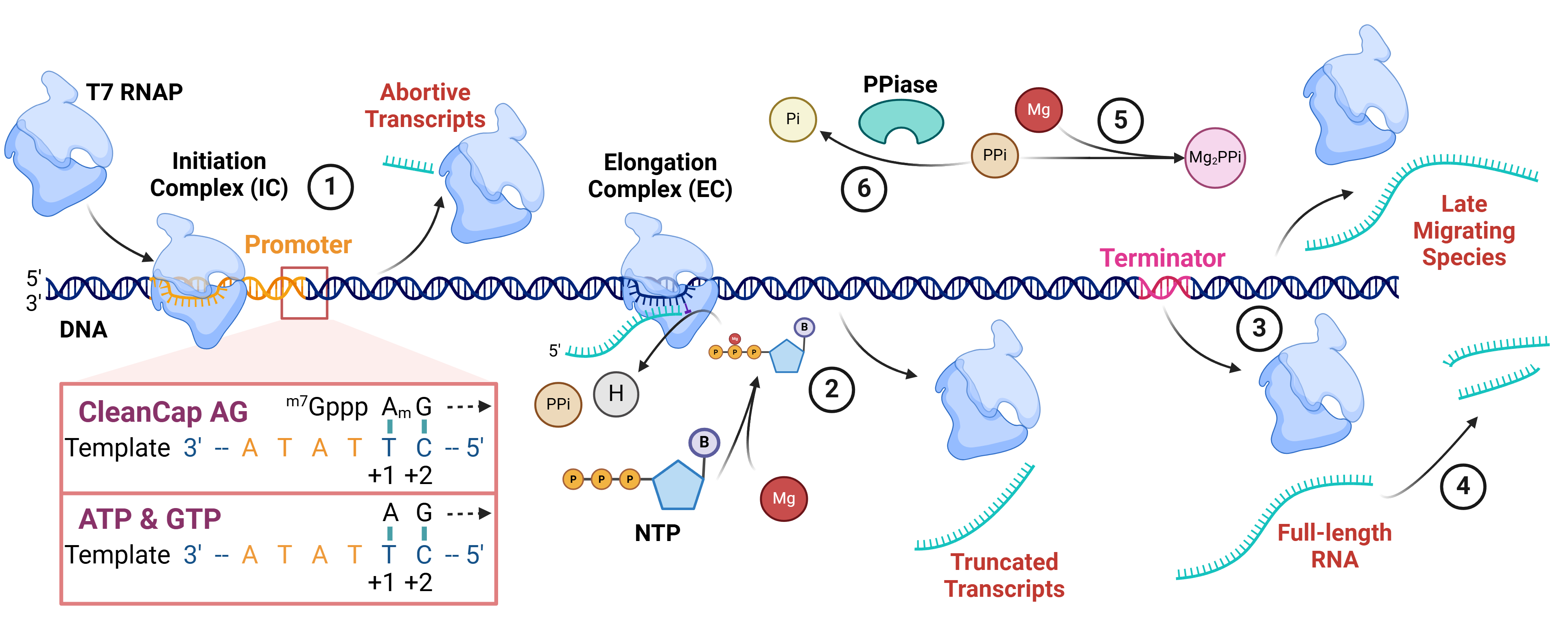}
    \caption{An overview of the molecular components and reaction pathways in the IVT network (Created with BioRender.com). This figure delineates the structured organization of reactions within the reconstructed IVT network, highlighting the following sequential steps: 
    \textcircled{\scalebox{0.9}{\raisebox{-0.5pt}{1}}} Initiation, Capping
    and Abortive Cycling; {\textcircled{\scalebox{0.9}{\raisebox{-0.5pt}{2}}} Elongation and Truncation;} \textcircled{\scalebox{0.9}{\raisebox{-0.5pt}{3}}} Termination and Read-through; \textcircled{\scalebox{0.9}{\raisebox{-0.5pt}{4}}} mRNA transcript degradation; \textcircled{\scalebox{0.9}{\raisebox{-0.5pt}{5}}} Mg$_2$PPi Precipitation; and \textcircled{\scalebox{0.9}{\raisebox{-0.5pt}{6}}} Enzymatic degradation of PPi. 
  {In this study, abortive cycling was excluded from the modeling focus because the abortive transcripts were not measured due to limitations in the analytical method.}
    }
    \label{fig:IVT_ReactionNetwork}
\end{figure*}

The IVT process primarily comprises three distinct steps: initiation, elongation, and termination \cite{dousis2023engineered}. Several parallel reactions occur simultaneously and interact with these main steps, including pyrophosphate (PPi) degradation (via pyrophosphatase), precipitation, and mRNA degradation, as depicted in Figure~\ref{fig:IVT_ReactionNetwork}. The mechanisms of each step and associated reaction were modeled separately, with their interactions captured through shared state variables. 
Depending on the progression of the process and environmental conditions, various branched pathways may be activated, leading to mRNA transcripts of varying lengths. These pathways include abortive cycling, truncation of elongating mRNA transcripts, and read-through of the termination region in the DNA template.

The IVT process parameters (input) and performance metrics (output) in this study are summarized in Table~\ref{tab:IVT_inputs_outputs} in Appendix~\ref{appendix:IVT_inputs_outputs}.
The inputs denoted by $\pmb{x}$ 
encompass the concentrations of Mg$^{2+}$ (mM), DTT (mM), ATP (mM), CTP (mM), GTP (mM), m$^1\Psi$TP (mM), CleanCap (mM), DNA (mg/mL), spermidine (mM), and Tris buffer (mM); enzyme activities of pyrophosphatase (PPase, U/mL) and T7 RNAP (U/mL); along with process parameters such as temperature ($^\circ$C), batch duration (min), initial pH, and DNA linearity (\%), defined as the percentage of linearized DNA template.
The multivariate outputs denoted by $\pmb{y}$, including yield (g/L) and several quality attributes, were modeled for the 
lab scale IVT process, based on whether the nascent transcript is properly capped at the 5$^\prime$ end, truncated during elongation, or correctly terminated without read-through at the termination stage. Specifically, except the yield, the quality attributes of interest include capping efficiency (\%), truncation proportion (\%), integrity (\%), and LMS proportion (\%); see the definitions 
in Section~\ref{subsec:outputsMeasure}.

Given any decision inputs $\pmb{x}$, the vector ${\pmb{y}}(\pmb{x})$ represents the IVT process outputs of interest. The mechanistic model predictions, $\hat{\pmb{y}}(\pmb{x}, \pmb{\theta}),$ depend on (1) the selection of inputs $\pmb{x}$, including: raw materials, process parameters, and other decision variables; and (2) mechanistic model parameters $\pmb{\theta}$, which characterize the 
state-dependent mechanisms governing the reaction rates and further impact overall process dynamics. For time-course data, the outputs capture the production trajectory. The mechanistic model structure and parameters are learned through the hybrid framework described in Section~\ref{subsec:HybridFramework} and via Gaussian process assisted Bayesian optimization presented in Section~\ref{sec:inference}.

\subsection{A Mechanism-Informed Machine Learning Framework for Model Development}
\label{subsec:HybridFramework}

\begin{figure}[h!]
    \centering
    \includegraphics[width=0.45\linewidth]{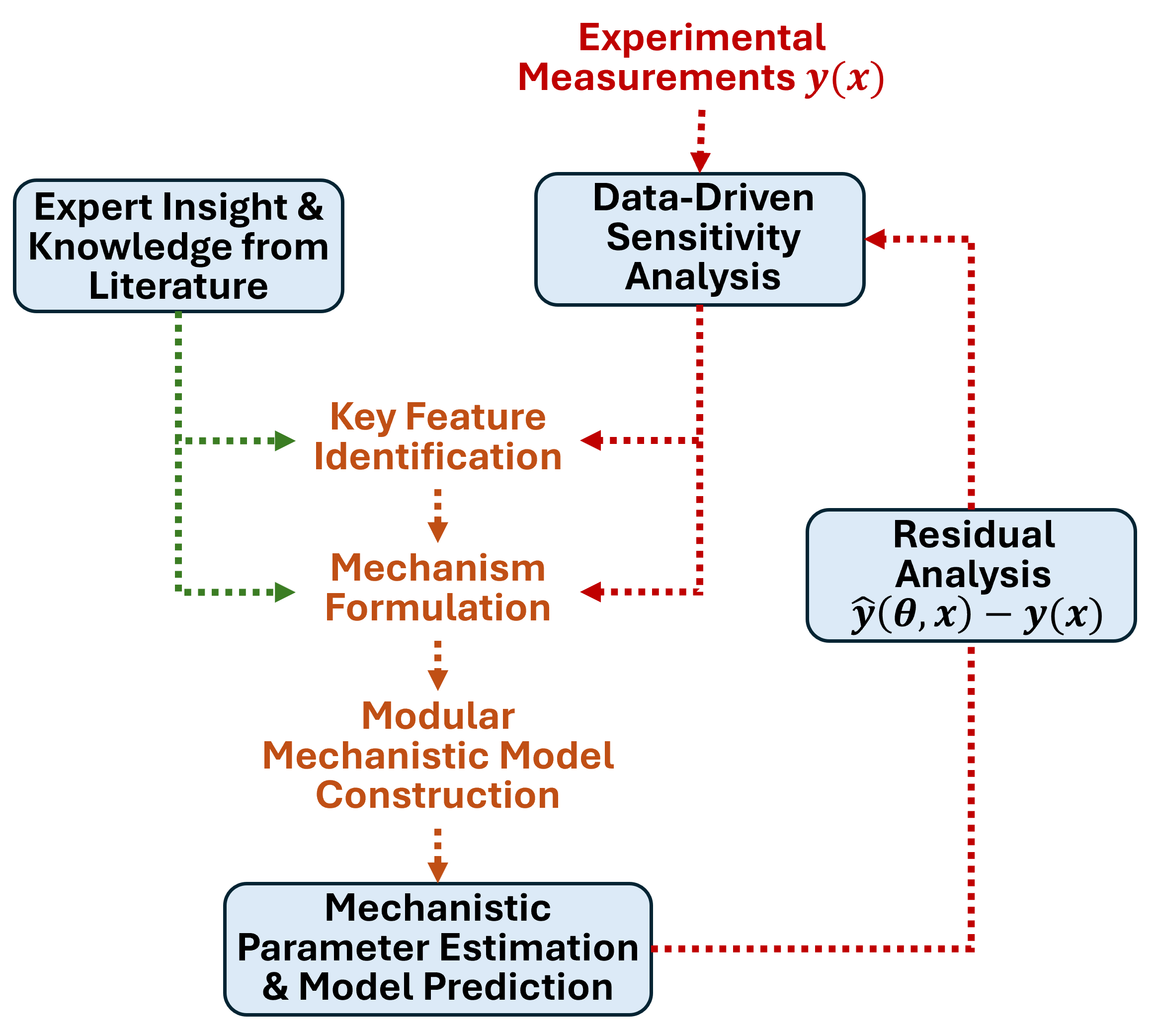}
    \caption{A workflow of the developed hybrid (mechanistic and machine learning) framework for data-informed, mechanism-driven model development and refinement.
    Experimental measurements $\pmb{y}(\pmb{x})$ are used in Shapley-based feature analysis to identify critical drivers of IVT process outcomes. These key features, together with expert knowledge and literature-derived mechanisms, guide formulation of plausible biochemical modules. A mechanistic model is constructed and used to simulate system behavior $\hat{\pmb{y}}(\pmb{x}, \pmb{\theta})$. Model prediction residuals $\hat{\pmb{y}}(\pmb{x}, \pmb{\theta})-\pmb{y}(\pmb{x})$ are analyzed to identify systematic deviations, which inform iterative refinement through targeted experiments and expert reinterpretation.
    }
    \label{fig:digital_twin}
\end{figure}

This section outlines the hybrid (mechanistic and machine learning) modeling and learning framework used to develop and refine the IVT process mechanistic \textit{in silico} model. The approach applies a mechanism-informed, data-guided methodology to progressively construct and enhance a modular mechanistic representation, as shown in Figure~\ref{fig:digital_twin}. The text details key steps for iterative improvement of the mechanistic IVT process model, including feature selection, formulation of mechanistic modules, mechanistic model construction, and analysis of prediction errors through sensitivity and residual analyses. 

\textbf{(1) Select the important features to identify key process parameters.} The procedure begins with feature importance analysis to identify key process parameters influencing mRNA yield and quality attributes—particularly capping efficiency and integrity, which remain underexplored in current literature. A Shapley value-based sensitivity analysis is applied to the XGBoost model \cite{chen2016xgboost} to quantify the contribution of each input $\pmb{x}$ to each output $\pmb{y}$, enabling the generation of scientific knowledge driven hypothesis about the potential underlying mechanisms. These hypotheses are then cross-referenced with subject matter expertise and literature findings, and, when necessary, validated through targeted experiments. Shapley value-based sensitivity analysis results about process parameters impacts on system performance metrics and detailed discussions can be found in Appendix~\ref{appendix:SV}.

\textbf{(2) Assemble the IVT process mechanistic model with interpretable mechanistic modules.}
Based on existing knowledge of the IVT process mechanisms and insights from data analytics, mechanistic modules are formulated upon fundamental kinetic models, such as Michaelis–Menten (M–M) kinetics \cite{kyriakopoulos2018kinetic} and the ordered sequential binding kinetic model \cite{leskovac2007comprehensive}, while taking into account potential competitive inhibition based on biochemical reactions within each module. These modules are implemented within a modular architecture to construct a mechanistic model that captures the underlying mechanisms and dynamic behaviors of the IVT reaction network, thereby enhancing interpretability and enabling conditionally independent refinement. A brief introduction to M–M kinetics and the ordered sequential binding kinetic models, accounting for molecular interactions, is provided in Appendix~\ref{appendix:ratemodel}.



\textbf{(3) Prediction error and residual analysis to identify critical missing mechanisms and iteratively improve the IVT process mechanistic model.} In the developed hybrid modeling and learning framework, the underlying mechanism of the IVT process is represented by two parts: (1) a mechanistic model with a modular design; and (2) random residuals due to the limitations of the current mechanistic model.
To iteratively improve model fidelity and prediction accuracy, the spatiotemporal dependence of the residuals, defined as the difference between model predictions and experimental measurements, denoted by $\pmb{e}(\pmb{x},\pmb{\theta})=\hat{\pmb{y}}(\pmb{x}, \pmb{\theta})-\pmb{y}(\pmb{x})$ given the same inputs $\pmb{x}$, is carefully analyzed using techniques similar to those applied during raw data exploration. This includes sensitivity-based diagnostics and feature attribution methods that reveal critical missing mechanisms contributing to systematic deviations or prediction errors, which then inform further updates of the mechanistic model. The iterative refinement process continues until the model adequately captures the observed dynamics and dependence of multivariate IVT process trajectory outputs, 
balancing mechanistic interpretability with predictive accuracy across a range of experimental conditions.

\subsection{Mechanistic Model with a Modular Design for IVT Process Dynamics}
\label{subsec:DynamicModel}






The IVT process dynamics, derived from the hybrid modeling and learning framework described in Section~\ref{subsec:HybridFramework}, are represented using a modular design, as depicted in Figure~\ref{fig:IVT_Process} in Appendix~\ref{appendix:IVT_Process}, with each of the six modules correspond to a specific step or parallel reaction.
The reactions and the corresponding kinetic rate models for $i$-th module ($i = 1, 2, \ldots, c$,  with $c=6$) are detailed in Sections~\ref{subsubsec:initiation}--\ref{subsubsec:PPi Enzymatic Degradation}, with the enzyme stability model—accounting for pH, temperature, and DNA linearity—discussed in Section~\ref{subsubsec:pH-enzyme}. Since the enzyme stability model does not involve any reaction or possess an associated stoichiometric matrix, it is treated as a supplementary factor rather than a separate module in this study.
These modules are interconnected through shared state variables and the transition rules that govern step changes in the IVT process. 
In addition, since the developed mechanistic model builds on RNA bases (i.e., A, U, C, G) and explicitly considers the length and composition of DNA templates, it can support the flexible manufacturing of different mRNAs.




Specifically, the dynamics of IVT process state $\{\pmb{s}_t\}$ is modeled by continuous-time differential equations, i.e.,
\begin{equation}
    \frac{d\pmb{s}_t}{dt} = \pmb{N} \pmb{v}(\pmb{s}_t;\pmb{\theta})
    = \sum_{i=1}^c\pmb{N}_i  \pmb{v}_i(\pmb{s}_t;\pmb{\theta}),   
    \label{eq.ODE_Dynamics}
\end{equation}
where $\pmb{N}$ is an $m \times n$ stoichiometry matrix representing the structure of the IVT reaction network, with $m$ denoting the number of molecular species and $n$ the number of reactions. The $(p,q)$-th element of $\pmb{N}$, denoted as $\pmb{N}(p,q)$, specifies the number of molecules of the $p$-th species consumed (negative value) or produced (positive value) in each occurrence of the $q$-th reaction. These values serve as the stoichiometric coefficients for the $p$-th species in the rate equation of the $q$-th reaction.  Vector $\pmb{v}$ is the corresponding $n$-dimensional reaction rate vector, which depends on the state variables $\pmb{s}_t$—including concentrations of relevant substrates, by-product inhibitors, and environmental factors such as pH—and it is specified by the mechanistic parameters $\pmb{\theta}$. 

To align with the modular design of the developed mechanistic framework, the stoichiometry matrix $\pmb{N}$ of the IVT reaction network is decomposed into smaller submatrices, denoted as $\pmb{N}_1, \pmb{N}_2, \dots, \pmb{N}_c$,  with $c=6$; see Figure~\ref{fig:IVT_Process}. Each submatrix corresponds to a specific module of the IVT process, including the (1) initiation, (2) elongation, and (3) termination steps, as well as the parallel reactions: (4) mRNA degradation, (5) Mg$_2$PPi precipitation, and (6) enzymatic PPi degradation.
The kinetic rate model describing the regulatory mechanism of the $i$-th module is denoted by $\pmb{v}_i(\pmb{s}_t;\pmb{\theta})$ for $i=1,2,\ldots,c$. 
The rate models presented in Sections~\ref{subsubsec:initiation}--\ref{subsubsec:pH-enzyme} are constructed based on established mechanistic principles and kinetic expressions from the literature \cite{akama2012multiphysics,arnold2001kinetic,van2021quality}, with 
further improvements informed by experimental data and domain expertise. For modules lacking direct experimental measurements in this study, the corresponding kinetic parameter values are adopted from existing literature and noted in Table~\ref{tab:parameterestimation} in Appendix~\ref{appendix:parameter}.

For simulation implementation and \textit{in silico} experiments of the developed mechanistic model, the Euler method \cite{braun1983differential} was employed to approximate solutions of these intractable ODEs in Equation~(\ref{eq.ODE_Dynamics}) 
due to its computational efficiency, which makes it well-suited for large-scale simulations across hundreds of batches in the IVT process model. However, higher-order solvers, such as the Runge–Kutta method, could also be used when higher numerical accuracy is required.

\subsubsection{Initiation, Capping, and Abortive Cycling} 
\label{subsubsec:initiation}

During the initiation step of transcription, the RNAP (E) binds to the DNA template (D) at the promoter region, forming an initiation complex that catalyzes the synthesis of mRNA \cite{cheetham1999structure,cheetham1999structural,martin1987kinetic,martin1988processivity,mcclure1980rate}. Rather than incorporating GTP as the initiator nucleotide at the +1 position of the synthesized mRNA molecule \cite{oakley1977structure,rong1998promoter,kuzmine2003binding,hengelbrock2024digital}, the sequence is typically modified to accommodate the co-transcriptional cap analog CleanCap AG \cite{CleanCap}. As shown in Figure~\ref{fig:IVT_ReactionNetwork}, the initial nucleotide incorporation step at the 5$^\prime$ end determines whether the mRNA transcript is capped or uncapped, depending on whether CleanCap AG or ATP and GTP bind first at the $+1$ and $+2$ positions.
  

If CleanCap AG binds first, simultaneously occupying both the +1 and +2 positions, the reaction proceeds as:
\begin{equation}
\begin{aligned}
\text{E} + \text{D} + & \text{MgCleanCap}
\rightleftharpoons \text{E}\cdot\text{D}\cdot\text{CleanCap} + \text{Mg}^{2+} 
\rightarrow \text{E}\cdot\text{D}\cdot \text{M}^{\text{capped}}_2 + \text{Mg}^{2+}
\nonumber 
\end{aligned}
\end{equation}
where $\text{M}^{\text{capped}}_2$ represents an mRNA transcript of length 2, already capped, and $\text{E} \cdot \text{D} \cdot \text{M}^{\text{capped}}_2$ denotes the enzyme-DNA-RNA complex at the early stage of transcription. 

Meanwhile, in the uncapped pathway, ATP and GTP compete with CleanCap AG for incorporation. If ATP binds first at the $+1$ position, followed by GTP at $+2$, the reaction leads to an uncapped mRNA transcript, i.e.,
\begin{equation*}
\begin{aligned}
\text{E} + \text{D} + & \text{MgATP}
\rightleftharpoons \text{E}\cdot\text{D}\cdot\text{ATP} + \text{Mg}^{2+}  \rightarrow \text{E}\cdot\text{D}\cdot \text{M}^{\text{uncapped}}_1 + \text{Mg}^{2+} \\
\text{E}\cdot \text{D}\cdot \text{M}^{\text{uncapped}}_1 + & \text{MgGTP}
\rightleftharpoons \text{E}\cdot\text{D}\cdot \text{M}^{\text{uncapped}}_1\cdot\text{GTP} + \text{Mg}^{2+} \rightarrow \text{E}\cdot\text{D}\cdot \text{M}^{\text{uncapped}}_2
+ \text{PPi} + \text{H}^+ + \text{Mg}^{2+}
\end{aligned}
\end{equation*}
where $\text{M}^{\text{uncapped}}_1$ and $\text{M}^{\text{uncapped}}_2$ represent an uncapped mRNA transcript of length 1 and 2, respectively. Because ATP binds first in this pathway, the mRNA lacks a 5$^\prime$ cap, which might 
lower mRNA stability, and compromise translational efficiency \cite{CleanCap}. To simplify notation, the superscripts distinguishing capped and uncapped mRNA will be omitted in the presentation of the following reactions.
  
  


The initiation complex is inherently unstable and frequently produces short mRNA transcripts, known as abortive transcripts, ranging from 2 to 10 nucleotides in length (the critical length may vary for different promoters), in a process termed abortive cycling \cite{martin1988processivity,dousis2023engineered}. Once the transcript exceeds approximately 10 nucleotides in length, the complex undergoes significant conformational changes, forming a stable and processive elongation complex \cite{dousis2023engineered}. During this transition, the complex escapes the DNA promoter region and releases the contracted upstream DNA \cite{xue2008kinetic,bandwar2007transition}.  
{In summary, the competition between ATP and GTP with CleanCap AG incorporation, along with abortive transcription and the transition to productive transcription, collectively determines mRNA quality.}

To accurately model the dynamics of capped and uncapped transcript synthesis and predict capping efficiency, the binding and initiation rates for both pathways, denoted as $v^\text{initial-cap}$ and $v^\text{initial-uncap}$, are formulated using distinct kinetic models. The capped pathway follows the M-M kinetic framework \cite{kyriakopoulos2018kinetic}, as CleanCap AG binds as a single molecule occupying both the $+1$ and $+2$ positions simultaneously, behaving as a single-substrate reaction. In contrast, the uncapped pathway follows an ordered sequential binding model \cite{leskovac2007comprehensive}, since ATP must bind first at $+1$, followed by GTP at $+2$ (see Figure~\ref{fig:IVT_ReactionNetwork}). The kinetic rate models governing these processes are provided in
Appendix~\ref{appendix:kinetic_models}, Equations~(\ref{eq.inital-cap})
and~(\ref{eq.inital-uncap}),
where $\prod_q act_{\text{T7},q}$ captures the dependency of T7 enzyme activity on critical factors, including pH ($act_\text{T7,pH}$), temperature ($act_\text{T7,Temp}$), and DNA linearity ($act_\text{T7,linearity}$). Each activity term is a dimensionless scaling factor ranging from 0 to 1. These enzyme activity terms follow the general model structures defined 
in Section~\ref{subsubsec:pH-enzyme}. 
The parameters $K_{\text{I}}$, $K_{\text{M}}$, and $k_{\text{T7}}$ correspond to the M–M inhibition constant, substrate affinity constant, and specific reaction rate, respectively. 
For the sequential binding of ATP and GTP, to account for competitive inhibition effects from other NTPs and CleanCap, the apparent Michaelis constants are defined
as Equations~(\ref{eq.Km_ATP_prime_init}) and~(\ref{eq.Km_GTP_prime_init}).

Therefore, the binding and initiation kinetic rate models presented in Equation~(\ref{eq.inital-cap}) are designed to account for several key factors: T7 RNAP binding to the promoter, modulation by the initial binding of ATP and GTP or CleanCap, competition among substrate nucleotides, the concentrations of 
free Mg$^{2+}$ cofactor, and the environmental impacts on T7 RNAP enzymatic activity (such as pH and temperature). 

Magnesium plays a critical role in the IVT process, as highlighted by Thomen et al. (2008) \cite{thomen2008t7} and Gunderson et al. (1987) \cite{gunderson1987interactions}. First, magnesium complexes with nucleotides to form substrates for the growing mRNA chain \cite{osumi1992asp537,young1997modeling}. Second, magnesium ions are believed to bind to the polymerase once the incoming magnesium-nucleotide complex has bound, as the interaction between metallic ions and the polymerase is facilitated by the triphosphate group of the nucleotide \cite{thomen2008t7}. However, Mg$^{2+}$ can also bind to existing MgNTP complexes to form species such as Mg$_2$NTP, thereby reducing the availability of functional reactants.

In addition, as reported by Young et al. (1997) \cite{young1997modeling} and Nikolaev et al. (2019) \cite{nikolaev2019systems}, the synthesis of abortive and full-length transcripts occurs at approximately the same order of magnitude when considered in terms of molarity. In other words, the larger the size of the full-length transcript, the less significant the relative contribution of nucleotides incorporated into abortive transcription products \cite{arnold2001kinetic}. 
Unlike the negligible waste of NTPs, the cost associated with the waste of CleanCap is a significant concern, as each abortive and full-length transcript consumes one CleanCap molecule. However, in this study, abortive cycling was excluded from the mechanistic model because the abortive transcripts were not measured due to limitations in the experimental assay method. 
Accurate monitoring and prediction of CleanCap concentration dynamics remain crucial for improving capping efficiency predictions and optimizing the IVT process. Future research leveraging advanced assay tools and molecular dynamics study will enable detailed modeling of abortive cycling, further enhancing the understanding of CleanCap dynamics.

\subsubsection{Elongation and Truncation} 
\label{subsubsec:elongation}

Throughout the elongation step, RNAP incorporates magnesium-complexed NTPs into the growing mRNA chain in accordance with the DNA coding sequence. Each incorporation forms a phosphodiester bond between the extending mRNA molecule and the incoming NTP. This reaction simultaneously releases a pyrophosphate ion (PPi or P$_2$O$_7^{4-}$) and a hydrogen ion (H$^+$), which increases the proton concentration in the solution, thereby decreasing pH \cite{young1997modeling,van2021quality}. Consequently, buffers such as Tris and HEPES are required to neutralize free H$^+$ ions and maintain pH within their effective buffering ranges.
Thus, the overall enzymatic mRNA synthesis reaction during the elongation step of the IVT process can be represented as:
\begin{equation*}
\begin{aligned}
\text{E}\cdot \text{D}\cdot \text{M}_j + & \text{MgNTP}
\rightleftharpoons \text{E}\cdot \text{D}\cdot \text{M}_j\cdot \text{NTP}
+ \text{Mg}^{2+} \rightarrow \text{E}\cdot \text{D}\cdot \text{M}_{j+1}
+ \text{PPi} + \text{H}^{+} + \text{Mg}^{2+}
\end{aligned}
\end{equation*}
where $\text{M}_j$ represents a nascent mRNA transcript of length $j$ with $j=1,2,\ldots, J$ and $J$ represents the full length of the target mRNA transcript.

To ensure generalizability across DNA templates with varying sequences and compositions, which allows flexible manufacturing of different mRNAs, the elongation rate model is based on a single-step polymerization process. The polymerization rate is individually modeled for each type of NTP, denoted as
$v_\text{ATP}^\text{elongate}$, $v_\text{UTP}^\text{elongate}$, $v_\text{CTP}^\text{elongate}$, and $v_\text{GTP}^\text{elongate}$, as defined by
Equations~(\ref{eq.elongate_ATP})--(\ref{eq.elongate_GTP}), 
where [EC] denotes the concentration of the active elongation complex formed by the T7 RNAP bound to the DNA template and the growing  transcript.
Therefore, the kinetic rate models describe the elongation rate influenced by the concentrations of the elongation complex, free Mg$^{2+}$ cofactor, and Mg-complexed NTP substrates, as well as competitive inhibition among substrate nucleotides and environmental factors affecting T7 RNAP enzymatic activity. 

The synthesis rate of the full-length transcript, denoted by $v_\text{full}^\text{elongate}$, is determined based on the NTP composition of the DNA template, which specifies the number of each type of NTP required during the polymerization process. This relationship is mathematically expressed as
Equation~(\ref{eq.elongate_full}),
where $C_\text{ATP}$, $C_\text{UTP}$, $C_\text{CTP}$, and $C_\text{GTP}$ represent the number of ATP, UTP, CTP, and GTP molecules required according to the DNA template sequence. 
By incorporating the NTP composition of the DNA template and mimicking the consumption dynamics of each type of NTP, the IVT process model enhances the prediction of yield and quality attributes. This approach also facilitates the identification of the limiting NTP, as well as the development of fed-batch processes and optimal control strategies (such as feeding strategy and pH control).

During the elongation step of transcription, particularly in batch-based IVT processes, RNAP progression can be hindered by obstacles such as the limitation or depletion of MgNTP, as reported by Hengelbrock et al. (2024) \cite{hengelbrock2024digital}, resulting in premature termination and the production of truncated mRNA transcripts. The production of truncated transcripts reduces the integrity of the final mRNA product. 
In addition, the formation of secondary structures and higher-order structures (3D conformations) in the nascent mRNA can influence the complex interactions of bio-molecules (e.g., RNA, DNA, and T7 enzyme) and further impact polymerase progression \cite{tadigotla2006thermodynamic}, as well as sequence-dependent kinetic factors, such as backtracking or pre-translocation pauses \cite{bai2004sequence,douglas2020bayesian}, which could destabilize the transcription elongation complex and potentially result in the production of truncated transcripts.

In this study, a baseline ratio parameter, $k_\text{ratio}$ in Equation~(\ref{eq.elongate_trunc}), is introduced to quantify the flux rate ratio between truncated transcripts and longer transcripts (i.e., full-length transcripts and LMS), capturing the potential influence of sequence and structural effects on energy barriers associated with different reaction pathways, which will be further investigated and explicitly modeled in future research.
In addition, building on the observations and analyses discussed in Appendix~\ref{appendix:SV}—which are also consistent with findings reported by Hengelbrock et al. (2024) \cite{hengelbrock2024digital}—substrate limitation, particularly the minimum concentration of Mg-complexed NTPs, is identified as a key factor contributing to transcript truncation. Thus, this factor is further incorporated into the truncated transcript synthesis rate model, as expressed in
Equation~(\ref{eq.elongate_trunc}),
where $[\mbox{MgNTP}]_{\min} = \min ([\mbox{MgATP}], [\mbox{MgUTP}], [\mbox{MgCTP}], [\mbox{MgGTP}])$ denotes the concentration of the most depleted MgNTP species in the reaction. 


\subsubsection{Termination and Read-through} 
\label{subsubsec:termination}

At the termination step of transcription, the interaction between RNAP and the DNA template is disrupted either upon encountering a terminator sequence \cite{yager1991thermodynamic,von1991transcript,mairhofer2015preventing} or upon reaching the end of a linearized DNA template, a process known as ``run-off transcription'' \cite{dousis2023engineered}. This disruption ceases the addition of complementary NTPs to the mRNA strand, resulting in the production of full-length mRNA and signifying the end of the \textit{in vitro} transcription process, i.e.,
\begin{equation}
    \text{E} \cdot \text{D} \cdot \text{M}_J {\rightarrow} \text{E} + \text{D} + \text{M}_J,
    \nonumber
\end{equation}
where $\text{M}_J$ stands for the full-length mRNA transcript with length $J$. 
However, when RNAP encounters DNA templates that are not fully linearized, it may extend transcription beyond the intended termination point, leading to the production of overextended transcripts. These transcripts, referred to as ``late migrating species (LMS)," can be detected using gel electrophoresis due to larger size.  

Depending on the linearity of the DNA template and key environmental conditions—particularly the concentration of free Mg$^{2+}$ ions, as discussed in Appendix~\ref{appendix:SV}—T7 RNAP may either terminate transcription at the designated stop site, yielding full-length transcripts of defined length $J$ at a rate denoted by $v_\text{full}^\text{terminate}$, or synthesize LMS at a rate $v_\text{LMS}^\text{terminate}$. 
A plausible mechanistic hypothesis is that elevated free Mg$^{2+}$ concentrations promote destabilization and dissociation of the T7 RNAP–DNA–mRNA ternary complex—particularly in the case of circular DNA templates—thereby facilitating proper transcription termination. This hypothesis is supported by the Shapley value-based sensitivity analysis shown in Figure~\ref{fig:shap_all}(c) in Appendix~\ref{appendix:SV}, where total Mg$^{2+}$ exhibits a strong negative effect on LMS formation, while NTP concentration exerts a significant positive influence. These opposing effects reflect an equilibrium-driven balance between total Mg$^{2+}$ and NTPs, which governs free Mg$^{2+}$ availability and, in turn, impacts LMS synthesis.
Additionally, a non-negligible level of LMS proportion is observed even when DNA linearity approaches 99\%, likely due to the formation of long loopback double-stranded RNA species or $3^\prime$-end hairpin structures \cite{dousis2023engineered}.

The full-length transcripts and LMS synthesis rates are expressed as
Equations~(\ref{eq.term_full}) and~(\ref{eq.term_LMS}),
where $[\text{EC}_J]$ denotes the concentration of the elongation complex associated with transcripts of length $J$. The effective linearity factor, $\alpha^{\text{terminate}}_\text{LMS}$, is expressed as a weighted combination of contributions from linear and circular DNA templates, where $\gamma_{\text{LMS}}^{\text{linear}}$ and $\gamma_{\text{LMS}}^{\text{circular}}$ correspond to LMS synthesis under linear and circular DNA templates, respectively.

\subsubsection{mRNA Degradation} 
\label{subsubsec:mRNA degradation}

In the IVT process, mRNA degradation is a potential challenge that could affect the yield and quality of the mRNA product under certain conditions. Degradation may arise from factors such as insufficient ribonuclease (RNase) inhibitor activity \cite{dickson2005ribonuclease}, suboptimal pH \cite{bernhardt2012primordial,chheda2024factors}, temperature excursion \cite{chheda2024factors}, or excessively high Mg$^{2+}$ concentration \cite{oivanen1998kinetics,li1999kinetics}. Such degradation reduces the integrity and lead to an accumulation of mRNA fragments, hereafter referred to as impurities (I$_d$). The chemical reaction illustrating the degradation of full-length mRNA (M$_J$) into mRNA fragments (I$_d$) can be expressed as follows:
\begin{equation}
    \text{M}_J {\rightarrow} \text{I}_d.
    \nonumber
\end{equation}

According to van de Berg et al. (2021) \cite{van2021quality}, the degradation rate of the mRNA product, denoted by $v^\text{degrade}$, is modeled as
Equation~(\ref{eq.degrade}),
where $k_\text{ac}$, $k_\text{ba}$, $k_\text{Mg}$, $n_\text{ac}$, $n_\text{ba}$, $n_\text{Mg}$, and $n_\text{RNA}$ are model-specific parameters. This formulation incorporates the effects of two critical environmental factors: pH, represented by the concentrations of hydrogen ions ($[\mbox{H}^+]$) and hydroxide ions ($[\mbox{OH}^-]$), and the concentration of magnesium ions ($[\mbox{Mg}^{2+}]$); the term $[\mbox{RNA}]$ denotes the concentration of full-length mRNA transcripts subject to degradation.

\subsubsection{Mg\texorpdfstring{$_2$}{2}PPi Precipitation}
\label{subsubsec:Precipitation}

The byproduct pyrophosphate (PPi) generated during elongation binds with Mg$^{2+}$ to form Mg$_2$PPi. This compound easily precipitates, leading to a significant drop in free Mg$^{2+}$ concentration, decreasing by several millimolar over time as mRNA production progresses \cite{young1997modeling,arnold2001kinetic,akama2012multiphysics}, i.e., 
\begin{equation}
    \text{Mg$_2$PPi} \xrightarrow{\text{precipitation}} \text{solid Mg}_2\text{PPi}.
    \nonumber
\end{equation}
Additionally, Mg$_2$PPi crystals may sequester DNA, inhibiting transcription and negatively impacting IVT efficiency \cite{shopsowitz2014rnai,stover2024mechanistic}.
After the onset of Mg$_2$PPi precipitation, the concentration of Mg$_2$PPi in the solution decreases. According to Quintana et al. (2005) \cite{quintana2005kinetics}, Akama et al. (2012) \cite{akama2012multiphysics}, and van de Berg et al. (2021) \cite{van2021quality}, the precipitation rate can be described by Equation~(\ref{eq.precip}),
where [Mg$_2$PPi] is the concentration of magnesium-bound pyrophosphate, [Mg$_2$PPi]$_\text{eq}$ is the equilibrium concentration, and $k^\text{precip}$ is the precipitation rate constant.

\subsubsection{PPi Enzymatic Degradation} 
\label{subsubsec:PPi Enzymatic Degradation}

To prevent the formation of Mg$_2$PPi precipitates, inorganic pyrophophatase (PPase) is commonly employed to hydrolyze PPi into orthophosphate (Pi or HPO$_4^{2-}$). Following the representation in Kern et al. (1997) \cite{kern1997application} and Stover et al. (2024) \cite{stover2024mechanistic}, this reaction can be written as:
\begin{equation}
    \text{MgP}_2\text{O}_7^{2-}~(\text{MgPPi}) + \text{H}_2\text{O} 
    \xrightarrow{\text{PPase}} 
    2~\text{HPO}_4^{2-}~(\text{Pi}) + \text{Mg}^{2+}.
    \nonumber
\end{equation}
This hydrolysis is crucial for maintaining transcription efficiency during the IVT process \cite{cunningham1990use,kern1997application}.  
Following the approaches of Chao et al. (2006) \cite{chao2006kinetic} and Stover et al. (2024) \cite{stover2024mechanistic}, the enzymatic degradation of PPi is
described by Equation~(\ref{eq.PPase}),
where [PPase] (U/mL) represents the enzyme activity per unit volume. The reaction rate depends on the concentration of magnesium-bound pyrophosphate ([MgPPi]) and the enzyme's M–M constant ($K_\text{M,PPase}$). The term $act_\text{PPase,pH}$ follows the general model structure defined in Equation~(\ref{eq:act_pH}), and captures the influence of pH on PPase stability and activity. Temperature-dependent effects were not included in this equation, as PPase activity is relatively stable within the typical temperature range of IVT processes \cite{lee2007activation}.

\subsubsection{
Enzyme Activity and Stability}
\label{subsubsec:pH-enzyme}

\textbf{(1) pH impact.} During the IVT process, the continuous release of free hydrogen ions (H$^+$) causes a gradual decrease in solution pH level, making it crucial to model the impact of pH on the activity of essential enzymes in the bioreactor, such as T7 RNAP and PPase. Enzyme activity typically exhibits a bell-shaped dependence on pH \cite{bisswanger2014enzyme}, meaning that there is an optimal pH range where the enzyme activity is highest, and deviations from this range can lead to decreased activity. To model enzyme pH stability, a generalized Gaussian function is employed in Equation~(\ref{eq:act_pH}),
where pH$_\text{opt}$ represents the optimal pH and the parameters, $\sigma_\text{pH}$ and $n_\text{pH}$, control the curve's width and sharpness, respectively. Compared to the Gaussian function used by Villiger et al. (2016) \cite{villiger2016controlling}, the introduction of the sharpness parameter $n_\text{pH}$ allows for greater flexibility in capturing different enzyme activity patterns. 

Experimental observations highlight the need for flexibility in the model structure. Kartje et al. (2021) \cite{kartje2021revisiting} reported that T7 RNAP reaction efficiency remains stable across a pH range of 7.3 to 8.3, with a slight increase near pH 7.9, aligning with the optimal pH of 8.1 reported by Milligan et al. (1987) \cite{milligan1987oligoribonucleotide}. This behavior indicates a broader tolerance for pH variation. In contrast, Lee et al. (2007) \cite{lee2007activation} demonstrated that PPase activity conforms to a Gaussian function with a sharp peak at an optimal pH of 8, showing a rapid decline as the pH deviates from this optimum.

\vspace{0.05in}
\noindent \textbf{(2) Temperature impact.} The enzyme stability with respect to temperature, characterized by $act_\text{Temp}$, is modeled using the temperature coefficient $Q_{10}$ \cite{ito2015thermodynamic}, which quantifies the change in the enzymatic reaction rate for every $\SI{10}{\degreeCelsius}$ deviation from the enzyme's optimal temperature Temp$_\text{opt}$; see Equation~(\ref{eq:act_Temp}).

\vspace{0.05in}
\noindent \textbf{(3) DNA Linearity impact.} Circular DNA templates potentially influence the structural conformation of the T7 RNAP–DNA complex, hindering promoter accessibility or polymerase processivity, and thereby reducing transcriptional efficiency relative to linear DNA. To account for the differential impact of template topology, an effective linearity factor ($act_\text{linearity}$) is introduced to represent the net transcriptional activity as a weighted combination of contributions from linear and circular DNA templates; see Equation~(\ref{eq:act_linear}).
Here, $\beta_{\text{activity}}^{\text{linear}}$ denotes the transcriptional activity associated with linear DNA, while $\beta_{\text{activity}}^{\text{circular}}$ represents the reduced activity attributed to circular DNA. The assumption $\beta_{\text{activity}}^{\text{linear}} > \beta_{\text{activity}}^{\text{circular}}$ reflects experimental observations and literature reports indicating that linear DNA templates typically support higher transcription efficiency than their circular counterparts due to more favorable enzyme–template interactions.

\subsection{Mass Balance and Equilibrium Equations}
\label{subsec:MassBalance}

\begin{sloppypar}
To accurately model the dynamics of the IVT process, the concentrations of key species or molecule components are systematically analyzed and categorized into three groups: (1) free species, represented as $\pmb{s}^{free}$; (2) complexes, represented as $\pmb{s}^{comp}$; and (3) total concentrations, represented as $\pmb{s}^{tot}$. The primary free species that significantly influence the kinetics of the IVT process are identified as [Mg$^{2+}$], [NTP] (including [ATP], [UTP], [CTP], and [GTP]), [CleanCap], [H$^{+}$], [Tris], [DTT], [spermidine], [Pi], and [PPi],  
i.e., 
\begin{equation*}
\pmb{s}^{free}_t =
\left\{
    [\mbox{Mg}^{2+}]_t, [\mbox{NTP}]_t, [\mbox{CleanCap}]_t, [\mbox{H}^{+}]_t,
    [\mbox{PPi}]_t, [\mbox{Pi}]_t, [\mbox{Tris}]_t, [\mbox{DTT}]_t,
    [\mbox{spermidine}]_t
\right\}^{\top}.
\end{equation*}

The free solution components can associate to form various complexes, $\pmb{s}^{comp}$, including [MgNTP], [Mg$_2$NTP], [MgPPi], [Mg$_2$PPi], [MgPi], etc., which may impact the dynamics of the IVT process.

The total concentration of each species, defined as the sum of its free ion concentrations and the concentrations of all its associated complexes, becomes:
\begin{equation*}
    \pmb{s}^{tot}_t = \left\{ 
    [\mbox{Mg}]_t^{tot}, [\mbox{NTP}]_t^{tot}, [\mbox{CleanCap}]_t^{tot}, [\mbox{H}]_t^{tot}, 
    [\mbox{PPi}]_t^{tot}, 
    [\mbox{Pi}]_t^{tot},[\mbox{Tris}]_t^{tot},
    [\mbox{DTT}]_t^{tot},
    [\mbox{spermidine}]_t^{tot} 
    \right\}^\top.
\end{equation*}
\end{sloppypar}

\begin{sloppypar}
Building on the studies by Kern et al. (1997) \cite{kern1997application}, Akama et al. (2012) \cite{akama2012multiphysics}, and Stover et al. (2024) \cite{stover2024mechanistic}, it is assumed that the key components in the IVT process achieve equilibrium due to the relatively fast reaction rates compared to the overall process dynamics of the key steps and the parallel reactions studied in Section~\ref{subsec:IVT}. 
Under this assumption, the mass balance equations (outlined in Table~\ref{tab:MassBalance} in Appendix~\ref{appendix:MBE}) and equilibrium equations (outlined in Table~\ref{tab:Equilibrium} in Appendix~\ref{appendix:MBE}) are used to define the concentration constraints for the chemical species in the system. 

The mass balance equations ensure the conservation of mass for each species within the IVT process. These equations establish the relationship between the total concentration of a species, its free ion concentration, and the concentrations of all its associated complexes. 
The equilibrium equations define the chemical equilibrium conditions for the reactions involving various species in the production system. Each equation specifies the equilibrium constant governing the relationship between the species on the reactant (left-hand) and product (right-hand) sides of the reaction. 

In this study, the initial total concentrations of all solution components are
directly specified or experimentally measured, except for
$[\mbox{H}]_t^{\text{tot}}$. However, the concentration of free hydrogen ions, $[\mbox{H}^+]_t$, can be calculated from the measured pH value using the relationship $\mbox{pH} = -\log([\mbox{H}^{+}])$.

Separately, macromolecular species such as the transcript pools (full-length, LMS, and truncated) and enzyme–template–transcript complexes (IC and EC) are modeled as dynamic kinetic states rather than equilibrium species, and are governed directly by the reaction-rate equations introduced in Section~\ref{subsec:DynamicModel}.
\end{sloppypar}


\section{Parameter Estimation and Performance Assessment}
\label{sec:inference}


The developed mechanistic model is trained using a criterion of minimizing prediction errors for yield and quality attributes based on data collected from the IVT batch process. The loss function, denoted by $\mathcal{L}$, is defined based on prediction errors across multivariate outputs, including yield, 5$^\prime$ capping efficiency, truncation proportion, mRNA integrity, and LMS proportion. Model inputs include action variables $\pmb{x}$ (e.g., initial conditions) and mechanistic parameter values $\pmb{\theta}$. For time-course datasets, the loss function also accounts for prediction errors in the state trajectory $\pmb{s}_t$ dynamics. The model assessment criteria (i.e., Mean Absolute Error (MAE), Weighted Absolute Percentage Error (WAPE), and rank
correlation \cite{hastie2009elements}) employed in this study are detailed in Appendix~\ref{appendix:goodness-of-fit}.

Due to the high computational cost of each \textit{in silico} simulation run, a Gaussian Process (GP)-based batch Bayesian optimization (Batch BO) approach was employed to accelerate the search for optimal parameter values while enabling parallel computing for model fitting and estimation. This approach captures interdependencies among the parameters $\pmb{\theta}$ and improves the
efficiency of parameter estimation, as described in Appendix~\ref{appendix:BOinference}.

\section{Results - Model Prediction Performance}
\label{sec:results}

The prediction performance of the modular mechanistic IVT model was systematically evaluated. 
To ensure robustness and mitigate overfitting, a repeated holdout cross-validation approach was employed. In each replication, the compiled datasets—comprising over 500 batches—were randomly divided into training (80\%) and testing (20\%) subsets using distinct random seeds. The optimal model parameters, denoted by $\pmb{\theta}^\star$, were inferred using Batch Bayesian Optimization (Appendix~\ref{appendix:BOinference}) based on the training data, and the model’s predictive accuracy was subsequently assessed on the corresponding test sets.

Comprehensive validation was conducted across five key performance indicators of the IVT process: mRNA yield, transcript integrity, capping efficiency, and the proportions of truncated transcripts and LMS. These outputs reflect the sequential progression of T7 RNAP during mRNA synthesis and maturation (Section~\ref{subsec:IVT}). Notably, the yield prediction was further validated using an external dataset from Boman et al. (2024) \cite{boman2024quality}, highlighting the model’s ability to generalize across distinct DNA templates and RNAs.
The final parameter estimates supported concurrent optimization across all five 
outputs, underscoring the model’s ability to mechanistically represent the complexity of the IVT process. These optimized parameters are summarized in Table~\ref{tab:parameterestimation} in Appendix~\ref{appendix:parameter}. Importantly, the model’s modular structure enables flexible reparameterization for specific analytical objectives—for example, refining $\pmb{\theta}^\star$ to prioritize yield—thereby enhancing its utility across both research and biomanufacturing applications.

The fitted mechanistic model exhibits promising predictive performance across IVT production outputs $\pmb{y}$, as evaluated on both training and testing datasets. The results of the model prediction performance metrics, that are reported as mean $\pm$ standard error (SE) estimated based on 10 independent replications, are recorded in Table~\ref{tab:prediction_performance}. Consistently, low MAE and WAPE across all outputs demonstrate the model’s robustness and predictive accuracy. These results suggest that repeated cross-validation yields stable error distributions, reinforcing the model’s reliability.
Furthermore, the high correlation coefficients between predicted and observed values confirm that the model effectively captures the mechanistic behavior governing the IVT process, supporting its broad applicability in mRNA synthesis modeling and process optimization. 

\begin{figure*}[htbp]
    \centering
    \includegraphics[width=0.9\linewidth]{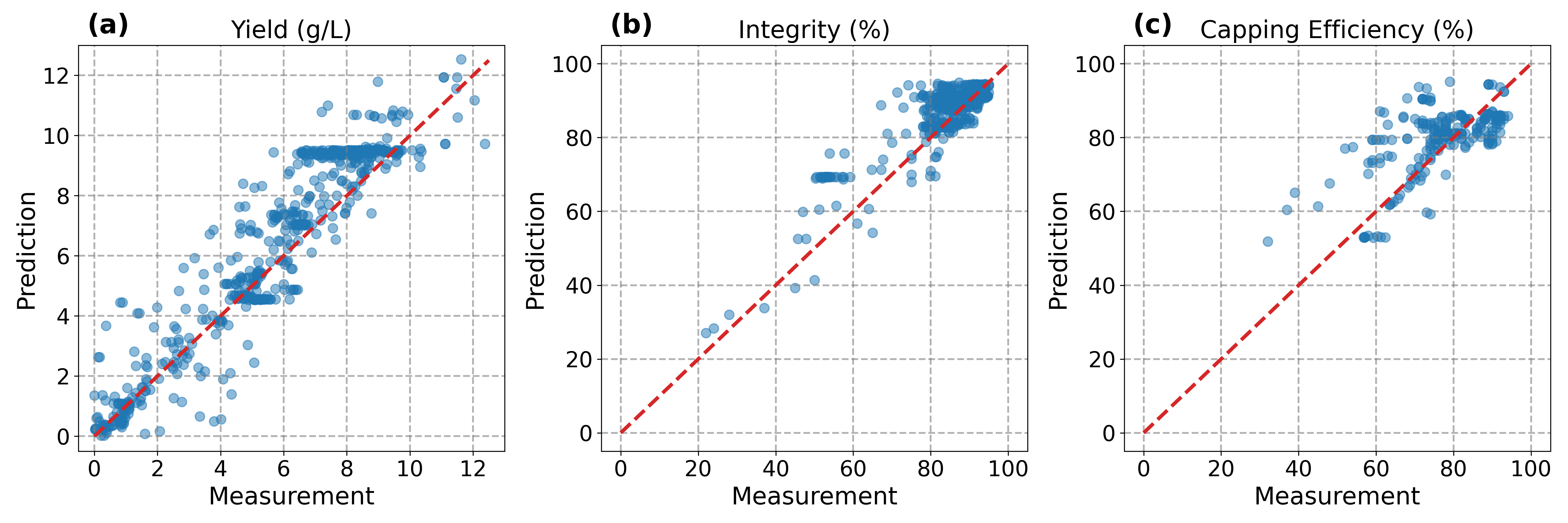}
    \caption{Comparison of model predictions versus experimental observations across the entire dataset for (1) mRNA yield (g/L), (2) integrity (\%), and (3) capping efficiency (\%) using the developed mechanistic \textit{in silico} model. The red dashed identity line represents perfect agreement between model predictions and experimental measurements. 
    }
    \label{fig:QQ_Prediction}
\end{figure*}

The Q–Q plots (quantile–quantile plots) comparing predicted and measured values for yield, integrity, and capping efficiency across the entire dataset are shown in Figure~\ref{fig:QQ_Prediction}. While a small number of data points show noticeable deviations, the majority of predictions align closely with the red dashed identity line, indicating strong overall agreement between the model predictions and experimental measurements. Considering measurement variability in the outputs and potential discrepancies between theoretical and actual input values, the \textit{in silico} model demonstrates robust predictive performance across a wide range of critical input conditions.  

In addition, representative time-course trajectories of yield, integrity, and capping efficiency, shown in Figure~\ref{fig:pred_trajectory} in
Appendix~\ref{appendix:Boman}, illustrate that the trained mechanistic \textit{in silico} model faithfully reproduces the dynamic behavior observed in the experimental IVT process.

The validation performance of yield prediction on the external dataset (30 batches) from Boman et al. (2024) \cite{boman2024quality}, using a different mRNA product, i.e., an eGFP sequence (995 nt), is also summarized in Table~\ref{tab:prediction_performance}, and the Q–Q plot (Figure~\ref{fig:QQ_Prediction_Boman}) is provided in Appendix~\ref{appendix:Boman}. The moderately elevated MAE and WAPE values are likely attributable to differences in reaction composition: whereas the present study utilized magnesium acetate (Mg(OAc)$_2$), the Boman dataset employed magnesium chloride (MgCl$_2$). Importantly, chloride ions (Cl$^-$) are known to strongly inhibit RNAP activity compared to acetate ions (OAc$^-$) \cite{kern1997application,maslak1994effects}, which may explain the slight upward bias observed in model predictions. Nevertheless, the strong correlation between predicted and observed yields supports the model’s ability to capture essential mechanistic dynamics and generalize effectively across variable reaction conditions.

\begin{table*}[htbp]
\fontsize{9.5}{11}\selectfont
\centering
\renewcommand{\arraystretch}{1.2}
\begin{tabular}{ll|cc|c|cc}
    \toprule
    \multirow{2}{*}{} & \multirow{2}{*}{\makecell{\textbf{Assessment} \\ \textbf{Criteria}}} 
    & \multicolumn{2{}}{c|}{\textbf{Mechanistic Model}} 
    & \textbf{Boman et al. (2024)} 
    & \multicolumn{2{}}{c}{\textbf{Hybrid Model}} \\
    \cmidrule(lr){3-4} \cmidrule(lr){5-5} \cmidrule(lr){6-7}
    & & \textbf{Train} & \textbf{Test} & \textbf{Test} & \textbf{Train} & \textbf{Test} \\
    \midrule
    \multirow{3}{*}{\makecell[l]{\textbf{mRNA} \\ \textbf{Yield}}} 
        & MAE (g/L)        & 0.95 $\pm$ 0.01 & 0.96 $\pm$ 0.04 & $1.88 \pm 0.00$ & $0.31 \pm 0.01$ & $0.51 \pm 0.03$ \\
        & WAPE (\%)        & 16.25 $\pm$ 0.02 & 16.73 $\pm$ 0.08 & $30.90 \pm 0.00$  & $5.64 \pm 0.17$ & $9.31 \pm 0.58$ \\
        & Correlation      & 0.94 $\pm$ 0.00 & 0.94 $\pm$ 0.01  & $0.81 \pm 0.00$ & $0.98 \pm 0.01$ & $0.96 \pm 0.04$ \\
    \midrule
    \multirow{3}{*}{\makecell[l]{\textbf{Truncation} \\ \textbf{Proportion}}} 
        & MAE (\%)         & $2.76 \pm 0.01$  & $2.73 \pm 0.04$  & --              & $0.90 \pm 0.03$ & $1.36 \pm 0.13$ \\
        & WAPE (\%)        & $32.52 \pm 0.18$ & $31.85 \pm 0.69$ & --              & $10.55 \pm 0.28$ & $16.19 \pm 1.26$ \\
        & Correlation      & $0.82 \pm 0.00$  & $0.79 \pm 0.02$  & --              & $0.96 \pm 0.01$ & $0.91 \pm 0.02$ \\
    \midrule
    \multirow{3}{*}{\makecell[l]{\textbf{mRNA} \\ \textbf{Integrity}}} 
        & MAE (\%)         & 4.01$\pm$ 0.01 & 4.01 $\pm$ 0.04 & --              & $1.74 \pm 0.06$ & $3.07 \pm 0.23$ \\
        & WAPE (\%)        & 4.69$\pm$ 0.01 & 4.73 $\pm$ 0.05 & --              & $2.04 \pm 0.07$ & $3.74 \pm 0.27$ \\
        & Correlation      & 0.86$\pm$ 0.01 & 0.87 $\pm$ 0.04 & --              & $0.97 \pm 0.00$ & $0.93 \pm 0.02$ \\
    \midrule
    \multirow{3}{*}{\makecell[l]{\textbf{LMS} \\ \textbf{Proportion}}} 
        & MAE (\%)         & $3.77 \pm 0.02$  & $3.73 \pm 0.07$  & --              & $1.60 \pm 0.06$ & $2.21 \pm 0.30$ \\
        & WAPE (\%)        & $40.90 \pm 0.23$ & $40.79 \pm 0.92$ & --              & $16.91 \pm 0.73$ & $27.42 \pm 2.25$ \\
        & Correlation      & $0.93 \pm 0.00$  & $0.93 \pm 0.00$  & --              & $0.98 \pm 0.01$ & $0.95 \pm 0.02$ \\
    \midrule
    \multirow{3}{*}{\makecell[l]{\textbf{Capping} \\ \textbf{Efficiency}}} 
        & MAE (\%)         & 7.06 $\pm$ 0.02 & 7.15 $\pm$ 0.09 & --              & $2.13 \pm 0.12$ & $3.72 \pm 0.30$ \\
        & WAPE (\%)        & 9.19 $\pm$ 0.03 & 9.39 $\pm$ 0.13  & --              & $2.93 \pm 0.16$ & $5.03 \pm 0.47$ \\
        & Correlation      & 0.84 $\pm$ 0.01 & 0.83 $\pm$ 0.06  & --              & $0.98 \pm 0.00$ & $0.95 \pm 0.01$ \\
    \bottomrule
\end{tabular}
\caption{Model prediction performance assessment criteria for mRNA yield, truncation proportion, integrity, LMS proportion, and capping efficiency. Criteria include MAE, WAPE, and rank correlation. Values are reported as mean $\pm$ SE over 10 independent replications. 
}
\label{tab:prediction_performance}
\end{table*}

A residual analysis (plotting residuals $\pmb{e} = \hat{\pmb{y}} - \pmb{y}$ against key input features $\pmb{x}$)—presented in Appendix~\ref{appendix:residual_analysis}—was conducted alongside the development of a hybrid model, where the residuals were further modeled as a function of $\pmb{x}$ using XGBoost \cite{chen2016xgboost} to ensure that no critical mechanistic dependencies were omitted.
While the hybrid model demonstrates superior predictive accuracy compared to the standalone mechanistic model (Table~\ref{tab:prediction_performance}), this improvement primarily reflects its ability to capture localized variations within a heterogeneous dataset. The dataset comprises over 500 experimental batches conducted by different operators for varied scientific objectives, utilizing diverse protocols, reagent sources (i.e., vendor variation), and assay methods. These uncontrolled contextual differences introduce non-generalizable, batch-specific effects that are more effectively captured by flexible statistical models such as XGBoost and GP regression, which are well-suited for learning localized, condition-dependent patterns. 

Nevertheless, this empirical improvement does not imply the presence of unmodeled mechanistic dependencies. As confirmed by the residual analysis (Appendix~\ref{appendix:residual_analysis}), the mechanistic model—grounded in biophysical and biochemical principles and designed for universal applicability across diverse process conditions—accurately captures the dominant biological mechanisms, with no significant systematic patterns left unaccounted for. Therefore, due to its interpretability, generalizability, and consistency with first‑principles understanding, the modular mechanistic model remains the preferred tool for \textit{in silico} experimentation and process optimization (Section~\ref{sec:sensivitiy}).

\section{\textit{In silico} Experiments and Insight Discussions}
\label{sec:sensivitiy}



Upon establishing a well-validated mechanistic \textit{in silico} model (Section~\ref{sec:results}), a series of \textit{in silico} experiments were conducted to evaluate the influence of process inputs on IVT performance metrics. This can provide interpretable predictions and insights to guide process optimization and wet experiments.
Specifically, 
the initial conditions of process inputs and the process time were systematically perturbed within defined ranges, and the resulting molecular concentration trajectories were visualized to facilitate comparison of output responses. This analysis facilitates the identification and validation of causal interdependencies within the IVT process, aligning well with established findings. Moreover, for previously unexplored relationships, the mechanistic \textit{in silico} model serves as an effective tool to narrow the hypothesis space, enabling more efficient identification of key factors that influence IVT production performance. 
Meanwhile, the sensitivity analysis of kinetic parameters in Appendix~\ref{apppendix:SA-parameter} quantifies each module’s influence on both IVT process predictions and associated prediction errors, thereby clarifying their contributions to production outcomes and overall model performance.

Without loss of generality, batch-based production was considered. 
Reference conditions for each input were determined based on commonly observed values in the dataset. Key inputs were perturbed individually or in combination, while holding other variables constant, to isolate their effects. Given the strong interaction between total Mg$^{2+}$ and NTPs, the Mg-to-NTP ratio was treated as a constrained input to maintain biologically relevant and practical IVT manufacturing concentration ranges.
Specifically, the reference conditions for each input were defined as follows: total Mg$^{2+}$ concentration of 36 mM, each NTP at 9 mM (corresponding to a Mg-to-NTP ratio of 1), DTT at 0 mM, CleanCap at 4 mM, DNA at 0.05 mg/mL with 95\% linearity, Spermidine at 0 mM, and Tris Buffer at 40 mM. Enzyme activities were set at 0.25 U/mL for PPase and 14,000 U/mL for T7 RNAP. Temperature, process time, and initial pH were fixed at 37$^\circ$C, 150 minutes, and pH 8.0, respectively. 

To elucidate the concentration dynamics of key components and their interdependencies with IVT process outputs, multivariate dynamic profiles were generated under the reference condition (C2: 36 mM total Mg$^{2+}$and 36 mM total NTP, i.e., 9 mM each), as well as under the following variations: C1 (18 mM Mg$^{2+}$, 36 mM NTP), C3 (72 mM Mg$^{2+}$, 36 mM NTP), C4 (36 mM Mg$^{2+}$, 18 mM NTP), and C5 (36 mM Mg$^{2+}$, 72 mM NTP) (Figure~\ref{fig:trajectory_Mg}). 
These visualizations, together with additional \textit{in silico} experiments, support the key findings regarding yield, integrity, and capping efficiency described below.

\subsubsection*{Total Mg$^{2+}$ and NTPs}
\label{subsubsec:SA_yield_input}

\textit{
(1) Effectively managing the equilibrium among free Mg$^{2+}$, MgNTP, and Mg$_2$NTP is crucial for optimizing the mRNA synthesis rate during the IVT process.} 
In IVT, free Mg$^{2+}$ ions interact with NTPs to form complexes such as MgNTP and Mg$_2$NTP. As noted by Thomen et al. (2008) \cite{thomen2008t7}, T7 RNAP requires both MgNTP and free Mg$^{2+}$ ions for optimal activity. Deviating from the optimal Mg-to-NTP ratio in either direction—too low or too high—leads to a significant decrease in mRNA yield, albeit for different reasons. 
When the Mg-to-NTP ratio is excessively high (C3: 72 mM total Mg$^{2+}$ with 36 mM NTP), the mRNA synthesis rate declines relative to the reference condition (C2: 36 mM Mg$^{2+}$ and 36 mM NTP), as shown in Figure~\ref{fig:trajectory_Mg}(a).
In this scenario, the excess Mg$^{2+}$ drives the formation of Mg$_2$NTP (Figure~\ref{fig:trajectory_Mg}(g)), a species that T7 RNAP cannot effectively utilize. This reduces the availability of MgNTP (Figure~\ref{fig:trajectory_Mg}(f)), thereby limiting the transcription rate. Consequently, the reduced availability of functional substrate, along with the accumulation of non-productive Mg$_2$NTP, appears to lead to a slightly lower final mRNA yield compared to the reference condition.
Conversely, when the Mg-to-NTP ratio is too low, as illustrated in Figure~\ref{fig:trajectory_Mg}(a), with a total of 36 mM NTP in the solution, decreasing Mg$^{2+}$ from 36 mM (C2) to 18 mM (C1) leads to a significant decrease in mRNA yield.  In this case, the concentrations of free Mg$^{2+}$ (Figure~\ref{fig:trajectory_Mg}(e)) and MgNTP (Figure~\ref{fig:trajectory_Mg}(f)) become insufficient to support T7 RNAP activity and facilitate transcript synthesis. 
At a fixed total Mg$^{2+}$ concentration of 36 mM, both C4 (18 mM NTP; limited MgNTP availability; Figure~\ref{fig:trajectory_Mg}(f)) and C5 (72 mM NTP; insufficient free Mg$^{2+}$; Figure~\ref{fig:trajectory_Mg}(e)) exhibit reduced mRNA synthesis rates compared to the reference condition (Figure~\ref{fig:trajectory_Mg}(a)). Nonetheless, the final mRNA yield at 150 minutes in C5 is higher, attributed to the increased substrate availability.

\textit{(2) The optimal Mg-to-NTP ratio is approximately 1 under commonly used IVT conditions.}
To determine the optimal Mg-to-NTP ratio for IVT, mRNA yield at 150 minutes was evaluated across batches with varying initial NTP concentrations, Mg-to-NTP ratios (and thus total Mg$^{2+}$ concentrations), and volume-based T7 RNAP activities, as shown in Figure~\ref{fig:Yield_MgNTP_NTP_T7}. Across different T7 RNAP activity levels and NTP concentrations, the optimal Mg-to-NTP ratio consistently centers around 1, consistent with the 0.9–1.2 range reported by Boman et al. (2024) \cite{boman2024quality}.
Notably, at a fixed T7 RNAP activity, the decline in mRNA yield becomes more pronounced for batches with higher initial NTP concentrations as the Mg-to-NTP ratio increases. For example, at 14,000 U/mL T7 RNAP, the batch with 48 mM NTP exhibits a more substantial decrease in yield than the batch with 24 mM NTP as the Mg-to-NTP ratio increases from 1.0 to 2.5. This trend is driven by the higher total Mg$^{2+}$ concentrations required to reach elevated Mg-to-NTP ratios, which promote the formation of Mg$_2$NTP complexes. As a result, the reduced availability of MgNTP limits mRNA synthesis, ultimately leading to lower final yields.

\textit{(3) Maintaining sufficient MgNTP concentrations throughout the IVT process helps minimize the synthesis of truncated transcripts.}
The dynamic trajectories of integrity and truncation proportion during the IVT process are shown in Figure~\ref{fig:trajectory_Mg}(b) and (c), respectively. As mRNA synthesis progresses, the depletion of MgNTP (Figure~\ref{fig:trajectory_Mg}(f)) corresponds to a gradual increase in the truncation proportion (Figure~\ref{fig:trajectory_Mg}(c)), consistent with observations reported by Hengelbrock et al. (2024) \cite{hengelbrock2024digital}. 
A similar conclusion emerges when comparing batches with different initial NTP concentrations: those starting at higher concentrations maintain greater MgNTP availability throughout the IVT process and exhibit better control over truncated transcript formation. Specifically, C5 (72 mM NTP) sustains higher MgNTP levels across the reaction, resulting in consistently lower truncation proportions compared with C2 (36 mM NTP) and C4 (18 mM NTP). The same pattern is observed between the C2 and C4, confirming the concentration-dependent effect of MgNTP availability on premature termination. This effect is likely driven by reduced polymerase stalling at higher MgNTP availability; when MgNTP levels become limiting, the elongation complex is more prone to pausing, which increases the probability of premature RNAP dissociation and truncated transcript formation.

\textit{(4) The LMS proportion gradually declines over the course of the IVT process.}
As discussed in Section~\ref{subsubsec:termination}, elevated free Mg$^{2+}$ concentrations may promote the destabilization and dissociation of the T7 RNAP–DNA–mRNA ternary complex, thereby facilitating proper transcription termination. Accordingly, the observed decline in LMS proportion during the IVT process (Figure~\ref{fig:trajectory_Mg}(d)) can be attributed to the increasing concentration of free Mg$^{2+}$ (Figure~\ref{fig:trajectory_Mg}(e)), which progressively reduces the read-through proportion over time.

\textit{(5) CleanCap binding is less dependent on free Mg$^{2+}$ compared to ATP and GTP, resulting in higher capping efficiency in low Mg-to-NTP ratio batches.}
The dynamic trajectories of capping efficiency under varying NTP and total Mg$^{2+}$ levels are shown in Figure~\ref{fig:trajectory_Mg}(i). With a fixed NTP concentration of 36 mM and 4 mM CleanCap, increasing total Mg$^{2+}$ from 18 mM (C1) to 36 mM (C2) and then to 72 mM (C3) progressively reduces capping efficiency, which aligns with the increasing free Mg$^{2+}$ levels shown in Figure~\ref{fig:trajectory_Mg}(e). This inverse trend supports the hypothesis that CleanCap incorporation is less Mg$^{2+}$-dependent than NTPs like ATP and GTP. The gradual increase in capping efficiency over the course of the process can be attributed to two factors discussed later: (1) competitive binding between CleanCap and ATP/GTP, and (2) the influence of pH.

\subsubsection*{T7 RNAP Activity}
\textit{(1) At the optimal Mg-to-NTP ratio, optimizing T7 RNAP activity is essential to balance material costs and maximize mRNA yield.}
Within the optimal Mg-to-NTP range, the required T7 RNAP activity to fully or nearly deplete the NTPs depends on the starting NTP concentration, as shown in Figure~\ref{fig:Yield_MgNTP_NTP_T7}. For example, when the initial NTP concentration is 24 or 36 mM, 10,000 U/mL of T7 RNAP is sufficient, as increasing the activity to 14,000 U/mL does not substantially improve yield. 
However, for batches with 48 mM NTP, the yield continues to increase as T7 RNAP activity rises from 10,000 to 14,000 U/mL, indicating that residual NTPs remain available for utilization. Conversely, 4,000 U/mL of T7 RNAP is insufficient to consume even 24 mM NTP at the optimal Mg-to-NTP ratio. Given the high cost of T7 RNAP \cite{raghuwanshi2024purification}, it is critical to balance enzyme input with marginal yield improvement to optimize process economics.



\subsubsection*{CleanCap and ATP/GTPs}
\textit{(1) Balancing the ratio between CleanCap and ATP/GTP concentrations is critical for optimizing capping efficiency.}
In addition to the effects of free Mg$^{2+}$, capping efficiency is strongly influenced by the relative concentrations of ATP and GTP compared to CleanCap. As shown in Figure~\ref{fig:trajectory_Mg}(i), under a fixed total Mg$^{2+}$ concentration of 36 mM and a CleanCap concentration of 4 mM, increasing the NTP concentration from 18 mM (C4) to 36 mM (C2) and then to 72 mM (C5) intensifies competition from ATP and GTP, leading to a decline in capping efficiency. This decline occurs despite a concurrent reduction in free Mg$^{2+}$ levels (Figure~\ref{fig:trajectory_Mg}(e)), which would otherwise be expected to enhance capping efficiency.
During the production process, ATP and GTP are consumed at significantly higher rates than CleanCap due to transcription stoichiometry—each transcript requires only a single CleanCap molecule, whereas hundreds to thousands of ATP and GTP molecules are incorporated depending on the sequence length and nucleotide composition. This disproportionate consumption reduces the competitive pressure from ATP and GTP over time, thereby favoring CleanCap incorporation and contributing to the increasing trend in capping efficiency observed throughout the IVT process (Figure~\ref{fig:trajectory_Mg}(i)).

\subsubsection*{DNA Template Linearity}

\textit{(1) Circular DNA potentially alters the conformation of the T7 RNAP–DNA complex and reduce transcription efficiency.}
The effects of DNA template linearity, Mg-to-NTP ratio, and NTP concentration on mRNA yield at 150 minutes are shown in the top row of Figure~\ref{fig:Linearity_MgNTP_Mg}. Increasing the DNA template linearity from 0.6 to 0.99 consistently enhances mRNA synthesis rate across various NTP concentrations and Mg-to-NTP ratios. However, this correlation diminishes under conditions where NTP becomes limiting, such as at 24 mM NTP or near the optimal Mg-to-NTP ratio, as the reaction approaches NTP depletion and additional template linearity offers little improvement in yield. 

\textit{(2) High linearity is critical for ensuring mRNA integrity and minimizing LMS synthesis.}
The effects of DNA template linearity, Mg-to-NTP ratio, and NTP concentration on mRNA integrity at 150 minutes are shown in the bottom row of Figure~\ref{fig:Linearity_MgNTP_Mg}. Among these factors, linearity 
plays a dominant role in determining integrity. Additionally, as hypothesized in Section~\ref{subsubsec:termination}, elevated free Mg$^{2+}$ levels may promote destabilization and dissociation of the complex formed among T7 RNAP, mRNA transcripts, and circular DNA templates, thereby facilitating proper transcription termination. As a result, even under conditions of low linearity—which include a high proportion of circular DNA templates—increasing the Mg-to-NTP ratio to maintain higher free Mg$^{2+}$ concentrations can help preserve mRNA integrity. This mechanism also explains the observed gradual decline in LMS proportion during the process (Figure~\ref{fig:trajectory_Mg}(d)), which coincides with the progressive release of free Mg$^{2+}$ (Figure~\ref{fig:trajectory_Mg}(e)).
Furthermore, even at 99\% linearity, truncated transcripts and undesired structures—such as long loopback double-stranded RNA or $3^\prime$-end hairpins—can still form during the IVT process \cite{dousis2023engineered}, limiting the maximum achievable integrity to approximately 90\%.

\subsubsection*{pH and Temperature}

\textit{(1) Maintaining pH near its optimal range maximizes mRNA yield without adversely affecting integrity or capping efficiency.}
The performance of the IVT process under varying initial pH conditions is illustrated in the top row of Figure~\ref{fig:pH_trajectory}. Consistent with findings from Milligan et al. (1987) \cite{milligan1987oligoribonucleotide} and Kartje et al. (2021) \cite{kartje2021revisiting}, the mRNA synthesis rate gradually declines as the pH deviates from the optimal range of approximately 7.5 to 8.0. 
However, no significant changes in integrity are observed within the pH range of 6.5 to 8.0, and integrity remains stable throughout the process across all batches.
In contrast, capping efficiency increases as pH decreases from 8 to 6.5. This trend may result from pH-dependent changes in the competitive dynamics between ATP/GTP and CleanCap for T7 RNAP binding, potentially favoring CleanCap incorporation at lower pH. Combined with the higher consumption rates of ATP and GTP compared to CleanCap, the gradual acidification of the system—caused by the continuous release of H$^+$ during the reaction—may further promote CleanCap binding. These factors collectively explain the upward trend observed in the capping efficiency trajectory (Figure~\ref{fig:trajectory_Mg}(i)). Further studies are needed to clarify the molecular mechanisms driving this effect.

\textit{(2) Optimal temperature control balances mRNA yield, integrity, and capping efficiency.}
The impact of temperature on IVT performance is illustrated in the bottom row of Figure~\ref{fig:pH_trajectory}. A pronounced reduction in mRNA synthesis rate is observed as the reaction temperature decreases from 40 $^\circ$C to 30 $^\circ$C, consistent with Rosa et al. (2022) \cite{rosa2022maximizing}, which reports an optimal reaction temperature between 37 $^\circ$C and 44 $^\circ$C. 
Integrity is not significantly affected by temperature variation, whereas capping efficiency decreases steadily as the temperature drops from 40$^\circ$C to 30$^\circ$C, likely due to reduced T7 RNAP catalytic activity and slower incorporation of CleanCap at lower temperatures. 

\section{Conclusion}
\label{sec:conclusion}

This work presents a modular mechanistic \textit{in silico} model for predicting mRNA yield, integrity, and capping efficiency in IVT processes. The model integrates biophysically interpretable submodels representing (1) initiation and capping, (2) elongation and truncation, (3) termination and read-through, (4) mRNA degradation, (5) Mg\textsubscript{2}PPi precipitation, and (6) enzymatic degradation of PPi. It further incorporates key environmental and template-specific factors—including pH, temperature, and DNA linearity—enabling a holistic evaluation of IVT process performance under a wide range of operating conditions.

Through systematic model calibration and evaluation across diverse experimental conditions, the developed model demonstrates robust predictive performance while maintaining mechanistic interpretability. A series of \textit{in silico} experiments were conducted to evaluate the individual and combined effects of key process parameters—including total Mg$^{2+}$ concentration, NTP concentration, T7 RNAP activity, CleanCap concentration, DNA template linearity, pH, and temperature—on IVT performance metrics. Results highlight the importance of maintaining an optimal Mg-to-NTP ratio near 1 to ensure efficient mRNA synthesis. Balancing the ratio between CleanCap and ATP/GTP concentrations is critical for maximizing capping efficiency. Optimizing T7 RNAP activity helps reduce material costs while sustaining high mRNA yield. Improving DNA template linearity enhances both yield and integrity, while precise control of pH and temperature enables a favorable balance among synthesis rate, integrity, and capping efficiency.

These insights provide actionable guidance for rational IVT process optimization, particularly in the design of feeding strategies (e.g., NTP supplementation) and the control of process parameters such as pH and temperature. For instance, implementing NTP feeding during production helps maintain the Mg-to-NTP ratio within the optimal range, thereby sustaining higher synthesis rates and improving final yield. To enhance capping efficiency, lowering the initial ATP/GTP concentrations can reduce competition with CleanCap; however, this may come at the cost of reduced yield and integrity. A controlled ATP/GTP feeding strategy throughout the IVT process provides a practical solution to this trade-off by ensuring sufficient NTP availability for transcription while minimizing CleanCap competition, thereby improving capping efficiency without compromising other critical quality attributes.

\section*{AUTHOR CONTRIBUTIONS}

\textbf{Conceptualization and methodology:} K.W., K.C., H.Z., J.P., A.D., G.D., B.C.M., W.X.  
\textbf{Experimental design and data acquisition:} K.W., E.R., J.P., G.D., B.C.M., W.X. 
\textbf{Model development, software, and visualization:} K.W., K.C., H.Z., J.P., F.C., W.X.  
\textbf{Interpretation:} All authors.  
\textbf{Writing:} K.W., K.C., E.R. (original draft); all authors (review \& editing).  
\textbf{Supervision and resources:} J.P., G.D., B.C.M., W.X.

\section*{Data and Code Availability}
Data and code used in this study were generated using internal computational models and experimental platforms. Please contact the corresponding author for additional information regarding data and code details.

\section*{Conflict of Interest}
The authors declare that there are no conflicts of interest.

\section*{Acknowledgments}
The authors thank their collaborators for insightful discussions and technical support related to this work.

\bibliographystyle{plain}
\bibliography{references}

\newpage
\begin{appendices}


\section{Appendix: Ranges of Input Parameters Used in the IVT Process}
\label{appendix:RangePara}

\begin{table}[htb!]
\centering
\fontsize{9.5}{12}\selectfont
\renewcommand{\arraystretch}{1.5} 
\caption{Input Parameters with Ranges}
\label{tab:RangePara}
\begin{tabular}{|l|c|c|c|}
\hline
\textbf{Input Parameter} & \textbf{Units} & \textbf{Minimum Value} & \textbf{Maximum Value} \\
\hline
Temperature Set Point & $^\circ$C & 30 & 44 \\
Agitation Rate\textsuperscript{$\dagger$} & RPM & 0 & 1400 \\
Total IVT Incubation Time & Minutes & 15 & 300 \\
Template DNA Linearity & \% & 60 & 100 \\
DNA Concentration in Reaction & mg/mL & 0.01 & 0.692 \\
Reaction Volume & mL & 0.5 & 100 \\
CleanCap AG & mM & 0 & 9 \\
ATP & mM & 0.3 & 20 \\
CTP & mM & 0.3 & 20 \\
GTP & mM & 0.3 & 20 \\
m$^1\Psi$TP & mM & 0 & 16 \\
UTP & mM & 0 & 20 \\
Spermidine & mM & 0 & 5 \\
DTT & mM & 0 & 40 \\
Magnesium Acetate & mM & 6 & 144 \\
PEG-8000\textsuperscript{$\dagger$} & \% & 0 & 4 \\
Tris-HCl Buffer & mM & 0 & 60 \\
Buffer pH & pH & 5.9 & 8.5 \\
Triton X-1000\textsuperscript{$\dagger$} & \% & 0 & 0.01 \\
DMSO\textsuperscript{$\dagger$} & \% & 0 & 1.67 \\
RNase Inhibitor\textsuperscript{$\dagger$} & U/mL & 0 & 1000 \\
Pyrophosphatase & U/mL & 0 & 25 \\
T7 Concentration & U/mL & 1.25 & 120000 \\
CaCl\textsubscript{2} Concentration\textsuperscript{$\dagger$} & mM & 0 & 2 \\
DNase Concentration\textsuperscript{$\dagger$} & U/mg DNA & 0 & 10000 \\
EDTA Concentration After Addition\textsuperscript{$\dagger$} & mM & 0 & 75 \\
ProK\textsuperscript{$\dagger$} & U/mL & 0 & 30 \\
\hline
\end{tabular}
\vspace{2mm}

\parbox{0.8\linewidth}{
\noindent\textsuperscript{$\dagger$} These parameters are not considered in the \textit{in silico} model based on expert knowledge and sensitivity analysis results.
}
\end{table}

\newpage
\section{Appendix: IVT Process Parameters and Performance Metrics}
\label{appendix:IVT_inputs_outputs}

\begin{table*}[h]
    \centering
    \renewcommand{\arraystretch}{1.2} 
    \setlength{\tabcolsep}{10pt} 
    \caption{Process Parameters and Performance Metrics of the IVT Process}
    \label{tab:IVT_inputs_outputs}
    \rowcolors{2}{gray!15}{white} 
    \begin{tabular}{p{3.5cm} p{1.5cm} | p{3.5cm} p{1.5cm}} 
        \hline
        \multicolumn{4}{c}{\textbf{Process Parameters}} \\
        \hline
        Total Mg$^{2+}$ conc. & mM & DTT conc. & mM \\
        ATP conc. & mM & CTP conc. & mM \\
        GTP conc. & mM & m$^1\Psi$TP conc. & mM \\
        CleanCap conc. & mM & DNA template conc. & mg/mL\textsuperscript{$\star$} \\
        DNA linearity & -- & Spermidine conc. & mM \\
        Tris buffer conc. & mM & PPase activity & U/mL \\
        T7 RNAP activity & U/mL\textsuperscript{$\dagger$} & Temperature & $^\circ$C \\
        Batch duration & min & Initial pH & -- \\
        \hline
        \multicolumn{4}{c}{\textbf{Performance Metrics}} \\
        \hline
        mRNA yield & g/L & Capping efficiency & \% \\
        Truncation proportion & \% & mRNA integrity & \% \\
        LMS proportion & \% & & \\
        \hline
    \end{tabular}
    \vspace{2mm}
    
    \raggedright
    \textsuperscript{$\star$} DNA Template concentration is converted to molarity based on its molecular weight. \\
    \textsuperscript{$\dagger$} T7 RNAP activity is converted to molarity based on its molecular weight and specific activity.
\end{table*}

\newpage
\section{Appendix: Shapley Value-based Sensitivity Analysis on Feature Selection}
\label{appendix:SV}

To support the development of a mechanistic model for predicting mRNA yield and quality attributes, a feature importance analysis was conducted using data-driven models—XGBoost \cite{chen2016xgboost}, Gaussian Process \cite{frazier2018tutorial}, and linear regression  \cite{hastie2009elements}. These models can flexibly capture complex, nonlinear input-output relationships without requiring prior mechanistic knowledge. Among them, XGBoost demonstrated superior predictive performance and was selected for Shapley value-based sensitivity analysis. Rooted in cooperative game theory, Shapley values provide a model-agnostic, mathematically rigorous framework to quantify the marginal contribution of each input feature, accounting for all interaction effects \cite{lundberg2017unified}. 

\begin{figure}[htb!]
  \centering
  \begin{subfigure}{0.45\textwidth}
    \centering
    \includegraphics[width=\textwidth]{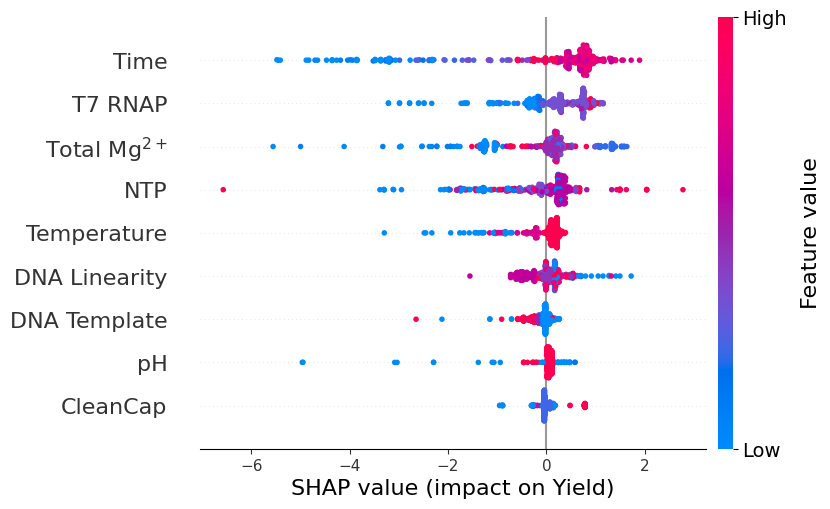}
    \caption{Shapley values across different substrates on mRNA yield.}
  \end{subfigure}
  \hfill
  \begin{subfigure}{0.45\textwidth}
    \centering
    \includegraphics[width=\textwidth]{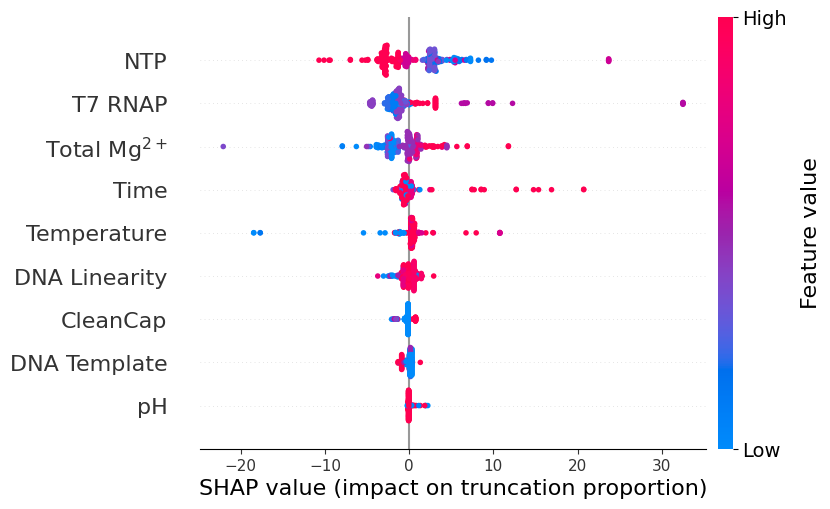}
    \caption{Shapley values across different substrates on truncation proportion.}
  \end{subfigure}

  \vspace{0.5cm} 

  \begin{subfigure}{0.45\textwidth}
    \centering
    \includegraphics[width=\textwidth]{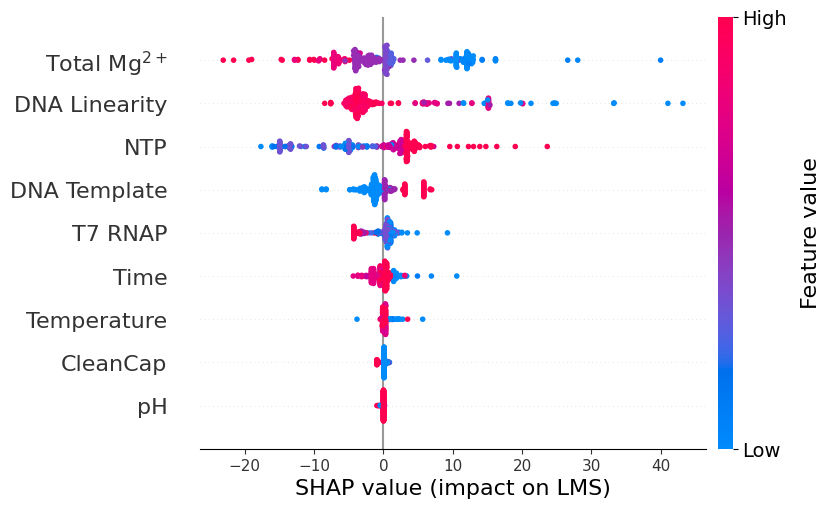}
    \caption{Shapley values across different substrates on LMS proportion.}
  \end{subfigure}
  \hfill
  \begin{subfigure}{0.45\textwidth}
    \centering
    \includegraphics[width=\textwidth]{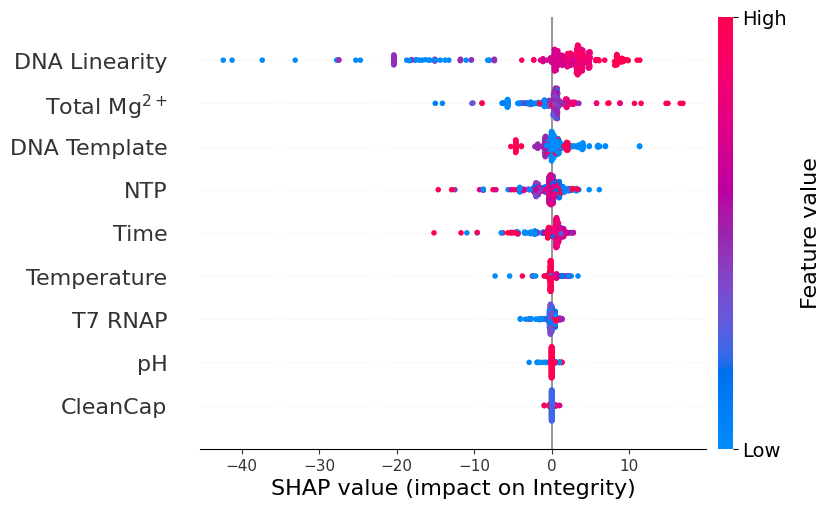}
    \caption{Shapley values across different substrates on mRNA Integrity.}
  \end{subfigure}

  \vspace{0.5cm} 

  \begin{subfigure}{0.45\textwidth}
    \centering
    \includegraphics[width=\textwidth]{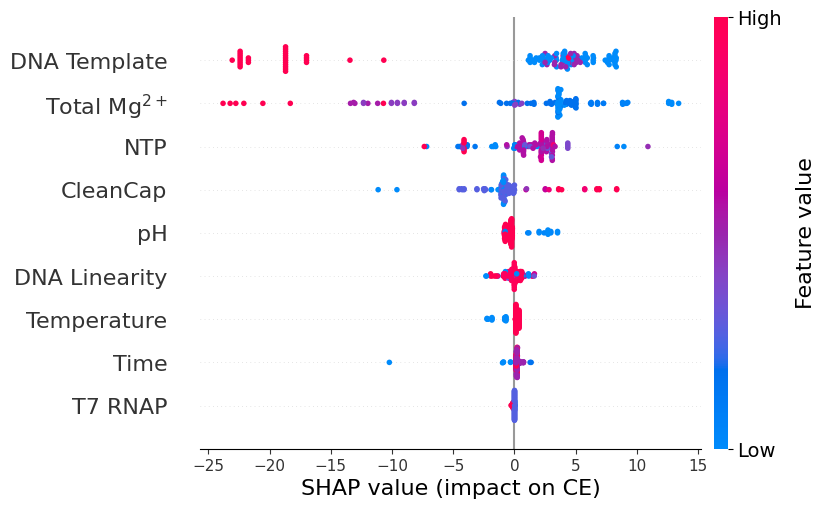}
    \caption{Shapley values across different substrates on mRNA capping efficiency.}
  \end{subfigure}
  \hfill
  \begin{subfigure}{0.45\textwidth}
    \centering
  \end{subfigure}

  \caption{Shapley values across different substrates on yield and quality attributes.}
  \label{fig:shap_all}
\end{figure}

Figure~\ref{fig:shap_all} presents the Shapley value-based sensitivity analysis of process inputs on (a) mRNA yield, (b) truncation proportion, (c) LMS proportion, (d) integrity, and (e) capping efficiency. Input features are ranked by overall importance along the y-axis, while the x-axis indicates the magnitude and direction of each feature’s contribution. Each point represents a single data instance, colored according to the corresponding feature value.

\textbf{Yield.} 
Figure~\ref{fig:shap_all}(a) presents the Shapley value analysis identifying the most influential process inputs on mRNA yield. The results highlight process time, T7 RNAP activity, total Mg$^{2+}$ concentration, NTP concentration, and temperature as key contributors. Specifically, insufficient process time may lead to incomplete consumption of the NTP substrates, resulting in reduced mRNA yield. In contrast, higher T7 RNAP activity and adequate NTP supply have a positive impact on yield. Temperature also plays a favorable role, as increasing it from suboptimal levels toward the enzyme's optimal range enhances T7 RNAP activity and thereby improves yield. Meanwhile, total Mg$^{2+}$ exerts a significant but complex effect; despite its importance, its relationship with yield is non-monotonic, as discussed in Section~\ref{sec:sensivitiy}. 

\textbf{Truncation proportion.} 
Figure~\ref{fig:shap_all}(b) shows the Shapley value analysis for factors influencing the mRNA truncation proportion. The most impactful variables identified are NTP concentration, total Mg$^{2+}$ concentration, T7 RNAP activity, process time, and temperature. Truncation has been observed to increase when the IVT system operates under unfavorable low MgNTP conditions, as reported by Hengelbrock et al. (2024)~\cite{hengelbrock2024digital}. Specifically, insufficient NTP levels, excessively high T7 RNAP activity, elevated total Mg$^{2+}$, prolonged process duration, and high temperature can collectively or individually shift the system toward low MgNTP availability, thereby increasing the risk of transcript truncation. Consequently, monitoring the limiting MgNTP species among all four types is critical for controlling truncation proportion.

\textbf{LMS proportion.} 
Figure~\ref{fig:shap_all}(c) presents the Shapley value analysis for factors influencing the proportion of long mRNA species (LMS). The three most impactful variables identified are total Mg$^{2+}$ concentration, DNA template linearity, and NTP concentration. Template linearity—defined as the percentage of linearized DNA—plays a critical role, as circularized or partially linearized templates may promote transcriptional readthrough beyond the intended termination site, leading to LMS formation. In addition, total Mg$^{2+}$ shows a strong negative influence on LMS, while NTP concentration exerts a significant positive effect. These opposing trends reflect an equilibrium-driven balance between total Mg$^{2+}$ and NTPs, which together regulate free Mg$^{2+}$ availability and, consequently, affect LMS synthesis.

\textbf{mRNA integrity.} 
Figure~\ref{fig:shap_all}(d) presents the Shapley value analysis for factors influencing mRNA integrity. The most significant contributors identified include DNA template linearity and total Mg$^{2+}$ concentration. As expected, higher template linearity promotes efficient transcription and reduces LMS formation, thereby enhancing overall integrity. Total Mg$^{2+}$ exhibits a positive effect on integrity, which contrasts with its negative impact on LMS formation—consistent with the notion that full-length transcripts and LMS species are competing outcomes. In contrast, NTP concentration shows no clear net effect on integrity. This is likely because NTPs positively influence truncation proportion but negatively impact LMS formation, resulting in an overall canceling effect on mRNA integrity.  Additionally, no consistent relationship is observed between DNA template concentration and mRNA integrity. This effect also disappears when analyzing the mechanistic model’s prediction residuals, as shown in Figure~\ref{fig:residual_integrity} in Appendix~\ref{appendix:residual_analysis}, suggesting that the observed influence may arise from data heterogeneity rather than true mechanistic behavior.

\textbf{Capping efficiency.} 
Figure~\ref{fig:shap_all}(e) presents the Shapley value analysis for various process inputs influencing mRNA capping efficiency. The most influential factors identified are NTP concentration, CleanCap concentration, and total Mg$^{2+}$ concentration. The concentration ratio of CleanCap to NTP directly affects the likelihood of successful cap formation. Interestingly, higher total Mg$^{2+}$—and consequently higher free Mg$^{2+}$—is associated with reduced capping efficiency. This observation supports the hypothesis discussed in Section~\ref{sec:sensivitiy}
: CleanCap binding is less dependent on free Mg$^{2+}$ compared to ATP and GTP, resulting in higher capping efficiency under low Mg-to-NTP ratio conditions. While a negative correlation between high DNA template concentration and capping efficiency is observed, no mechanistically validated explanation currently exists. Accordingly, this effect is not captured by the current mechanistic \textit{in silico} model and will be investigated further in future work.

\newpage
\section{Appendix: Kinetic Models}
\label{appendix:ratemodel}

\textbf{Michaelis-Menten kinetics} 

For a typical single-substrate enzymatic reaction,
$$
E + A \underset{k_R}{\overset{k_F}{\rightleftarrows}} EA \overset{k_{cat}}{\rightarrow} E + P,
$$
the substrate $A$ binds to the enzyme $E$ to form the enzyme--substrate complex $EA$, which then converts into the product $P$ and regenerates the enzyme $E$. The kinetic parameters include:  
(1) $k_F$ and $k_R$, the forward and reverse rate constants for complex formation and dissociation;  
(2) $k_{cat}$, the catalytic rate constant, which may depend on environmental factors such as pH and temperature.

Under the quasi-steady-state assumption, the reaction rate follows Michaelis--Menten kinetics:
$$
v = \frac{V_{\text{max}} \cdot [A]}{K_M + [A]},
$$
where $V_{\text{max}} = k_{cat} \cdot E_0$ is the maximum rate and $K_M = \frac{k_R + k_{cat}}{k_F}$ is the Michaelis constant, with $E_0$ denoting the total enzyme concentration.

In the presence of a competitive inhibitor $I$, which competes with $A$ for the active site on $E$, the apparent Michaelis constant becomes
$$
K_M^{\text{app}} = K_M \left(1 + \frac{[I]}{K_i} \right),
$$
where $K_i$ is the inhibition constant. The overall rate becomes:
$$
v = \frac{V_{\text{max}} \cdot [A]}{K_M^{\text{app}} + [A]},
$$
with $V_{\text{max}}$ unchanged.

\vspace{1em}
\noindent \textbf{Ordered sequential binding kinetics} 

For a two-substrate reaction following a sequential ordered mechanism:
$$
E + A \underset{k_{RA}}{\overset{k_{FA}}{\rightleftarrows}} EA 
\underset{k_{RB}}{\overset{k_{FB}}{\rightleftarrows}} EAB 
\overset{k_{cat}}{\rightarrow} E + P + Q.
$$
The enzyme first binds substrate $A$, then $B$, forming the ternary complex $EAB$, which is subsequently converted into products $P$ and $Q$, while regenerating the enzyme $E$.

Assuming rapid equilibrium for complex formation, the reaction rate is given by:
$$
v = \frac{V_{\text{max}} \cdot [A][B]}{K_{iA} K_B + K_B [A] + [A][B]},
$$
where $K_{iA} = \frac{k_{RA}}{k_{FA}}$ is the dissociation constant for $A$, $K_B = \frac{k_{RB} + k_{cat}}{k_{FB}}$ is the effective constant for $B$, and $V_{\text{max}}$ is defined analogously to Michaelis–Menten kinetics.

Ordered sequential binding kinetics can also be extended to account for competitive inhibition by introducing an apparent Michaelis constant, similar to the approach used in Michaelis–Menten kinetics.

\newpage
\section{Appendix: Modular Structure of the IVT Reaction Network}
\label{appendix:IVT_Process}

\begin{figure*}[h]
    \centering
    \includegraphics[width=0.9\linewidth]{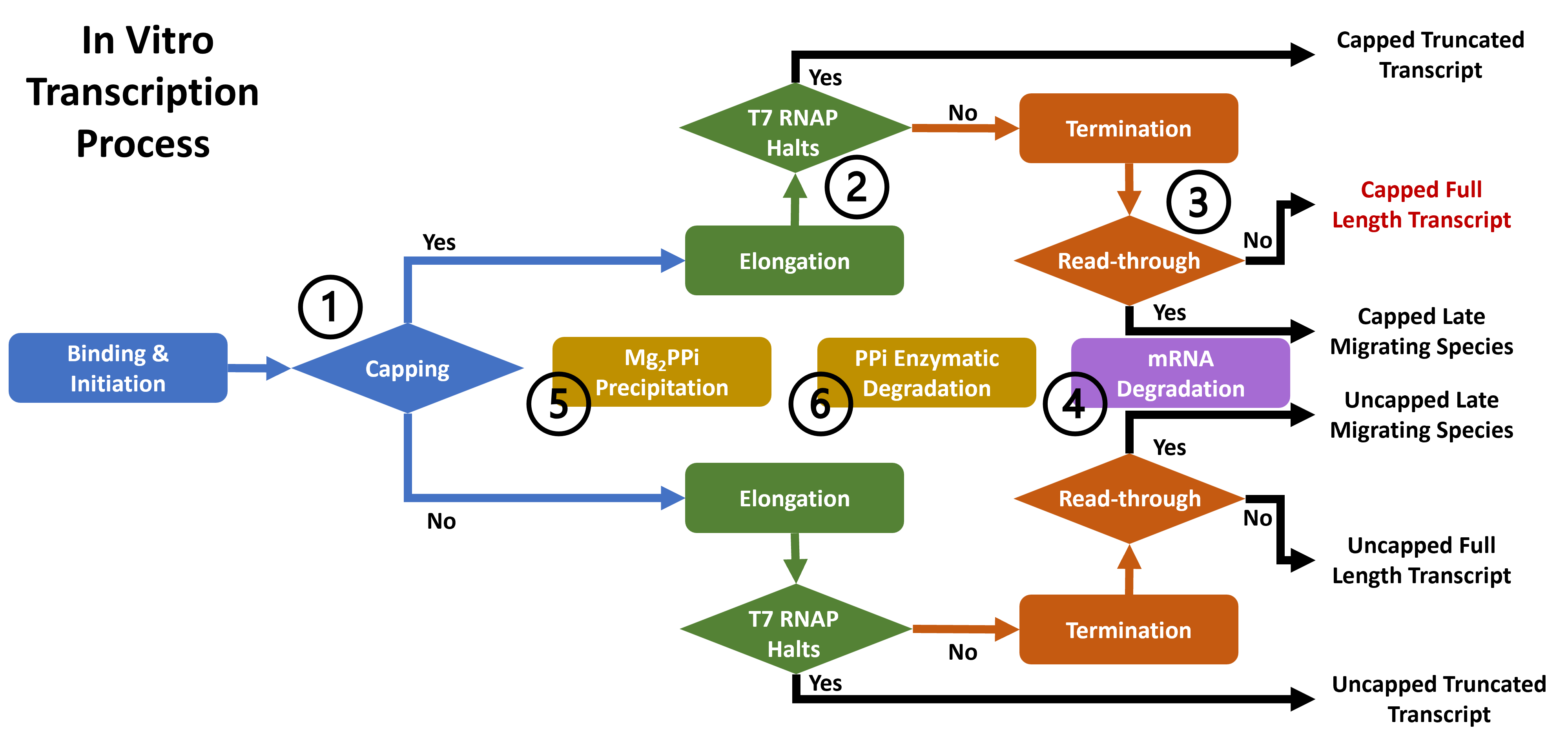}
    \caption{The dynamic model of the enzymatic IVT reaction network consists of the following interconnected modules:
    \textcircled{\scalebox{0.9}{\raisebox{-0.5pt}{1}}} Initiation and Capping, \textcircled{\scalebox{0.9}{\raisebox{-0.5pt}{2}}} Elongation and Truncation, \textcircled{\scalebox{0.9}{\raisebox{-0.5pt}{3}}} Termination and Read-through, \textcircled{\scalebox{0.9}{\raisebox{-0.5pt}{4}}} mRNA Transcript Degradation, \textcircled{\scalebox{0.9}{\raisebox{-0.5pt}{5}}} Mg$_2$PPi Precipitation, and \textcircled{\scalebox{0.9}{\raisebox{-0.5pt}{6}}} Enzymatic Degradation of PPi.  
    }
    \label{fig:IVT_Process}
\end{figure*}

\newpage
\section{Appendix: Kinetic Rate Models of the IVT Process}
\label{appendix:kinetic_models}

\subsection*{Initiation, Capping, and Abortive Cycling}

\begin{align}
v^{\text{initial-cap}} \;=\;&
\prod_{q\in\{\text{pH, Temp, linearity}\}} act^\text{initial-cap}_{\text{T7},q} \times k^{\text{initial-cap}}_\text{T7} \times [\mbox{T7RNAP}] \times \frac{[\mbox{DNA}]}{[\mbox{DNA}] + K_\text{M,DNA}} \times \frac{[\mbox{Mg}^{2+}]}{[\mbox{Mg}^{2+}] + K^{\text{initial-cap}}_\text{M,Mg}} \nonumber
\\ &
\times \frac{[\mbox{MgCleanCap}]}{
[\mbox{MgCleanCap}] + K^\text{initial}_\text{M,MgCleanCap} \times \left(1 + \frac{[\mbox{MgATP}]}{K_\text{I,MgATP}} + \frac{[\mbox{MgUTP}]}{K_\text{I,MgUTP}} + \frac{[\mbox{MgCTP}]}{K_\text{I,MgCTP}} + \frac{[\mbox{MgGTP}]}{K_\text{I,MgGTP}}\right)}
\label{eq.inital-cap}
\\[15pt]
v^\text{initial-uncap} \;=\;&
\prod_{q\in\{\text{pH, Temp, linearity}\}} act^\text{initial-uncap}_{\text{T7},q} \times k^{\text{initial-uncap}}_\text{T7} \times [\mbox{T7RNAP}] \times \frac{[\mbox{DNA}]}{[\mbox{DNA}] + K_\text{M,DNA}} 
\times \frac{[\mbox{Mg}^{2+}]}{[\mbox{Mg}^{2+}] + K^{\text{initial-uncap}}_\text{M,Mg}} \nonumber
\\ & 
\times \frac{[\mbox{MgATP}][\mbox{MgGTP}]}{{K_\text{M,MgATP}^\text{initial}}^\prime{K_\text{M,MgGTP}^\text{initial}}^\prime +
{K_\text{M,MgGTP}^\text{initial}}^\prime[\mbox{MgATP}] + [\mbox{MgATP}][\mbox{MgGTP}]}
\label{eq.inital-uncap} 
\\[15pt]
{K_\text{M,MgATP}^\text{initial}}^\prime \;=\;& K_\text{M,MgATP}^\text{initial} \times \left(1 + \frac{[\mbox{MgGTP}]}{K_\text{I,MgGTP}} + \frac{[\mbox{MgUTP}]}{K_\text{I,MgUTP}} + \frac{[\mbox{MgCTP}]}{K_\text{I,MgCTP}} + \frac{[\mbox{MgCleanCap}]}{K_\text{I,MgCleanCap}}\right)
\label{eq.Km_ATP_prime_init}
\\[15pt]
{K_\text{M,MgGTP}^\text{initial}}^\prime \;=\;&  K_\text{M,MgGTP}^\text{initial} \times \left(1 + \frac{[\mbox{MgATP}]}{K_\text{I,MgATP}} + \frac{[\mbox{MgUTP}]}{K_\text{I,MgUTP}} + \frac{[\mbox{MgCTP}]}{K_\text{I,MgCTP}} + \frac{[\mbox{MgCleanCap}]}{K_\text{I,MgCleanCap}}\right)
\label{eq.Km_GTP_prime_init}
\end{align}

\subsection*{Elongation and Truncation}

\begin{align}
v_\text{ATP}^\text{elongate} \;=\;&
\prod_{q\in\{\text{pH, Temp, linearity}\}} act^\text{elongate}_{\text{T7},q} 
\times k^\text{elongate}_\text{T7,ATP} \times [\mbox{EC}] 
\times \frac{[\mbox{Mg}^{2+}]}{[\mbox{Mg}^{2+}] + K^\text{elongate}_\text{M,Mg}}   \nonumber \\
&\times \frac{[\mbox{MgATP}]}{
[\mbox{MgATP}] + K^\text{elongate}_\text{M,MgATP} \times \left(1 + \frac{[\mbox{MgUTP}]}{K_\text{I,MgUTP}} + \frac{[\mbox{MgCTP}]}{K_\text{I,MgCTP}} + \frac{[\mbox{MgGTP}]}{K_\text{I,MgGTP}}\right)}
\label{eq.elongate_ATP}
\\[15pt]
v_\text{UTP}^\text{elongate} \;=\;&
\prod_{q\in\{\text{pH, Temp, linearity}\}} act^\text{elongate}_{\text{T7},q}
\times k^\text{elongate}_\text{T7,UTP} \times [\mbox{EC}] 
\times \frac{[\mbox{Mg}^{2+}]}{[\mbox{Mg}^{2+}] + K^\text{elongate}_\text{M,Mg}}   \nonumber \\
&\times \frac{[\mbox{MgUTP}]}{
[\mbox{MgUTP}] + K^\text{elongate}_\text{M,MgUTP} \times \left(1 + \frac{[\mbox{MgATP}]}{K_\text{I,MgATP}} + \frac{[\mbox{MgCTP}]}{K_\text{I,MgCTP}} + \frac{[\mbox{MgGTP}]}{K_\text{I,MgGTP}}\right)}
\label{eq.elongate_UTP} 
\\[15pt]
v_\text{CTP}^\text{elongate} \;=\;&
\prod_{q\in\{\text{pH, Temp, linearity}\}} act^\text{elongate}_{\text{T7},q} \times k^\text{elongate}_\text{T7,CTP} \times [\mbox{EC}] 
\times \frac{[\mbox{Mg}^{2+}]}{[\mbox{Mg}^{2+}] + K^\text{elongate}_\text{M,Mg}}   \nonumber \\
&\times \frac{[\mbox{MgCTP}]}{
[\mbox{MgCTP}] + K^\text{elongate}_\text{M,MgCTP} \times \left(1 + \frac{[\mbox{MgATP}]}{K_\text{I,MgATP}} + \frac{[\mbox{MgUTP}]}{K_\text{I,MgUTP}} + \frac{[\mbox{MgGTP}]}{K_\text{I,MgGTP}}\right)}
\label{eq.elongate_CTP}
\\[15pt]
v_\text{GTP}^\text{elongate} \;=\;&
\prod_{q\in\{\text{pH, Temp, linearity}\}} act^\text{elongate}_{\text{T7},q} \times k^\text{elongate}_\text{T7,GTP} \times [\mbox{EC}] 
\times \frac{[\mbox{Mg}^{2+}]}{[\mbox{Mg}^{2+}] + K^\text{elongate}_\text{M,Mg}}   \nonumber \\
&\times \frac{[\mbox{MgGTP}]}{
[\mbox{MgGTP}] + K^\text{elongate}_{\text{M,MgGTP}} \times \left(1 + \frac{[\mbox{MgATP}]}{K_\text{I,MgATP}} + \frac{[\mbox{MgUTP}]}{K_\text{I,MgUTP}} + \frac{[\mbox{MgCTP}]}{K_\text{I,MgCTP}}\right)}
\label{eq.elongate_GTP}
\\[15pt]
v_\text{full}^\text{elongate} \;=\;& \frac{1}{ \frac{C_\text{ATP}}{v_\text{ATP}^\text{elongate}} + \frac{C_\text{UTP}}{v_\text{UTP}^\text{elongate}}  + \frac{C_\text{CTP}}{v_\text{CTP}^\text{elongate}} + \frac{C_\text{GTP}}{v_\text{GTP}^\text{elongate}} }
\label{eq.elongate_full}
\\[15pt]   
v_\text{trunc}^\text{elongate} \;=\;&
\frac{K^\text{elongate}_{\text{M,MgNTP}_{\min}}}{[\mbox{MgNTP}]_{\min} + K^\text{elongate}_{\text{M,MgNTP}_{\min}}} \times k_\text{ratio} \times v_\text{total}^\text{elongate}
\label{eq.elongate_trunc}
\\[15pt]
v_\text{total}^\text{elongate} \;=\;&
v_\text{trunc}^\text{elongate} + v_\text{full}^\text{elongate}
\nonumber
\end{align}

\subsection*{Termination and Read-through}

\begin{align}
v_\text{full}^\text{terminate} \;=\;& k_\text{full}^\text{terminate} \times [\text{EC}_J]
\label{eq.term_full}
\\[15pt]
v_\text{LMS}^\text{terminate}  \;=\;& \alpha^{\text{terminate}}_\text{LMS} \times \frac{K^\text{terminate}_{\text{M},\text{Mg}}}{[\mbox{Mg}^{2+}] + K^\text{terminate}_{\text{M},\text{Mg}}} \times v_\text{full}^\text{terminate}
\label{eq.term_LMS} 
\\[15pt]
\alpha^{\text{terminate}}_\text{LMS} \;=\;&  \gamma_{\text{LMS}}^{\text{linear}} \times \text{linearity} + \gamma_{\text{LMS}}^{\text{circular}} \times (1 - \text{linearity}) 
\nonumber
\end{align}

\subsection*{mRNA Degradation}
\begin{equation}
    v^\text{degrade} = (k_\text{ac}[\mbox{H}^+]^{n_\text{ac}} + k_\text{ba}[\mbox{OH}^-]^{n_\text{ba}} +k_\text{Mg}[\mbox{Mg}^{2+}]^{n_\text{Mg}}) \times [\mbox{RNA}]^{n_\text{RNA}}
\label{eq.degrade}
\end{equation}

\subsection*{Mg$_2$PPi Precipitation}
\begin{equation}
    v^\text{precip} = \max\{0, k^\text{precip} \times ([\mbox{Mg$_2$PPi}] - [\mbox{Mg$_2$PPi}]_\text{eq})\}
\label{eq.precip}
\end{equation}

\subsection*{PPi Enzymatic Degradation}
\begin{equation}
    v^\text{PPase} = 
    act_\text{PPase,pH} \times 
    k^\text{PPase} \times \text{[PPase]} \times \frac{[\text{MgPPi}]}{[\text{MgPPi}] + K_\text{M,PPase}}
\label{eq.PPase}
\end{equation}

\subsection*{Enzyme Activity and Stability}

\begin{align}
act_\text{pH} \;=\;& e^{-\big({\frac{|\text{pH} - \text{pH$_{\text{opt}}$}|}{\sigma_\text{pH}}}\big)^{n_\text{pH}}}
\label{eq:act_pH}
\\[15pt]
act_\text{Temp} \;=\;& Q^{\frac{|\text{Temp}-\text{Temp}_\text{opt}|}{10}}_{10}
\label{eq:act_Temp}
\\[15pt]
act_\text{linearity} \;=\;& \beta_{\text{activity}}^{\text{linear}} \times \text{linearity} + \beta_{\text{activity}}^{\text{circular}} \times (1 - \text{linearity})
\label{eq:act_linear}
\end{align}

\newpage
\section{Appendix: Kinetic Parameters Estimation}
\label{appendix:parameter}

\begin{table}[htb!]
\centering
\fontsize{9.5}{12}\selectfont
\caption{Estimated Kinetic Parameters of the Developed Mechanistic \textit{in silico} Model}
\label{tab:parameterestimation}
\begin{tabular}{lll|lll}
\toprule
\multicolumn{6}{c}{\textbf{Initiation}} \\
\midrule
Name & Units & Fitted Value & Name & Units & Fitted Value \\
\midrule
$n^{\text{initial-cap}}_{\text{pH}}$ & -- & 2.11 & $n^{\text{initial-uncap}}_{\text{pH}}$ & -- & 8.82 \\
$\text{pH}^\text{initial-cap}_\text{opt}$ & -- & 6.30 & $\text{pH}^\text{initial-uncap}_\text{opt}$ & -- & 7.50 \\
$\sigma^\text{initial-cap}_\text{pH}$ & -- & 1.29 & $\sigma^\text{initial-uncap}_\text{pH}$ & -- & 1.42 \\
$Q_{10}^\text{initial-cap}$ & -- & 0.11 & $Q_{10}^\text{initial-uncap}$ & -- & 0.27 \\
$\text{Temp}^\text{initial-cap}_\text{opt}$ & $^\circ\text{C}$ &  42.34 & $\text{Temp}^\text{initial-uncap}_\text{opt}$ & $^\circ\text{C}$ &  41.20 \\
$k^\text{initial-cap}_\text{T7}$ & $\mathrm{min}^{-1}$ & $5.19 \times 10^{3}$ & $k^\text{initial-uncap}_\text{T7}$ & $ \mathrm{min}^{-1}$ & $1.01 \times 10^{4}$ \\
$K^\text{initial-cap}_\text{M,Mg}$ & M & $1.19 \times 10^{-4}$ & $K^\text{initial-uncap}_\text{M,Mg}$ & M & $1.37 \times 10^{-3}$ \\
$K_\text{M,MgCleanCap}^\text{initial}$ & M & $1.88 \times 10^{-2}$ & $K_\text{M,MgATP}^\text{initial}$ & M & $2.71 \times 10^{-2}$ \\
$K_\text{M,MgGTP}^\text{initial}$ & M & $2.71 \times 10^{-2}$ & & \hphantom{$\cdot \mathrm{min}^{-1} \cdot (\mathrm{U/mL})^{-1}$} & \\
\midrule
\multicolumn{6}{c}{\textbf{Initiation \& Elongation}} \\
\midrule
$\beta^\text{linear}_\text{activity}$ & -- &  1 & $\beta^\text{circular}_\text{activity}$ & --  & 0.18 \\
$K_\text{M,DNA}$ & M &  $6.74 \times 10^{-8}$ & $K_\text{I,MgCleanCap}$ & M & $2.60 \times 10^{-2}$ \\
$K_\text{I,MgATP}$ & M & $7.80 \times 10^{-2}$ & $K_\text{I,MgUTP}$ & M & $9.00 \times 10^{-2}$ \\
$K_\text{I,MgCTP}$ & M & $9.80 \times 10^{-2}$ & $K_\text{I,MgGTP}$ & M & $3.70 \times 10^{-2}$ \\
\midrule
\multicolumn{6}{c}{\textbf{Elongation}} \\
\midrule
$Q_{10}^\text{elongate}$ & -- & 0.07 & $\text{Temp}^\text{elongate}_\text{opt}$ & $^\circ\text{C}$ &  38.75 \\
$n^{\text{elongate}}_{\text{pH}}$ & -- &  2.58  & $\text{pH}^\text{elongate}_\text{opt}$ & -- &  7.07 \\
$\sigma^\text{elongate}_\text{pH}$ & -- & 0.97 & 
$k^\text{elongate}_\text{T7,ATP}$ & $\mathrm{min}^{-1}$ & $3.58 \times 10^{3}$ \\
$k^\text{elongate}_\text{T7,UTP}$ & $\mathrm{min}^{-1}$ & $1.41 \times 10^{4}$ & $k^\text{elongate}_\text{T7,CTP}$ & $\mathrm{min}^{-1}$ & $4.57 \times 10^{3}$ \\
$k^\text{elongate}_\text{T7,GTP}$ & $\mathrm{min}^{-1}$ & $6.39 \times 10^{3}$ & $K^\text{elongate}_\text{M,Mg}$ & M & $5.67 \times 10^{-5}$ \\
$K^\text{elongate}_\text{M,MgATP}$ & M & $7.00 \times 10^{-3}$ & $K^\text{elongate}_\text{M,MgUTP}$ & M & $8.00 \times 10^{-3}$\\
$K^\text{elongate}_\text{M,MgCTP}$ & M & $3.40 \times 10^{-3}$ &
$K^\text{elongate}_\text{M,MgGTP}$ & M & $2.90 \times 10^{-3}$ \\ $K_{\text{M,MgNTP}_{min}}^\text{elongate}$ & M & $1.40 \times 10^{-3}$ &
$k_\text{ratio}$ & -- & $2.50 \times 10^{-1}$ \\
\midrule
\multicolumn{6}{c}{\textbf{Termination}} \\
\midrule
$k_\text{full}^\text{terminate}$ & $\mathrm{min}^{-1}$ & $2.00 \times 10^{2}$ &
$K_\text{M,Mg}^\text{terminate}$ & M & $5.76 \times 10^{-4}$ \\ 
$\gamma_{\text{LMS}}^{\text{linear}}$ & \% &  $3.76$ &
$\gamma_{\text{LMS}}^{\text{circular}}$ & \% &  84.14 \\
\midrule
\multicolumn{6}{c}{\textbf{mRNA Degradation}\cite{van2021quality}} \\
\midrule
$k_\text{ac}$ & -- & $2.69 \times 10^{-2}$ & $n_\text{ac}$ & -- &  1 \\
$k_\text{ba}$ & -- & $4.21 \times 10^{-5}$ & $n_\text{ba}$ & -- &  1 \\
$k_\text{Mg}$ & -- & $3.65 \times 10^{-4}$ & $n_\text{Mg}$ & -- &  1 \\
$n_\text{RNA}$ & -- & 1 \\
\midrule
\multicolumn{6}{c}{\textbf{Mg\texorpdfstring{$_2$}{2}PPi Precipitation}\cite{akama2012multiphysics,van2021quality}} \\
\midrule
$k^\text{precip}$ & $\mathrm{min}^{-1}$ &  $0$ & $[\text{Mg$_2$PPi}]_\text{eq}$ & M & $1.40 \times 10^{-5}$ \\
\midrule
\multicolumn{6}{c}{\textbf{PPi Enzymatic Degradation}} \\
\midrule
$n^{\text{PPase}}_{\text{pH}}$ & -- & 1.21 & $\text{pH}^\text{PPase}_\text{opt}$ & -- & 7.77 \\
$\sigma^\text{PPase}_\text{pH}$ & -- & 1.89 & $K_\text{M,PPase}$ & M &  $2.14 \times 10^{-4}$ \\
$k^\text{PPase}$ & $\mathrm{M} \cdot \mathrm{min}^{-1} \cdot (\mathrm{U/mL})^{-1}$ &  $1.48 \times 10^{-3}$ \\
\bottomrule
\end{tabular}
\end{table}

\newpage
\section{Appendix: Mass Balance Equation}
\label{appendix:MBE}

\begin{table}[htb!]
\centering
\renewcommand{\arraystretch}{1.5} 
\caption{The mass balance equations (adapted from \protect\cite{kern1997application,akama2012multiphysics,stover2024mechanistic})}
\label{tab:MassBalance}

\begin{tabular}{|l|l|}
\hline
\textbf{M1} & $[\mbox{Mg}]^{tot} = [\mbox{Mg}] + [\mbox{MgNTP}] + 2 \times [\mbox{Mg$_2$NTP}] + [\mbox{MgHNTP}] +$  $ [\mbox{MgPPi}] + 2 \times [\mbox{Mg$_2$PPi}] + [\mbox{MgHPPi}] + { [\mbox{MgPi}]}$ \\ \hline
\textbf{M2} & $[\mbox{NTP}]^{tot} = [\mbox{NTP}] + [\mbox{MgNTP}] + [\mbox{Mg$_2$NTP}] + [\mbox{MgHNTP}] + [\mbox{HNTP}]$ \\ \hline
\textbf{M3} & $[\mbox{H}]^{tot} = [\mbox{H}] + [\mbox{MgHNTP}] + [\mbox{HNTP}] + [\mbox{HPPi}] + 2 \times [\mbox{H$_2$PPi}] + [\mbox{MgHPPi}] + [\mbox{HPi}]$ \\
            & \quad  \quad  \quad  \quad  \quad $ {[\mbox{HTris}] + [\mbox{HDTT}] + [\mbox{Hspermidine}]}$ \\ \hline
\textbf{M4} & $[\mbox{PPi}]^{tot} = [\mbox{PPi}] + [\mbox{MgPPi}] + [\mbox{Mg$_2$PPi}] + [\mbox{HPPi}] + [\mbox{H$_2$PPi}] + [\mbox{MgHPPi}]$ \\ \hline
\textbf{M5} & ${[\mbox{Tris}]^{tot} = [\mbox{Tris}] + [\mbox{HTris}]}$ \\ \hline
\textbf{M6} & ${[\mbox{DTT}]^{tot} = [\mbox{DTT}] + [\mbox{HDTT}]}$ \\ \hline
\textbf{M7} & $
{[\mbox{spermidine}]^{tot} = [\mbox{spermidine}] + [\mbox{Hspermidine}]}$ \\ \hline
\textbf{M8} & $[\mbox{Pi}]^{tot} = [\mbox{Pi}] + [\mbox{HPi}] + [\mbox{MgPi}]$ \\ \hline
\end{tabular}
\end{table}

\begin{table}[htb!]
\centering
\renewcommand{\arraystretch}{1.5} 
\caption{The equilibrium equations (adapted from \protect\cite{kern1997application,akama2012multiphysics,stover2024mechanistic})}
\label{tab:Equilibrium}

\begin{tabular}{|c|l||c|l|}
\hline
\textbf{E1}  &  $[\mbox{H}][\mbox{NTP}] = K_\text{HNTP}[\mbox{HNTP}]$   & \textbf{E2}  & $[\mbox{Mg}][\mbox{NTP}] = K_\text{MgNTP}[\mbox{MgNTP}]$    \\ \hline
\textbf{E3}  & $[\mbox{Mg}][\mbox{MgNTP}] = K_\text{Mg$_2$NTP}[\mbox{Mg$_2$NTP}]$ & \textbf{E4}  & $[\mbox{Mg}][\mbox{HNTP}] = K_\text{MgHNTP}[\mbox{MgHNTP}]$ \\ \hline
\textbf{E5} & $[\mbox{Mg}][\mbox{PPi}] = K_\text{MgPPi}[\mbox{MgPPi}]$   & \textbf{E6} & $[\mbox{Mg}][\mbox{MgPPi}] = K_\text{Mg$_2$PPi}[\mbox{Mg$_2$PPi}]$  \\ \hline
\textbf{E7} & $[\mbox{H}][\mbox{PPi}] = K_\text{HPPi}[\mbox{HPPi}]$   & \textbf{E8} & $[\mbox{H}][\mbox{HPPi}] = K_\text{H$_2$PPi}[\mbox{H$_2$PPi}]$   \\ \hline
\textbf{E9} & $[\mbox{Mg}][\mbox{HPPi}] = K_\text{MgHPPi}[\mbox{MgHPPi}]$  & \textbf{E10} &  {$[\mbox{H}][\mbox{Tris}] = K_\text{Tris$^+$} [\mbox{Tris$^+$}]$} \\ \hline
\textbf{E11} & {$[\mbox{H}][\mbox{DTT}] = K_\text{DTT$^+$}[\mbox{DTT$^+$}]$}  & \textbf{E12} & {$[\mbox{H}][\mbox{spermidine}] = K_\text{spermidine$^+$}[\mbox{spermidine$^+$}]$} \\ \hline
\textbf{E13}  &  {$[\mbox{Mg}][\mbox{Pi}] = K_\text{MgPi}[\mbox{MgPi}]$}   & \textbf{E14}  &  {$[\mbox{H}][\mbox{Pi}] = K_\text{HPi}[\mbox{HPi}]$}\\ \hline
\end{tabular}
\end{table}

Due to limited information on its equilibrium dissociation constants in the existing literature, CleanCap is considered a specialized NTP and is assumed to share the same equilibrium dissociation constants as other NTPs in this study.
According to Kern et al. (1997) \cite{kern1997application}, Akama et al. (2012) \cite{akama2012multiphysics}, and Stover et al. (2024) \cite{stover2024mechanistic}, the equilibrium association constants used in the calculations are as follows: $pK_\text{HNTP} = 6.95$, $pK_\text{MgNTP} = 4.42$, $pK_\text{Mg$_2$NTP} =1.69$, $pK_\text{MgHNTP} = 1.49$, $pK_\text{HPPi} = 8.94$, $pK_\text{MgPPi} = 5.42$, $pK_\text{Mg$_2$PPi} = 2.33$, $pK_\text{H$_2$PPi} = 6.13$, $pK_\text{MgHPPi} = 3.05$, $pK_\text{Tris$^+$} = 8.1$, $pK_\text{DTT$^+$} = 9.21$, $pK_\text{spermidine$^+$} = 8.34$,  $pK_\text{MgPi} = 1.88$ and $pK_\text{HPi} = 6.92$. 

\newpage
\section{Appendix: Model Performance Assessment Criteria}
\label{appendix:goodness-of-fit}


Suppose there are $K$ batches of experimental observations denoted as $\pmb{s}_{t_h}^{k}$  with $k=1, 2,\ldots, K$, including both discrete and time-course data. Let $t_h$ with $h=1,2,\ldots,H^k$ represent the observation times for each $k$-th batch, which vary by batch. The predictions of the mechanistic model specified by parameters $\pmb{\theta}$ are denoted by $\{\hat{\pmb{s}}_{t_h}^{k}\}$. Given any inputs $\pmb{x}$, the prediction loss function, denoted by $\mathcal{L}(\pmb{x},\pmb{\theta})$, is evaluated based on the model prediction errors. 
Mean Absolute Error (MAE), Weighted Absolute Percentage Error (WAPE), and rank correlation \cite{hastie2009elements} were considered in defining the loss function and measuring the model prediction performance in this study.

\textbf{(1) MAE} evaluates the prediction accuracy by quantifying 
the expected absolute prediction errors, i.e.,
\begin{equation}
    \text{MAE} = \frac{1}{\sum_{k=1}^K H^k}\sum_{k=1}^K\sum_{h=1}^{H^k} \Big|\pmb{s}^k_{t_{h}} - \hat{\pmb{s}}^k_{t_{h}}\Big|.
\label{eq:MAE}
\end{equation}

\textbf{(2) WAPE} is a unit-free measure evaluating the prediction accuracy 
in term of the total absolute relative error, i.e., 
\begin{equation}
    \text{WAPE} = \frac{\sum_{k=1}^K\sum_{h=1}^{H^k} \Big|\pmb{s}^k_{t_{h}} - \hat{\pmb{s}}^k_{t_{h}}\Big|}{\sum_{k=1}^K\sum_{h=1}^{H^k}\pmb{s}^k_{t_{h}} } \times 100\%.
\label{eq:WAPE}
\end{equation}

\textbf{(3) Rank correlation} evaluates the strength and direction of the relationship between two variables (i.e., experimental observations and model predictions)  
based on their ranks rather than their actual values. 
Unlike traditional correlation measures, such as Pearson correlation, which assess linear relationships, rank correlation evaluates how well the variables maintain a monotonic relationship, where one consistently increases or decreases with the other. This makes it particularly useful for assessing the similarity of 
non-linear relationships  
as it prioritizes the relative order of values rather than their magnitudes. 
In this study, Spearman's Rank Correlation Coefficient \cite{spearman1904proof} was implemented and denoted as $\rho$,
\begin{equation}
    \rho = 1 - \frac{6 \sum_{k=1}^{K}\sum_{h=1}^{H^k} (d_h^k)^2}
    {\sum_{k=1}^K H^k \left[(\sum_{k=1}^K H^k)^2 - 1\right]},
\end{equation}
where $d_h^k \equiv R[\pmb{s}_{t_h}^{k}]-R[\hat{\pmb{s}}_{t_h}^{k}]$ represents the difference between the ranks, denoted by $R[\pmb{s}_{t_h}^{k}]$ and $R[\hat{\pmb{s}}_{t_h}^{k}]$, of the sequences of experimental observations and model predictions, i.e., $\{\pmb{s}_{t_h}^{k}\}$ and $\{\hat{\pmb{s}}_{t_h}^{k}\}$ with $h=1,2,\ldots,H^k$ for each $k$-th batch.

\newpage
\section{Appendix: Model Fitting through Batch Bayesian Optimization (BO)}
\label{appendix:BOinference}

The optimal mechanistic model parameters are determined by minimizing the average prediction loss, i.e.,
\begin{equation}
    \pmb{\theta}^\star \equiv \underset{\pmb{\theta}}{\operatorname{argmin}}\,
   \bar{\mathcal{L}}(\pmb{\theta}),
    \label{eq.objective}
\end{equation}
where $\bar{\mathcal{L}}(\pmb{\theta})\equiv\int_{\mathcal{X}} \mathcal{L}(\pmb{x},\pmb{\theta})d\pmb{x}$ represents the average prediction loss on the design space $\mathcal{X}$ of input decisions $\pmb{x}$ as shown in Table~\ref{tab:IVT_inputs_outputs}; this objective can be extended to allocate more weights to more critical areas of the decision space $\pmb{x}$. 
However, solving the optimization problem for the IVT process model fitting presents several challenges.
First, each simulation run is time-consuming. 
Second, the parameter space is high-dimensional, encompassing more than 30 parameters, and the objective has complex nonlinear dependence.
Third, the state transition dynamics are jointly governed by mass balance equations and mechanism rate models, resulting in yield and quality attributes that are non-convex functions, making gradient calculation for each parameter challenging.

Therefore, a Gaussian process (GP) surrogate model characterizing the belief over the objective function $\bar{\mathcal{L}}(\pmb{\theta})$ in Equation~(\ref{eq.objective})
is used to guide the sequential selection of simulation experiments and speed up the search for the optimal model parameters $\pmb{\theta}^\star$.  
To further accelerate the parameter search, Batch BO \cite{hunt2020batch} is employed, enabling parallel evaluation of multiple candidate mechanistic parameter sets. Here, “batch” refers to the Bayesian optimization strategy and is unrelated to batch IVT process operations.
Compared with standard Bayesian Optimization, Batch BO visits multiple candidate points simultaneously for parallel evaluation to efficiently explore the parameter space. 

In specific, at each $\ell$-th iteration, Sobol quasi-Monte Carlo sampling \cite{caflisch1998monte} is used to generate $B$ batches of candidate points, denoted by $\pmb{\Theta}_1, \pmb{\Theta}_2, \ldots, \pmb{\Theta}_B$. Each $b$-th batch $\pmb{\Theta}_b$ contains $q$ candidates, i.e.,
$
\pmb{\Theta}_b = \{ \pmb{\theta}^{(b,1)}, \pmb{\theta}^{(b,2)}, \ldots, \pmb{\theta}^{(b,q)} \}$ for $b = 1,2, \ldots, B$.
Batch BO then uses the q-acquisition function and the surrogate model to guide the selection of each batch for evaluation. The surrogate is iteratively updated with new simulation observations, and the process continues until a specified stopping criterion is met.
By the $\ell$-th iteration of the GP-assisted optimization, the design points visited so far is denoted by $\mathcal{H}_\ell$; this forms the historical dataset $\mathcal{D}_{\ell}\equiv \{(\pmb{\theta}_z, \hat{\bar{\mathcal{L}}}(\pmb{\theta}_z))~\mbox{with}~ \pmb{\theta}_z \in \mathcal{H}_\ell\}$, where $\hat{\bar{\mathcal{L}}}(\pmb{\theta}_z)$ represents the simulation estimated prediction loss of the candidate model specified by parameters $\pmb{\theta}_z$. 
The prior belief of the objective function is modeled as a Gaussian process:
\begin{equation*}
  \bar{\mathcal{L}}(\pmb{\theta}) \sim\mathcal{N}(\mu(\pmb{\theta}),\kappa(\pmb{\theta}, \pmb{\theta})),
\end{equation*}
with $\mu(\cdot)$ and $\kappa(\cdot, \cdot)$ denoting the mean and covariance functions,
which encode prior beliefs about the objective function and the spatial correlations between evaluations \cite{frazier2018tutorial}.
Given historical observations $\mathcal{D}_{\ell}$ at visited design points $\mathcal{H}_{\ell}$, the posterior distribution of $\bar{\mathcal{L}}(\pmb{\Theta})$ at any new selected batch of prediction points denoted by $\pmb{\Theta}$ becomes, 
\begin{eqnarray}
{\bar{\mathcal{L}}}(\pmb{\Theta})|\mathcal{D}_{\ell}&\sim &\mathcal{N}(\mu_{\ell}(\pmb{\Theta}),\Sigma_{\ell}(\pmb{\Theta})),
\label{posterior_dist}\\
\mu_{\ell}(\pmb{\Theta}) &=& \kappa(\pmb{\Theta},\mathcal{H}_\ell)\kappa(\mathcal{H}_\ell,\mathcal{H}_\ell
)^{-1}(\hat{\bar{\mathcal{L}}}(\mathcal{H}_\ell)-\mu(\mathcal{H}_\ell))+\mu(\mathcal{H}_\ell),
\nonumber\\
\Sigma_{\ell}(\pmb{\Theta})&=&\kappa(\pmb{\Theta},\pmb{\Theta})-\kappa(\pmb{\Theta},\mathcal{H}_\ell)\kappa(\mathcal{H}_\ell,\mathcal{H}_\ell)^{-1}\kappa(\mathcal{H}_\ell,\pmb{\Theta}),\nonumber
\end{eqnarray}
where $\hat{\bar{\mathcal{L}}}(\mathcal{H}_\ell)$ represents the vector of simulation estimated model prediction losses at the design points $\mathcal{H}_\ell$.
This GP surrogate model is used to speed up the search for the optimal model parameters minimizing the prediction loss.

In addition, the q-acquisition function helps decide where to visit next in the search space by balancing exploration (checking uncertain areas) and exploitation (focusing on more promising areas). This balance makes the search for the best solution more efficient. Various q-acquisition functions can be employed, such as q-Expected Improvement (q-EI) \cite{ginsbourger2010kriging,janusevskis2012expected}, q-Probability of Improvement (q-PoI) \cite{yang2022parallel}, and q-log Expected Improvement (q-logEI) \cite{ament2024unexpected}; see Balandat et al. (2020) \cite{balandat2020botorch} for a more comprehensive discussion on q-acquisition functions.


In this study, q-EI was selected as the q-acquisition function and it allows multiple parameter sets to be evaluated simultaneously at each $\ell$-th iteration, thereby accelerating the search for optimal values. 
Formally, the q-EI is defined as the expected improvement over the current best observation $\bar{\mathcal{L}}(\pmb{\theta}_\ell^\star)$, considering the joint distribution of the objective function values at these candidate points as modeled by the GP:
\begin{equation}
    \text{q-EI}_{\ell}(\pmb{{\Theta}}_b)=\E_{\ell}\max_{j=1,2,\ldots,q}\left[\max\left(0, \bar{\mathcal{L}}(\pmb\theta^{(b,j)})-\hat{\bar{\mathcal{L}}}(\widehat{\pmb{\theta}}_{\ell}^\star)\right)\right], \mbox{~for~} b=1,2,\ldots,B,
    \label{eq.q-EI}
\end{equation}
where $\widehat{\pmb{\theta}}_{\ell}^\star$ represents the optimal model parameter estimate obtained at the $\ell$-th iteration, i.e.,
$
    \widehat{\pmb{\theta}}_{\ell}^\star={\operatorname{argmin}}_
    {\pmb{\theta}_z \in \mathcal{H}_\ell}
    \hat{\bar{\mathcal{L}}}(\pmb{\theta}_z) $,
and the expectation $\mathbb{E}_{\ell}$ is taken with respect to the posterior distribution of $\bar{\mathcal{L}}(\pmb\Theta)$ conditional on the historical observations $\mathcal{D}_{\ell}$ as shown in Equation~(\ref{posterior_dist}). 
Thus, the next batch of model parameters $\pmb{\Theta}$ for evaluation is chosen by selecting the batch with maximum of q-EI among $B$ batches:
\[
\pmb{\Theta}_{\ell+1}^\star = {\underset{b=1,\ldots,B}{\operatorname{argmax}}\, \text{q-EI}_{\ell}(\pmb{\Theta}_b)}.
\]
Then, let $\mathcal{H}_{\ell+1} = \mathcal{H}_\ell \cup  \pmb{\Theta}_{\ell+1}^\star$, run simulations to estimate the prediction loss at each candidate in 
\(\pmb{\Theta}_{\ell+1}^\star\), and update 
\(\mathcal{D}_{\ell+1}\). 
Once the stopping criterion of optimal search is met at certain $L$-th iteration, 
the optimal estimate with the smallest prediction loss $\widehat{\pmb{\theta}}_{L}^\star$ is returned.
Since analytically computing the expected improvement, q-EI in Equation~(\ref{eq.q-EI}),
is challenging, 
it is often approximated using Monte Carlo simulations or other numerical methods that can efficiently estimate the expectation under the GP posterior \cite{balandat2020botorch}.
Refer to Algorithm~\ref{alg:BOParameterEstimation} on the mechanistic parameters inference in Appendix~\ref{appendix:BOalgorithm} for detailed implementation.

\newpage
\section{Appendix: Batch Bayesian Optimization Algorithm}
\label{appendix:BOalgorithm}

Bayesian optimization is a strategy for the global optimization of an unknown objective function $\mathcal{L}(\pmb\theta)$ that is \textbf{expensive to evaluate}. It typically involves two main components:
\begin{itemize}
    \item \textbf{Surrogate Model}: A probabilistic model, often a Gaussian Process (GP), is used to model the objective function $\mathcal{L}$. Given a set of observations $\mathcal{D}=\{(\pmb{\theta}_n,y_n)\}$, where $\pmb{\theta}_i$ is a point in the input space and $y_i=\mathcal{L}(\pmb\theta_i)$ is its corresponding observation (possibly noisy), the GP provides a posterior distribution over the objective function $p(\mathcal{L}|\mathcal{D})$.
    \item \textbf{q-Acquisition Function}: A function that uses the surrogate model's posterior to score each point in the input space based on its potential to improve over the current best observation. The acquisition function guides the selection of the next point(s) to evaluate. 
\end{itemize}

The Batch Bayesian Optimization (BBO) algorithm was implemented using BoTorch, a PyTorch-based library designed for Bayesian optimization. Refer to Balandat et al. (2020) \cite{balandat2020botorch} for details.

\begin{algorithm}[!h]
\caption{Batch Bayesian Optimization with Parallel Gaussian Processes}
\label{alg:BOParameterEstimation}
\begin{algorithmic}[1]

\Require Objective function $\bar{\mathcal{L}}(\pmb{\theta})$ to minimize
\Require Number of batches $B$
\Require Batch size $q$ for parallel evaluations
\Require Initial dataset $\mathcal{D}_{0} = \{(\pmb{\theta}_z, \hat{\bar{\mathcal{L}}}(\pmb{\theta}_z)) : \pmb{\theta}_z \in \mathcal{H}_0\}$,  
\hspace*{1.2em} where $\mathcal{H}_0$ is generated via Latin Hypercube Sampling
\State Initial iteration index $l \gets 0$
\Ensure Optimized parameters $\pmb{\theta}^\star$

\While{stopping criterion not met}
    \State Update the GP posterior using $\mathcal{D}_l$.
    \State Compute batch acquisition function $\text{q-EI}_l(\pmb{\Theta}_b)$ based on the GP posterior.
    \State Generate $B$ candidate batches $\pmb{\Theta}_1, \dots, \pmb{\Theta}_B$.
    \State Select the next batch:
    \[
    \pmb{\Theta}_{l+1}^\star = \underset{b=1,\dots,B}{\operatorname{argmax}}~\text{q-EI}_l(\pmb{\Theta}_b).
    \]
    \State Update $\mathcal{H}_{l+1} = \mathcal{H}_l \cup \pmb{\Theta}_{l+1}^\star$ and $\mathcal{D}_{l+1} \equiv \{(\pmb{\theta}_z, \hat{\bar{\mathcal{L}}}(\pmb{\theta}_z)) \mid \pmb{\theta}_z \in \mathcal{H}_{l+1}\}$.
    \State Increment $l \gets l+1$.
\EndWhile

\State $\pmb{\theta}^\star \gets 
    \argmin_{\pmb{\theta}_z \in \mathcal{H}_l}\ \hat{\bar{\mathcal{L}}}(\pmb{\theta}_z)$
\State \Return $\pmb{\theta}^\star$

\end{algorithmic}
\end{algorithm}

\newpage
\section{Appendix: Prediction Performance}
\label{appendix:Boman}

\begin{figure}[htbp]
    \centering
    \includegraphics[width=0.9\linewidth]{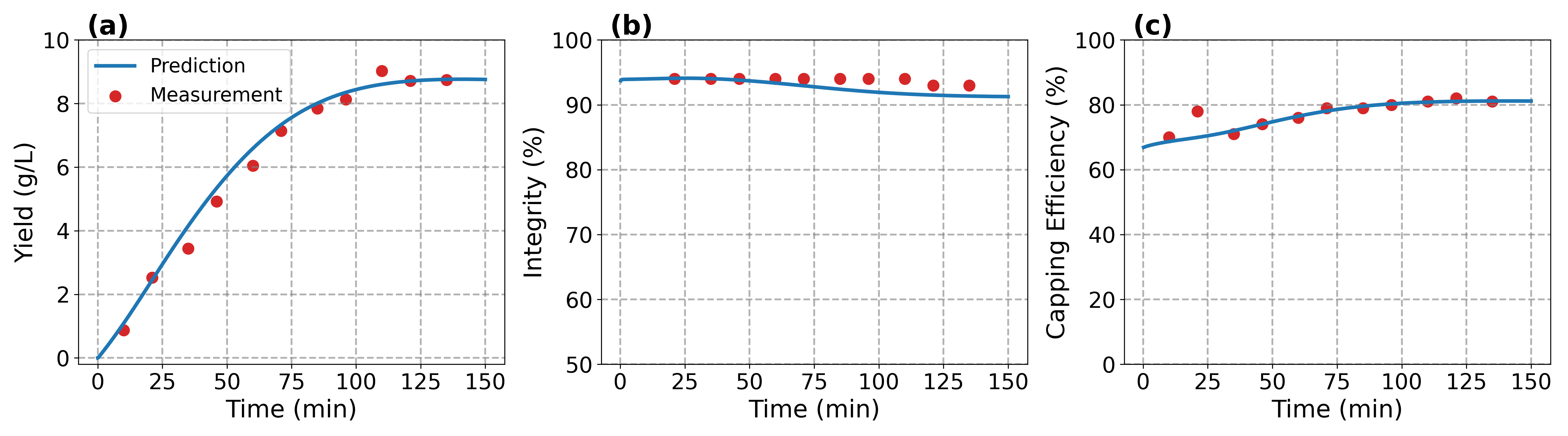}
    \caption{Predicted dynamic profiles of the IVT process for a representative experimental batch. Model predictions are shown for (a) mRNA yield (g/L), (b) integrity (\%), and (c) capping efficiency (\%). Red dots indicate experimental measurements, while blue lines denote model predictions. 
    }
    \label{fig:pred_trajectory}
\end{figure}

\begin{figure}[htbp]
    \centering
    \includegraphics[width=0.6\linewidth]{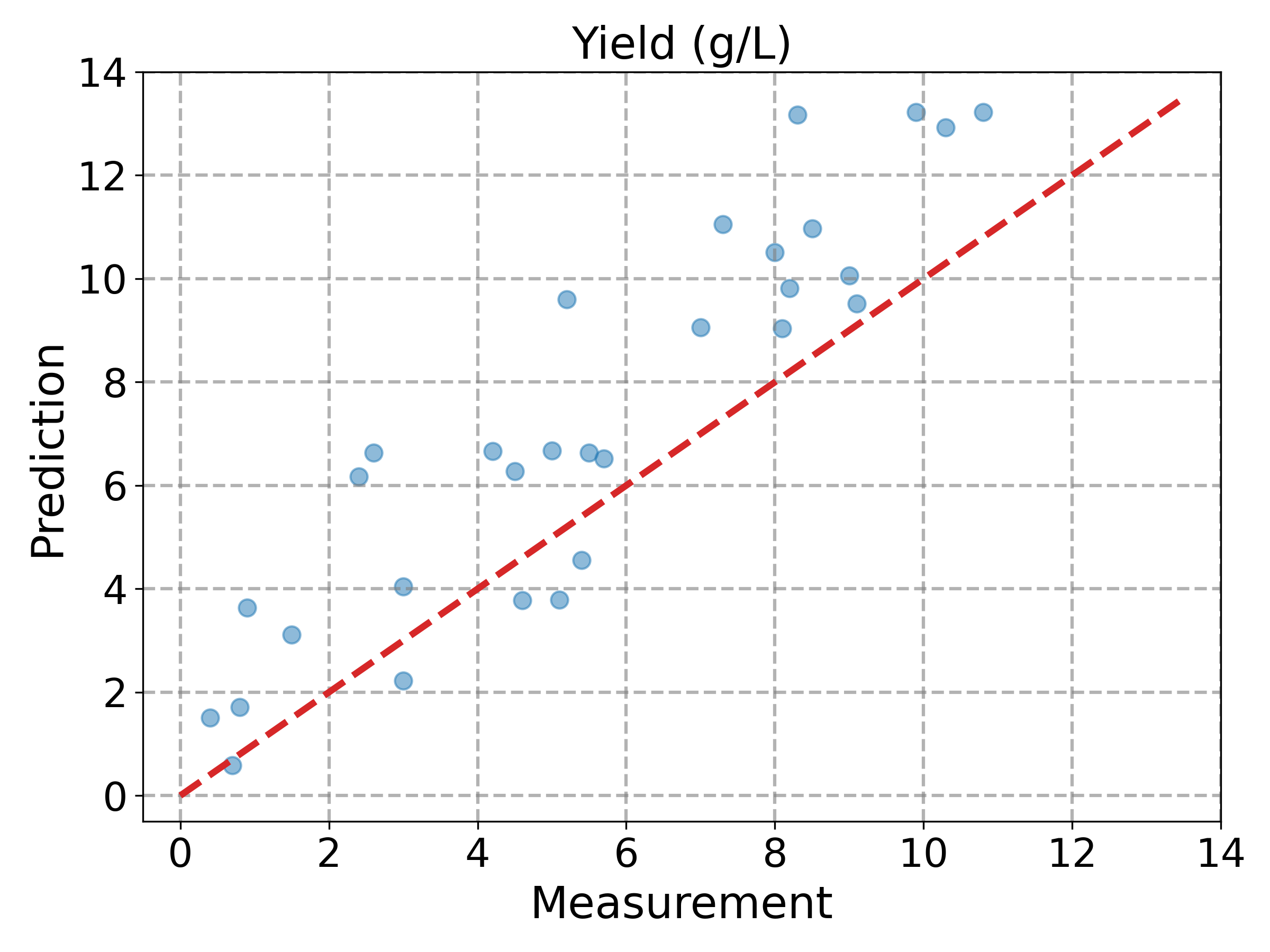}
    \caption{Comparison of model predictions versus experimental observations on the external dataset from Boman et al. (2024) \cite{boman2024quality} for mRNA yield (g/L) using the developed mechanistic \textit{in silico} model. The red dashed identity line represents perfect agreement between model predictions and experimental measurements. 
    }
    \label{fig:QQ_Prediction_Boman}
\end{figure}

\newpage
\section{Appendix: Residual Analysis}
\label{appendix:residual_analysis}


\begin{figure}[htb!] 
    \centering
    \includegraphics[width=\textwidth]{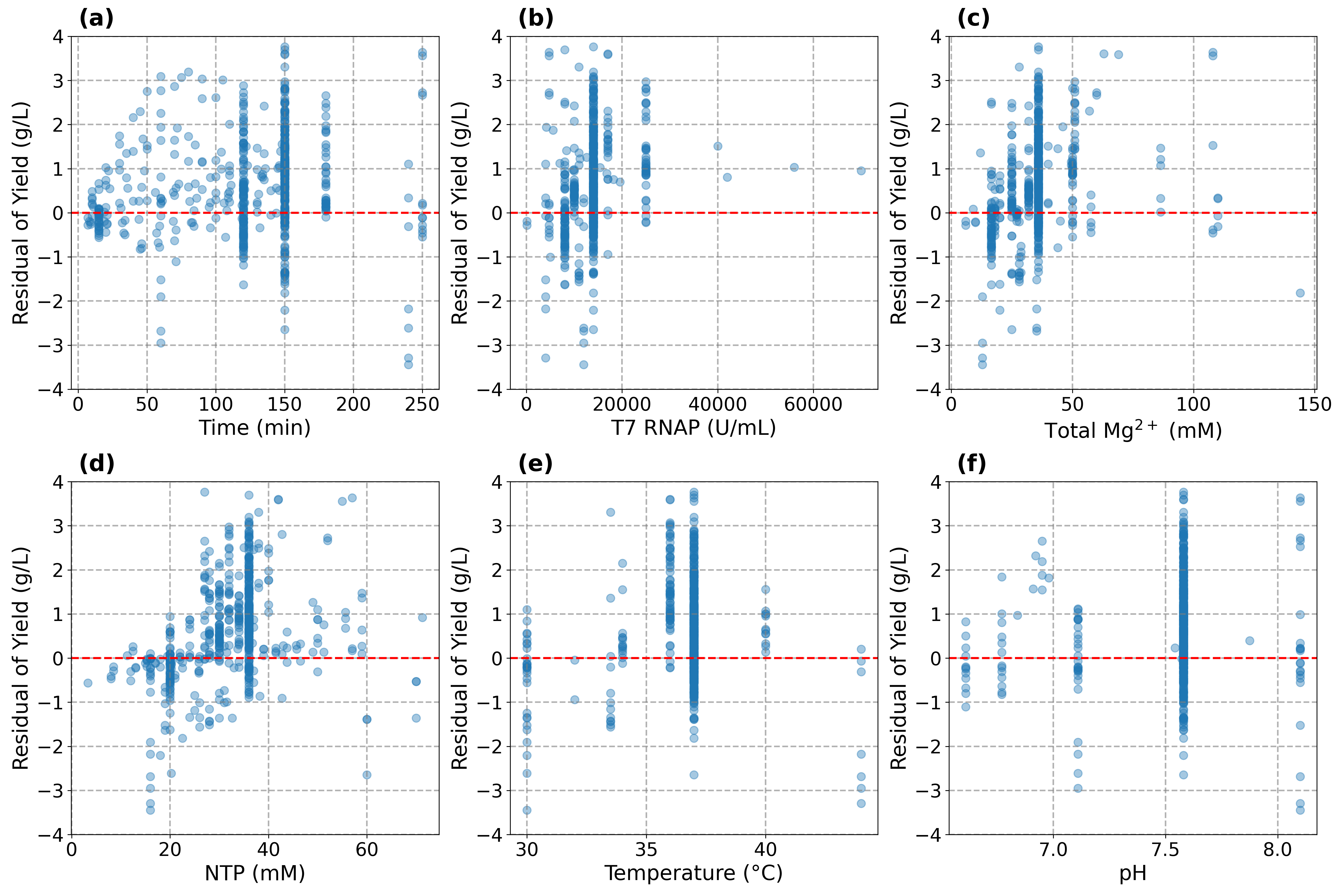}
    \caption{Residual distribution of mRNA yield (g/L) with respect to key process parameters:
    (a) Process time (min);
    (b) T7 RNAP activity (U/mL);
    (c) Total Mg$^{2+}$ concentration (mM);
    (d) NTP concentration (mM);
    (e) Temperature ($^\circ$C);
    (f) pH.
    The red dashed line represents zero residual, serving as a reference for model fit assessment.
    }
    \label{fig:residual_yield}
\end{figure}

\newpage
\begin{figure}[htb!] 
    \centering
    \includegraphics[width=\textwidth]{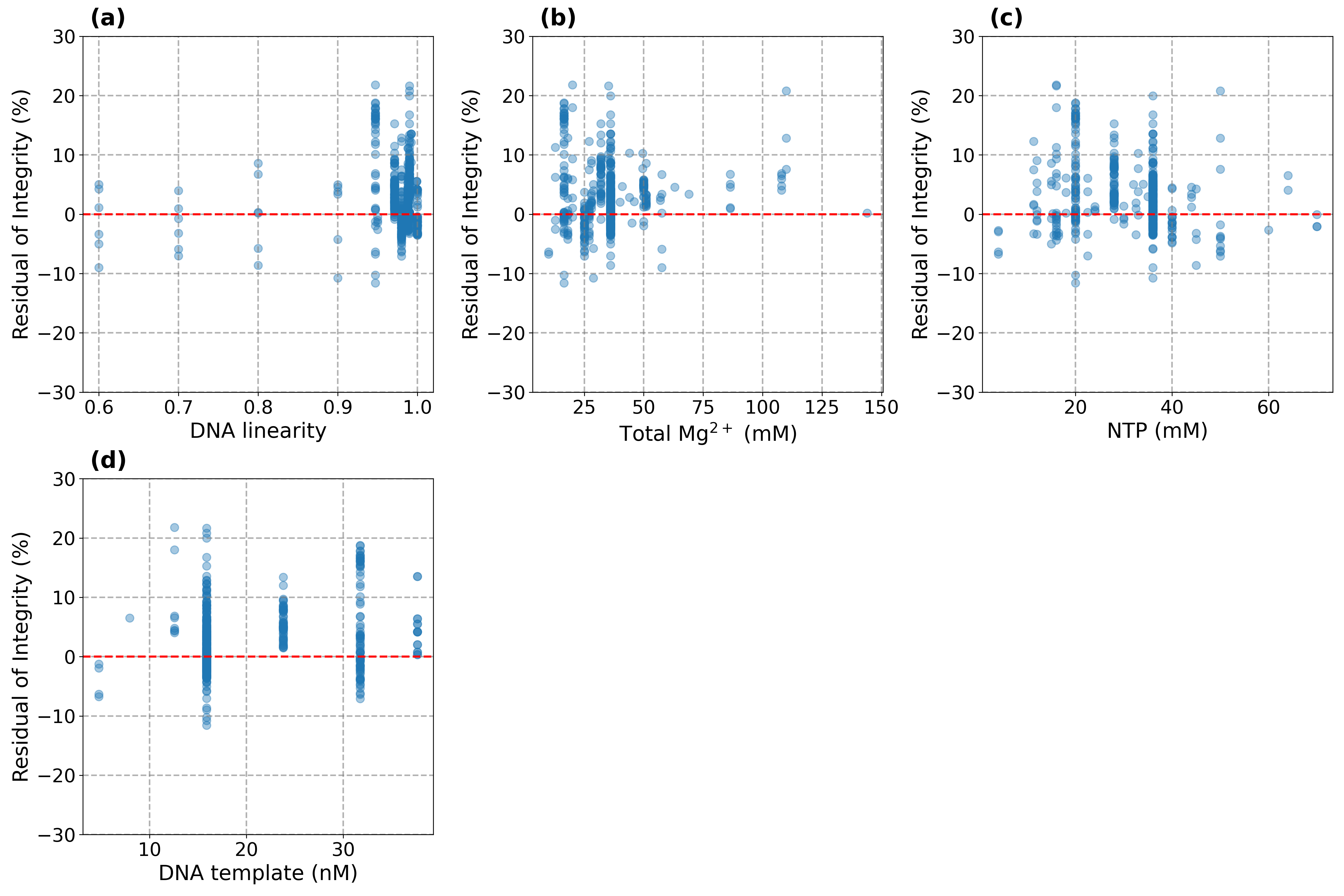}
    \caption{Residual distribution of mRNA integrity (\%) with respect to key process parameters:
    (a) DNA linearity;
    (b) Total Mg$^{2+}$ concentration (mM);
    (c) NTP concentration (mM);
    (d) DNA template concentration (nM) 
    The red dashed line represents zero residual, serving as a reference for model fit assessment.
    }
    \label{fig:residual_integrity}
\end{figure}

\newpage
\begin{figure}[htb!] 
    \centering
    \includegraphics[width=\textwidth]{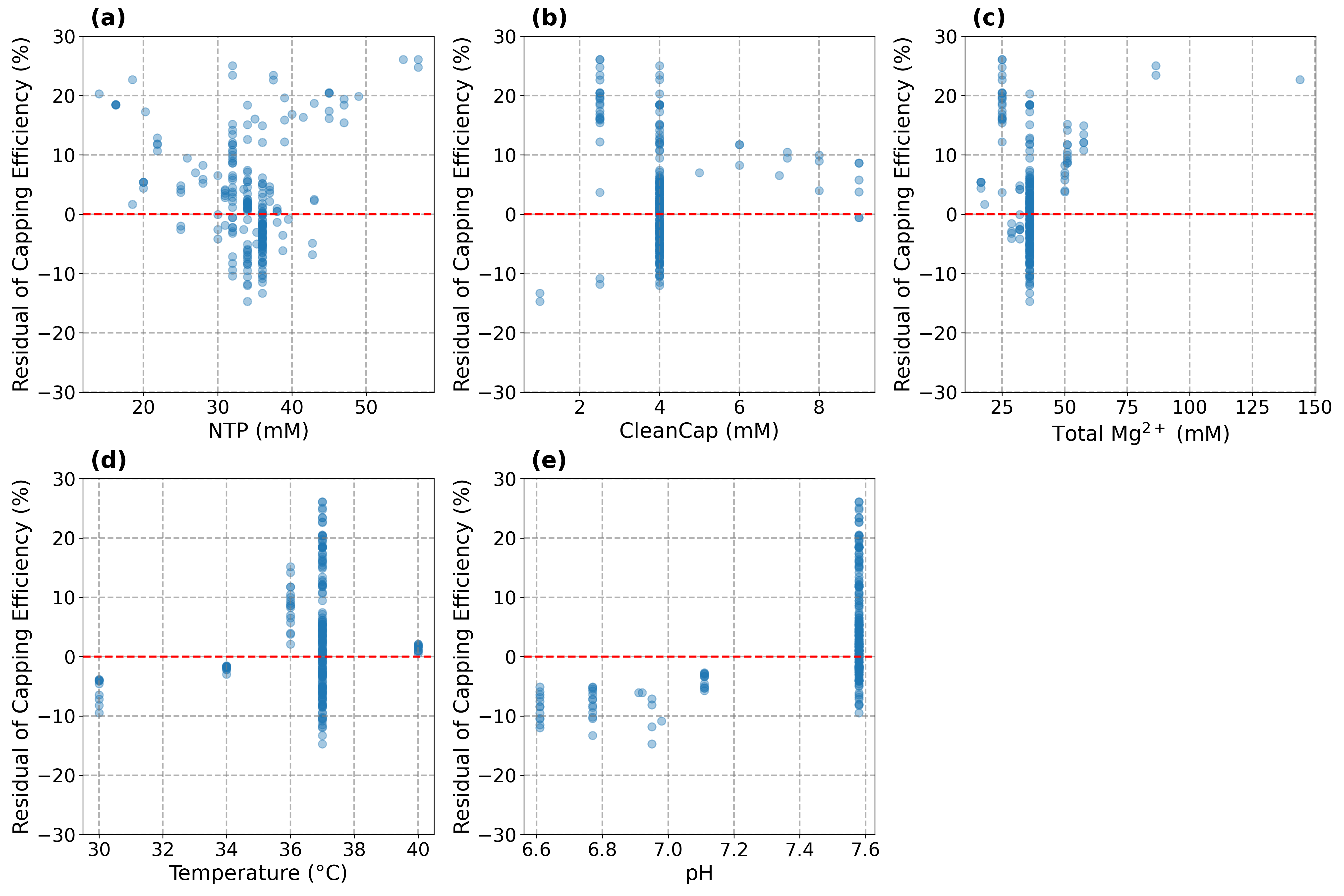}
    \caption{Residual distribution of capping efficiency (\%) with respect to key process parameters:
    (a) NTP concentration (mM);
    (b) CleanCap concentration (mM);
    (c) Total Mg$^{2+}$ concentration (mM);
    (d) Temperature ($^\circ$C);
    (e) pH.
    The red dashed line represents zero residual, serving as a reference for model fit assessment.
    }
    \label{fig:residual_ce}
\end{figure}

\newpage
\section{Appendix: In Silico Simulation Results}
\label{appendix:in-silico}

\begin{figure*}[htb!]
    \centering
    \includegraphics[width=0.9\linewidth]{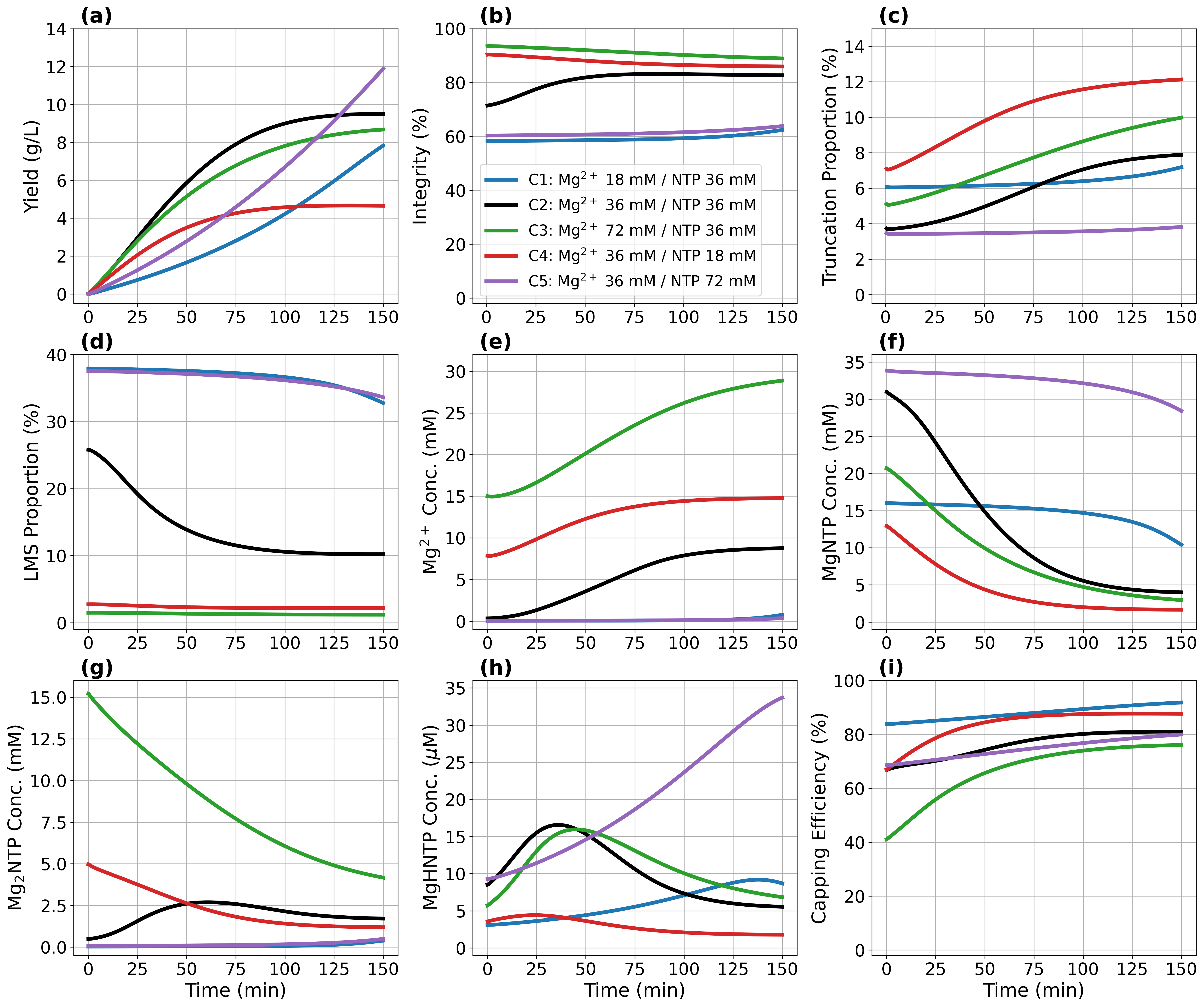}
    \caption{Predicted dynamic profiles of the IVT process under varying total Mg$^{2+}$ and NTP concentrations. Shown trajectories include: (a) mRNA yield (g/L), (b) integrity (\%), (c) truncation proportion (\%), (d) LMS proportion (\%), (e) free Mg$^{2+}$ concentration (mM), (f) MgNTP complex concentration (mM), (g) Mg$_2$NTP complex concentration (mM), (h) MgHNTP complex concentration ($\mu$M), and (i) capping efficiency (\%). 
    }
    \label{fig:trajectory_Mg}
\end{figure*}

\newpage
\begin{figure*}[htb!] 
    \centering
    \includegraphics[width=0.9\textwidth]{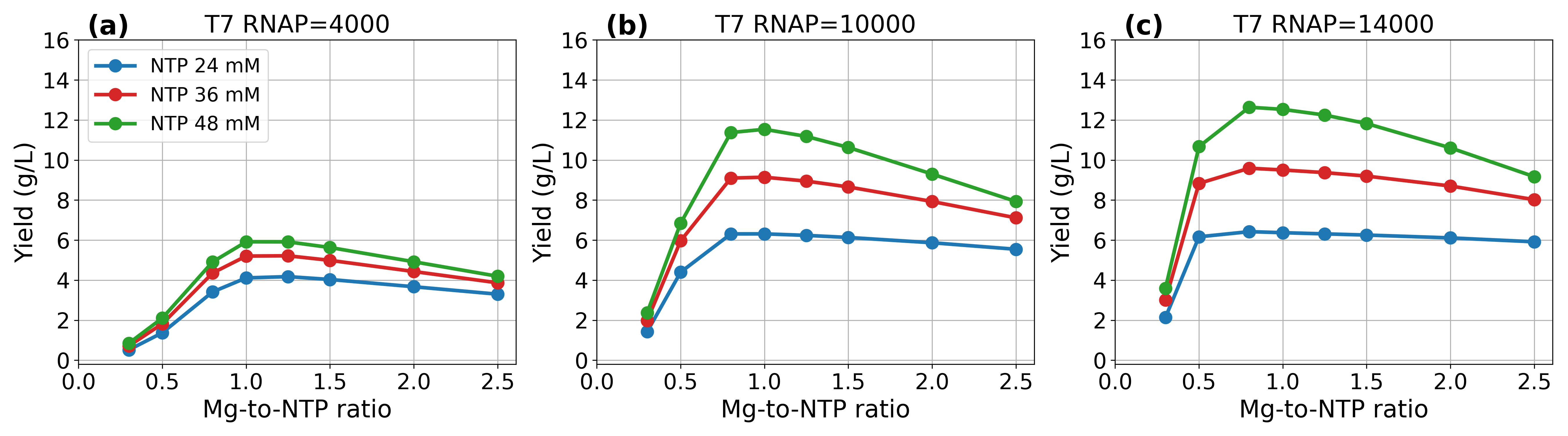}
    \caption{The effects of NTP concentration, Mg-to-NTP ratio, and T7 RNAP activity on mRNA yield at 150 minutes. 
    }
    \label{fig:Yield_MgNTP_NTP_T7}
\end{figure*}

\begin{figure*}[htb!] 
    \centering
    \includegraphics[width=0.9\textwidth]{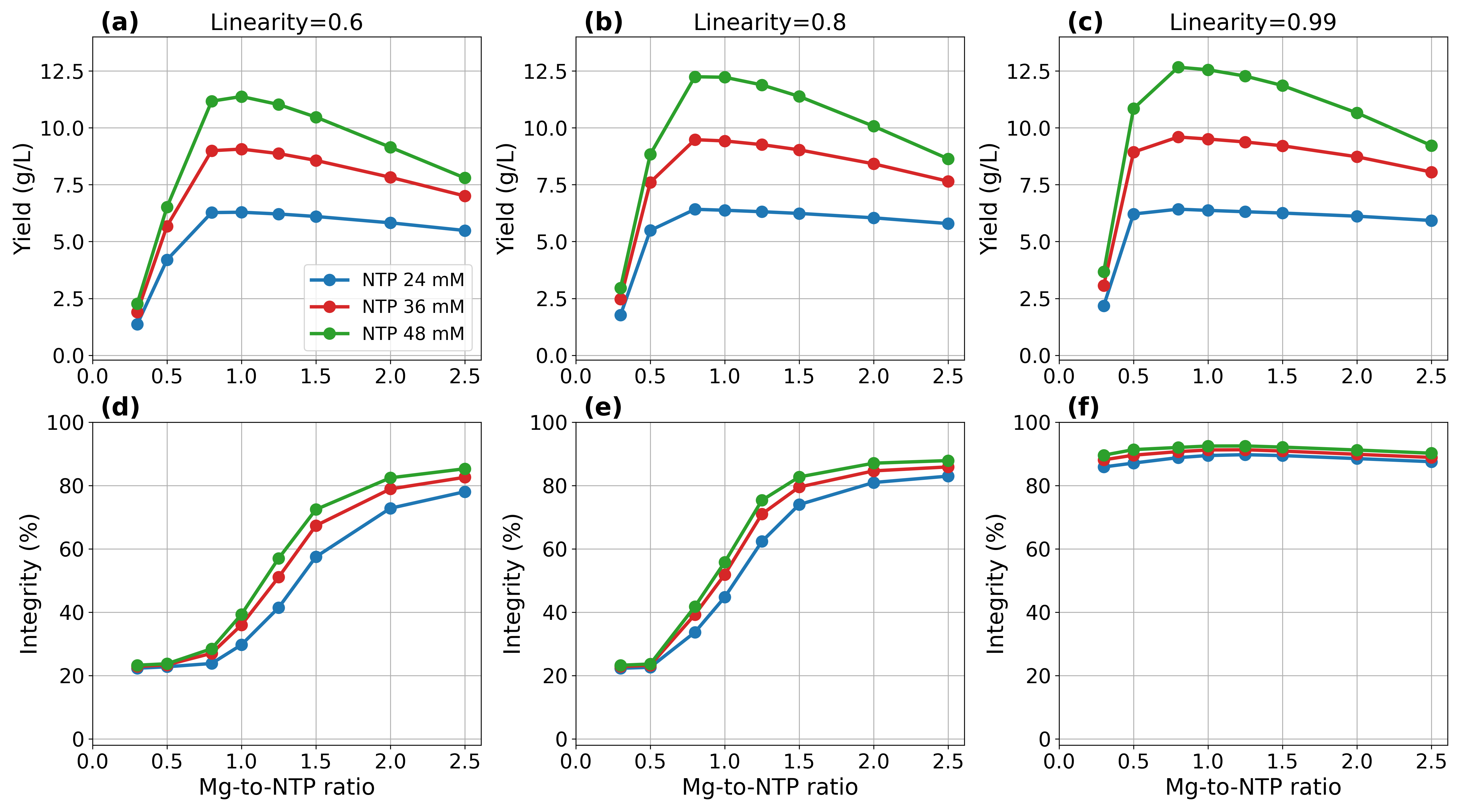}
    \caption{
    The effects of DNA linearity, Mg-to-NTP ratio, and NTP concentration (mM) on mRNA yield and integrity at 150 minutes. 
    Each column represents a different linearity level: (a),(d) 0.6; (b),(e) 0.8; and (c),(f) 0.99. 
    Top row: mRNA yield (g/L); Bottom row: mRNA integrity (\%).
    }
    \label{fig:Linearity_MgNTP_Mg}
\end{figure*}

\newpage
\begin{figure*}[htb!] 
    \centering
    \includegraphics[width=0.9\textwidth]{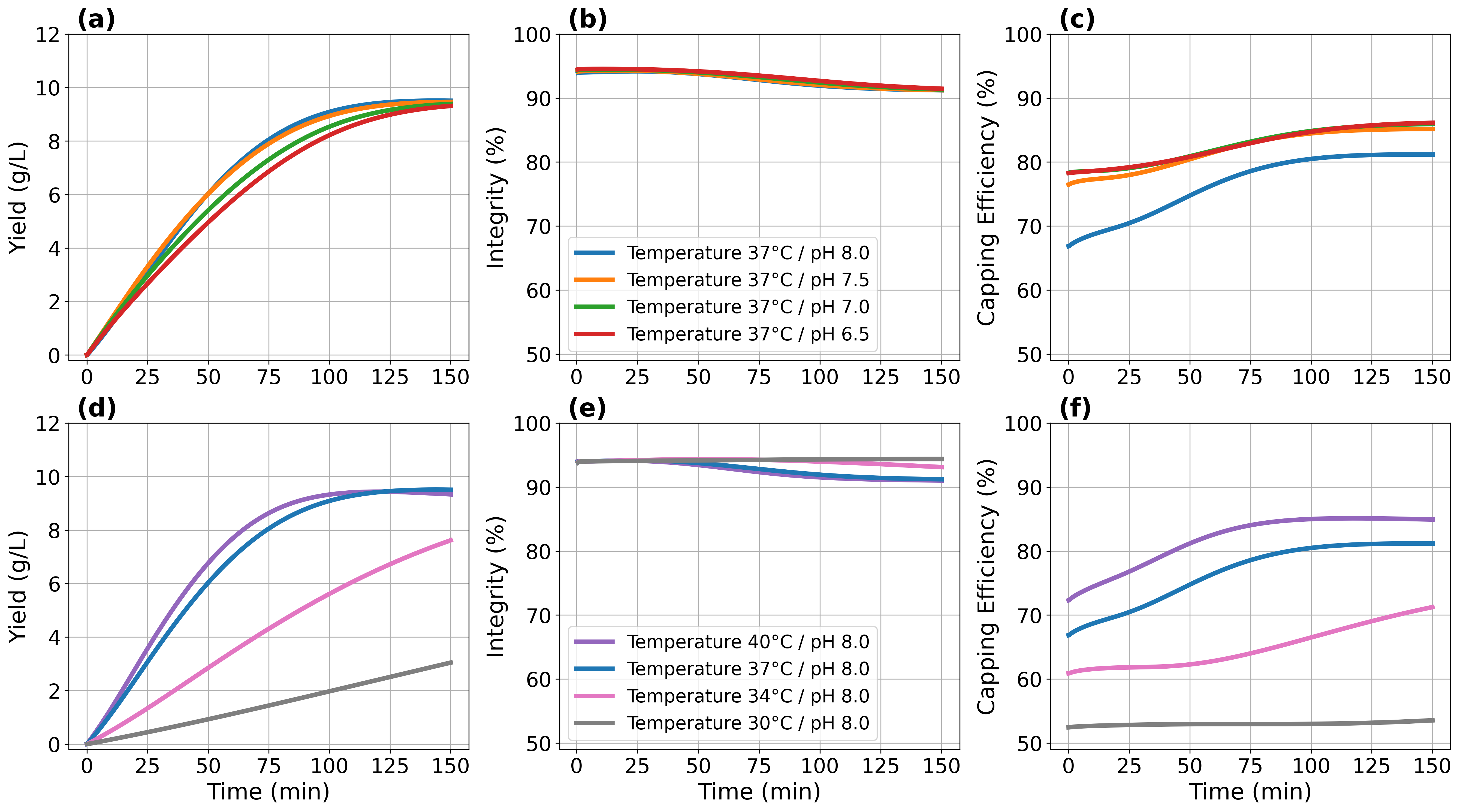}
    \caption{Predicted dynamic Profiles of the IVT Process Across Different pH and temperature. Illustrated trajectories include: (a) Yield, (b) Integrity and (c) Capping Efficiency. Top row: pH; Bottom row: Temperature ($^\circ$C). 
    }
    \label{fig:pH_trajectory}
\end{figure*}

\newpage
\section{Appendix: Sensitivity Analysis of Mechanistic Parameters}
\label{apppendix:SA-parameter}

Given the optimal $K$-dimensional mechanistic parameter vector $\pmb{\theta}^\star$, the sensitivity of each parameter with respect to the prediction error of process performance metrics was assessed by individually perturbing each parameter while holding the remaining parameters fixed at their optimal values. This analysis provides insight into the functional contribution of each module within the model.
Specifically, for the $k$-th parameter $\theta_k$, a symmetric perturbation of $\pm30$\% was applied to obtain perturbed values $\theta_k^{+}$ and $\theta_k^{-}$. The sensitivity of $\theta_k$ was quantified by the change in prediction error: $ \Delta_k = \pmb{e}(\pmb{x}; \theta_k^{+}, \pmb{\theta}^\star_{-k}) - \pmb{e}(\pmb{x}; \theta_k^{-}, \pmb{\theta}^\star_{-k})$, where $\pmb{\theta}^\star_{-k}$ denotes all components of $\pmb{\theta}^\star$ excluding $\theta_k$. A parameter $\theta_k$ was deemed critical if its relative sensitivity exceeds 10\%, that is, $\Delta_k \big/ \pmb{e}(\pmb{x}; \pmb{\theta}^\star) > 10\%$.
Table~\ref{tab:sensitivity_analysis} summarizes the critical mechanistic parameters influencing mRNA yield, integrity, and capping efficiency, evaluated at 150 minutes into the IVT process.

\textbf{Yield.} As noted by Xue et al. (2008) \cite{xue2008kinetic}, the RNA synthesis rate is primarily governed by the time T7 RNAP spends in the initiation and elongation steps. For long DNA templates, elongation becomes the rate-limiting step, and elongation-related parameters exert greater influence on mRNA yield. Nevertheless, accurate estimation of initiation kinetics remains essential, as initiation governs the formation rate of new elongation complexes (see Section~\ref{subsec:IVT}). Key mechanistic parameters include $k^\text{initial-cap}_\text{T7}$, $k^\text{initial-uncap}_\text{T7}$, $k^\text{elongate}_\text{T7,ATP}$, $k^\text{elongate}_\text{T7,UTP}$, $k^\text{elongate}_\text{T7,CTP}$, and $k^\text{elongate}_\text{T7,GTP}$. Additionally, the strong sensitivity of $K^\text{initial-uncap}_\text{M,Mg}$, $K^\text{initial-cap}_\text{M,Mg}$, and $K^\text{elongate}_\text{M,Mg}$ highlights the dependence of transcription on free Mg$^{2+}$ levels \cite{thomen2008t7,rosa2022maximizing}. The high sensitivity of temperature-related parameters—$Q_{10}^\text{elongate}$ and $\text{Temp}^\text{elongate}_\text{opt}$—underscores the importance of accurately representing T7 RNAP’s thermodynamic response, which significantly influences the overall transcription rate. Finally, the sensitivity of DNA template linearity-related parameters—$\beta^\text{linear}_\text{activity}$ and $\beta^\text{circle}_\text{activity}$—further emphasizes the impact of template structure on mRNA synthesis.

\begin{table*}[h]
\centering
\caption{Key mechanistic parameters influencing the prediction accuracy of IVT performance metrics.}
\label{tab:sensitivity_analysis}
\renewcommand{\arraystretch}{1.3}
\setlength{\tabcolsep}{8pt}
\begin{tabular}{p{3.5cm} | p{11cm}}
\toprule
\textbf{Performance Metric} & \textbf{Critical Mechanistic Parameters} \\
\midrule
\textbf{Yield (g/L)} & 
$k^\text{initial-cap}_\text{T7}$, $k^\text{initial-uncap}_\text{T7}$, $k^\text{elongate}_\text{T7,ATP}$, $k^\text{elongate}_\text{T7,UTP}$, $k^\text{elongate}_\text{T7,CTP}$, $k^\text{elongate}_\text{T7,GTP}$; 
$K^\text{initial-cap}_\text{M,Mg}$, $K^\text{initial-uncap}_\text{M,Mg}$, $K^\text{elongate}_\text{M,Mg}$;
$Q_{10}^\text{elongate}$, $\text{Temp}^\text{elongate}_\text{opt}$; $\beta^\text{linear}_\text{activity}$, $\beta^\text{circle}_\text{activity}$ \\
\midrule
\textbf{Integrity (\%)} & 
$K_{\text{M,MgNTP}_{min}}^\text{elongate}$, $k_\text{ratio}$; $\gamma_{\text{LMS}}^{\text{linear}}$, $\gamma_{\text{LMS}}^{\text{circle}}$; $K^\text{terminate}_{\text{M},\text{Mg}}$  \\
\midrule
\textbf{Capping Efficiency (\%)} & 
$k^\text{initial-cap}_\text{T7}$, $k^\text{initial-uncap}_\text{T7}$; $K^{\text{initial-cap}}_\text{M,Mg}$, $K^{\text{initial-uncap}}_\text{M,Mg}$; $\text{pH}^\text{initial-cap}_\text{opt}$, $\text{pH}^\text{initial-uncap}_\text{opt}$; $Q_{10}^\text{initial-cap}$, $\text{Temp}^\text{initial-cap}_\text{opt}$, $Q_{10}^\text{initial-uncap}$, $\text{Temp}^\text{initial-uncap}_\text{opt}$ \\
\bottomrule
\end{tabular}
\end{table*}



\textbf{mRNA integrity.} It is primarily governed by the most influential parameters including those associated with the truncated transcript synthesis model—namely, $K_{\text{M,MgNTP}_{min}}^\text{elongate}$ and $k_\text{ratio}$—as well as those from the read-through proportion model, such as $\gamma_{\text{LMS}}^{\text{linear}}$ and $\gamma_{\text{LMS}}^{\text{circle}}$. This aligns with mechanistic expectations, as these parameters directly govern the distribution of T7 RNAP activity among truncated, full-length, and LMS transcript synthesis. Furthermore, the high sensitivity of $K^\text{terminate}_{\text{M},\text{Mg}}$ underscores the importance of accurately quantifying the effect of free Mg$^{2+}$ on transcription termination and overall mRNA integrity.

\textbf{Capping efficiency.} It is primarily governed by the competitive dynamics between ATP/GTP and CleanCap during the initiation step. Accordingly, kinetic parameters associated with the capped and uncapped initiation pathways—namely $k^\text{initial-cap}_\text{T7}$ and $k^\text{initial-uncap}_\text{T7}$—exhibit high sensitivity. Additionally, the parameters regulating differential free Mg$^{2+}$ affinity between these pathways ($K^{\text{initial-cap}}_\text{M,Mg}$ and $K^{\text{initial-uncap}}_\text{M,Mg}$), as well as pathway-specific optimal pH values ($\text{pH}^\text{initial-cap}_\text{opt}$ and $\text{pH}^\text{initial-uncap}_\text{opt}$) and thermal stability factors ($Q_{10}^\text{initial-cap}$, $\text{Temp}^\text{initial-cap}_\text{opt}$, $Q_{10}^\text{initial-uncap}$, $\text{Temp}^\text{initial-uncap}_\text{opt}$), also strongly influence capping efficiency predictions.

\end{appendices}

\end{document}